\documentclass[reprint, amsmath,amssymb,aps,prd,unsortedaddress,superscriptaddress]{revtex4-2}
\usepackage{amsmath,amstext,amssymb, wasysym}
\usepackage{apjfonts} 
\usepackage{aas_macros}
\usepackage{xcolor, soul}
\usepackage{appendix}
\usepackage{orcidlink}
\usepackage{tabularray}
\usepackage{subcaption}
\usepackage{caption}
\DeclareMathOperator{\sinc}{sinc}

\begin{document}
\title{Localization of binary neutron star mergers with a single Cosmic Explorer}
\author{Pratyusava Baral\,\orcidlink{0000-0001-6308-211X
}}\email{pbaral@uwm.edu}\affiliation{University of Wisconsin-Milwaukee, Milwaukee, WI 53201, USA}
\author{Soichiro Morisaki\,\orcidlink{0000-0002-8445-6747}}
\affiliation{Institute for Cosmic Ray Research, The University of Tokyo, 5-1-5 Kashiwanoha, Kashiwa, Chiba 277-8582, Japan}
\author{Ignacio Maga\~na Hernandez\,\orcidlink{0000-0003-2362-0459}}
\affiliation{University of Wisconsin-Milwaukee, Milwaukee, WI 53201, USA}
\author{Jolien Creighton\,\orcidlink{0000-0003-3600-2406}}
\affiliation{University of Wisconsin-Milwaukee, Milwaukee, WI 53201, USA}

\begin{abstract}
    Next-generation ground-based gravitational-wave detectors, such as Cosmic Explorer (CE), are expected to be sensitive to gravitational-wave signals with frequencies as low as 5 Hz, allowing signals to spend a significant amount of time in the detector frequency band. As a result, the effects caused by the rotation of the Earth become increasingly important for such signals. Additionally, the length of the arms of these detectors can be comparable to the wavelength of detectable gravitational waves, which introduces frequency-dependent effects that are not significant in current-generation detectors. These effects are expected to improve the ability to localize compact binary coalescences in the sky even when using only one detector. This study aims to understand how much these effects can help in localization. We present the first comprehensive Bayesian parameter estimation framework that accounts for all these effects using \textsc{Bilby}, a commonly used Bayesian parameter estimation tool. We focus on sky localization constraints for binary neutron star events with an optimal signal-to-noise ratio of 1000 with one detector at the projected CE sensitivity. We find that these effects help localize sources using one detector with sky areas as low as 10 square degrees. Moreover, we explore and discuss how ignoring these effects in the parameter estimation can lead to biases in the inference. 
    
\end{abstract}
\maketitle

\section{Introduction}
The LIGO-Virgo-KAGRA (LVK) \citep{Aasi_2015, Acernese_2015, https://doi.org/10.48550/arxiv.2008.02921} collaboration has confidently detected around 90 compact binary coalescences (CBCs) which include binary black hole (BBH) \citep{LVKO3, 4OGC}, binary neutron star (BNS) \citep{PhysRevLett.119.161101, Abbott_2020} and neutron star black hole (NSBH) \citep{Abbott_2021} mergers. One BNS known as GW170817, had an observed electromagnetic (EM) counterpart, opening the door to the unexplored world of multimessenger astronomy \citep{Abbott_2017a, Abbott_2017b} with GWs allowing us to test our understanding of gravity, cosmology, and astrophysics \citep{P1,P2, P3}.

Given the success of current generation GW detectors, several new ground-based next-generation (3G/XG) GW detectors have been proposed, including the Cosmic Explorer (CE) \citep{Reitze2019Cosmic} and the Einstein Telescope (ET) \citep{Punturo_2010}, which are expected to be operational post-2030. Over the next decade, technological advancements are expected to significantly enhance the sensitivity of ground-based detectors, enabling them to detect frequencies as low as a few hertz. This would enable us to detect $\mathcal{O}(10^5-10^6)$ CBCs \citep{PhysRevD.100.064060} and in particular, signals with extremely high signal-to-noise ratios (SNRs) in the order of $\mathcal{O}(1000)$ within one year of observation. 

Increased sensitivity at lower frequency means that loud gravitational-wave signals from BNS will last in the detector band for about an hour allowing the source to move across the sky relative to Earth's rotation. Long detector arms compel us to calculate the travel time of a GW across the detector beyond the static limit where the wavelength of a gravitational wave is assumed to be much longer than the arms of the detector \citep{malik2008, malik2009}. These effects make the antenna response time and frequency-dependent, which breaks certain degeneracies that otherwise exist between extrinsic parameters (those relating to the relative position and orientation of the detector and source). This enables us to localize sources using only one detector. Locating a source in the sky is extremely important to facilitate EM follow-up. Given the length of the signal, it might be feasible to localize the source before the merger which is essential for observing prompt afterglows \cite{Sachdev_2020}. However, in this paper, we work with the full bandwidth of signals lasting up to the merger.

A few localization studies using Fisher Matrices in XG detectors exist in the literature \citep{PhysRevD.97.064031, PhysRevD.97.123014}. Such an approximation, though accurate at some regions of the parameter space, may not generally be valid even at high SNRs \citep{PhysRevD.77.042001}. For a single detector, we expect multimodalities in the right ascension ($\rm{RA}$) and declination ($\rm{dec}$) which is completely neglected by Fisher matrix estimates and hence inadequate. Recent work by \citet{Nitz2021} and \citet{PhysRevLett.127.081102} performs Bayesian parameter estimation (PE) for BNS mergers in XG detectors. To make the problem computationally feasible, the former work constructs a heterodyned likelihood  taking into account all effects due to the rotation of the Earth to study early warning capabilities. The latter work constructs reduced order models taking into account only the amplitude modulations due to Earth-rotation using BNS signals lasting 90 minutes in-band from 5 Hz to 2048 Hz for a network of two Cosmic Explore detectors and a single Einstein Telescope (a proposed triangular ground-based detector). Both of these studies 
ignore the high-frequency effects due to the size of the detector. It is not clear how ignoring some effects play a role in parameter recovery and so we include these effects in our analysis. For space-based detectors like the Laser Interferometer Space Antenna (LISA) \citep{https://doi.org/10.48550/arxiv.2008.02921}, similar studies have been performed \citep{PhysRevD.103.083011}. The physics of finite size effects remains the same, and the rotation of Earth effect is replaced by similar effects due to the revolution of LISA around the Sun. However, the implementation varies as LISA operates in a very different frequency range and the detector shapes and sizes are vastly also different. 

This work does a proof of concept localization study using comprehensive Bayesian parameter estimation for BNS mergers at an SNR of 1000 using simulated data with a single CE detector.  All the effects due to the detector size and the rotation of the Earth have been taken into account for the construction of the simulated dataset. Unlike previous studies, we implement all effects relevant to CE, allowing all parameters other than spin to vary during the parameter estimation.  Approximations to the likelihood \citep{PhysRevD.104.044062} are made so that the PE can be performed in reasonable timescales. It's worth stressing that ignoring the effects due to the detector size and the Earth rotation in PE will not only cause poor localization but also lead to the biased estimation of parameters. This study aims to understand the localization capabilities of a single CE due to the above-mentioned effects and develop a complete Bayesian framework to study these effects. We use one detector to prevent the formation of a baseline that can triangulate sources and thus infer their position in the sky.

\section{Instrument Response}
The beam pattern function{\color {blue}s} ($F_{+,\times}$) of a present-generation detector depend on the unit vector $\textbf{n}$ describing the direction of propagation of the GW, and the polarization angle $\psi$. In CE the Earth-rotation effects make $\textbf{n}$ time-dependent and the detector size makes the beam pattern function depend on the gravitational-wave frequency $f$ and the unit vector $\textbf{e}$ along the arms of the detector. Each of these effects are explained in the next sections. The beam pattern function can further be factorized into two terms, the detector tensor $D^{ij}$ and the polarization tensor ($\varepsilon_{+,\times}$)
\begin{equation}\label{BeamPatternb}
F_{+,\times}(\textbf{n}(t), \textbf{e}, \psi, f) \equiv \varepsilon_{+,\times ij}(\textbf{n}(t),\psi) D^{ij}({{\bf n}(t)\cdot{\bf e}},f)
\end{equation}
\subsection{Effects due to Earth's rotation}
The Earth rotates 15 degrees in a stellar hour, which means the position of the source changes as observed in a frame that co-rotates with the Earth, for example, the Earth-Centered Earth-Fixed (ECEF) frame. This makes the unit vector to the source $\textbf{n}(\theta, \phi(t))$ change with time in this frame. The azimuthal angle $\phi$ of a source changes with time while the latitude $\theta$ is constant. The polarization tensor is given by \citep{PhysRevD.63.042003},
\begin{eqnarray}
({\varepsilon}_+)_{ij} &=& ( {\mathbf{X}} \otimes {\mathbf{X}} -
{\mathbf{Y}} \otimes {\mathbf{Y}})_{ij} \\ ({\varepsilon}_\times)_{ij} &=&
( {\mathbf{X}} \otimes {\mathbf{Y}} + {\mathbf{Y}} \otimes
{\mathbf{X}})_{ij}  .
\end{eqnarray}
where $\mathbf{X}$ and $\mathbf{Y}$ are the axes of the wave frame, given by
\begin{eqnarray}
{\mathbf{X}} &=& (\sin \phi \, \cos \psi - \sin \psi \, \cos \phi \, \cos
\theta)\; {\mathbf{i}} \nonumber \\
&& 
-(\cos \phi \, \cos \psi  + \sin \psi \, \sin \phi \, \cos \theta)\;
{\mathbf{j}} +
\sin \psi \, \sin \theta \; {\mathbf{k}} \label{e:waveX}\\
{\mathbf{Y}} &=& (- \sin \phi \, \sin \psi - \cos \psi \, \cos \phi \, \cos
\theta)\;  {\mathbf{i}} \nonumber \\
&& 
  +( \cos \phi \, \sin \psi - \cos \psi \, \sin \phi \, \cos \theta) \;
  {\mathbf{j}}
 + \sin \theta \, \cos \psi\;  {\mathbf{k}} \;  \label{e:waveY}
\end{eqnarray}
Here $\psi$ is the polarization angle and \textbf{i}, \textbf{j}, and \textbf{k} are unit vectors along $x$, $y$, and $z$ axis respectively in the Earth-fixed frame. The temporal dependence on the azimuthal angle makes the polarization tensor ${({\varepsilon}_{+,\times})}_{ij}$ time-dependent.     

The rotation of the Earth also affects the term $2\pi f t_{\rm{CE}}$ in the phase of CBC signal as the time of arrival at the detector ($t_{\rm{CE}} \equiv t_{\oplus} - \Delta t (\theta, \phi(t))$) changes with Earth's rotation, which appears in the phase of a CBC signal \citep{PhysRevD.85.122006}. Here $t_{\oplus}$ is the signal arrival time at the center of the Earth and $\Delta t (\theta, \phi(t))$ is the GW traveling time from the detector to the center of the Earth,
\begin{equation}
\Delta t (\theta, \phi(t)) \equiv -\frac{\bf{n}(\theta, \phi(t)) \cdot \bf{d}}{c}
\end{equation}
where $\bf{d}$ is the vector pointing from the center of the Earth to the detector. 

The likelihood is easier to compute in the frequency domain as we assume that the noise is stationary and Gaussian and so the time dependences need to be converted to frequency dependences. We assume a quasimonochromatic GW (restricting our attention to dominant quadruple (2,2) mode) for which there exists an invertible function $t(f)$ relating the instantaneous frequency of the signal to the time before the end of the signal, as would be obtained using the stationary phase approximation of a frequency-domain waveform. With such a relation, we can associate the time-dependent change in the direction of propagation (in an Earth-fixed frame) to a particular frequency of the signal.  The frequency for every time is computed up to the second post-Newtonian order \citep{PhysRevD.52.848}. A multi-harmonic signal would require each harmonic to be considered separately in this formalism. For parameter estimation purposes we parameterize the sky using the right ascension (RA $\equiv \phi+\mbox{GMST}$), declination ($\rm{dec} \equiv \pi/2-\theta$) and GMST. Thus the rotation of the Earth makes $\textbf{n}$ a function of time and hence a function of frequency. 
\subsection{Effects due to the finite detector size}
Any interferometric gravitational-wave detector measures the relative change in its arm length. For a detector with long arms, the temporal dependence of a GW signal along the arm of a detector becomes important. Ignoring these effects in the detector response and using the static limit, under the assumption that the wavelength of a GW is long compared to the arm length as used in the present-generation detectors, introduces systemic biases in localization larger than the statistical error for a typical BNS \citep{PhysRevD.96.084004}. This effect adds an external phase shift proportional to the detector scalar $D(\textbf{n}\cdot {\bf e},f)$ given by \citep{malik2008, malik2009, doi:https://doi.org/10.1002/9783527636037.ch6}, 
\begin{equation}\label{equation: D}
\begin{aligned}
D(\textbf{n}\cdot {\bf e},f) &=\exp\{{\pi i f L(1-\textbf{n} \cdot \textbf{e})/c}\} \sinc\{\pi f L(1+\textbf{n} \cdot \textbf{e})/c\}\\
&+ \exp\{{-\pi i f L(1+\textbf{n} \cdot \textbf{e})/c}\} \sinc\{\pi f L(1-\textbf{n} \cdot \textbf{e})/c\}
\end{aligned}
\end{equation}
where $\sinc(x) \equiv \sin(x)/x$, $L$ is the arm length of the detector, and $c$ is the speed of light. Present generation detectors operate in the limit of $fL/c \to 0$, resulting in $D \to \frac{1}{2}$ which is consistent with the static long wavelength approximation. Note that $\textbf{n}$ changes with time as we adopt an Earth-fixed coordinate, thereby putting additional frequency dependence. For a collection of arms, it's convenient to define a detector tensor $D_{ij}$ as
\begin{equation}
    D^{ij} \equiv D(\textbf{n},f) e_x^i e_x^j - D(\textbf{n},f) e_y^i e_y^j
\end{equation}
where $e_x^i$ and $e_y^j$ are the components of the unit vectors $\textbf{e}$ pointing along the two arms of the detector. 

\subsection{The noise spectral density}
We use the low-frequency optimized PSD obtained from \citep{Srivastava_2022} for a CE with arm-length 40 km. {The detector is placed at the LIGO-Hanford site.} 

\section{Methodology}

\subsection{Construction of Simulated Injections}
Simulated signals are calculated with the TaylorF2 waveform model with tidal corrections \citep{PhysRevD.80.084043}. Distances to simulated sources are determined such that their optimal SNRs are 1000. Choosing lower SNRs leads to poor sky localizations (>10000 square degrees) for some injections. Spins of simulated sources are assumed to be vanishing. For random instrumental noise, we assume the so-called zero-noise realization, where random noise happens to be vanishing.
The observed signal is given by,
\begin{equation}
\begin{aligned}
\Tilde{h}(f) = &\Big[  F_+({\rm RA},{\rm dec},\psi,f) h_+(f, \theta_{\rm{JN}}, d_{\rm L}, \phi_c, {\Theta}_I) \\
& ~~ + F_\times({\rm RA},{\rm dec},\psi,f) h_\times(f, \theta_{\rm{JN}}, d_{\rm L}, \phi_c,{\Theta}_I) \Big] \\
&\times \exp\left[-2\pi i f \left(t_{\oplus} - \Delta t ({\rm RA}, {\rm dec},f)\right)\right]
\end{aligned}
\end{equation}
${\Theta}_I$ refers to the set of intrinsic parameters comprising of the redshifted chirp mass $\mathcal{M}_z$, the mass ratio $q$, tidal deformability parameters, $\tilde{\Lambda}$ and $\delta \tilde{\Lambda}$. The angle, $\theta_{\rm JN}$ is between the line of sight to the binary and its total angular momentum vector, $d_{\rm L}$ is the luminosity distance and $\phi_c$ is the phase at coalescence. We do two sets of 48 injections corresponding to $\theta_{JN}$ equal to 0.2 and 1.0 for a source frame 1.4-1.44 $M_\odot$ BNS system. For $\theta_{JN}$ equal to 0.2 the values of injected $D_L$ vary from 38 Mpc to 154 Mpc while for $\theta_{JN}$ equal to 1.0 $D_L$ takes values from 22 Mpc to 88 Mpc. Details about the injections and the priors used for the results are summarized in Table \ref{fig:table}. We use the Planck15 \cite{P15} fiducial cosmology for conversion between $d_{\rm L}$ and redshift ($z$) throughout. The antenna response has been computed for every frequency assuming a single CE detector with a 40 km arm length placed at the LIGO-Hanford site with the same orientation.

\subsection{Accelerating the likelihood evaluations}
Repeated generation of waveforms for likelihood computation is highly time-consuming due to the long duration of the signal, making it computationally infeasible to perform Bayesian parameter estimation without any approximate methods. A multibanding technique, a form of adaptive sampling of the waveform, developed by \citet{PhysRevD.104.044062} was employed to reduce the computational cost roughly by a factor of 500. In this method the total frequency range is divided into several overlapping frequency bins. Each bin is sampled appropriately to construct the noise-averaged inner product of the data with the template. Using downsampled frequencies implies fewer waveform evaluations speeding up the whole process. The noise averaged inner product of the template with itself is computed on downsampled waveform values and linearly interpolated. The absence of higher multipoles to the waveform ensures that this function is smooth enough to be approximated by a linear interpolator. A typical likelihood evaluation on an Intel(R) Xeon(R) Gold 6152 x86\_64 CPU with a clock rate of  2.10GHz takes 20 ms with waveform evaluations at around 33000 frequency points. The errors of log-likelihood introduced by this approximation are much less than unity. 

\subsection{Sampling techniques for PE}
\begin{table*}
\centering
\begin{tblr}{
  cell{6}{3} = {r=2}{},
  hline{1,13} = {-}{0.08em},
  hline{2} = {-}{},
}
Parameters                                                                                                                                                                       & Unit                            & {Injected Value}                                   & Prior   & Minimum                                        & Maximum                      \\
$\mathcal{M}_z$                                                                          &M$_{\odot}$   & 1.24 (1+z$_{\rm,inj}$) & Uniform & $\mathcal{M}_{z,{\rm inj}}-10^{-5}$                          & $\mathcal{M}_{z,{\rm inj}}+10^{-5}$         \\
q                                                                                                                                                                                & -                               & 0.972                                                 & Uniform & 0.125                                      & 1                        \\
$\tilde{\Lambda}$                                             & -                               & obtained using a SLY                                          & Uniform & $\tilde{\Lambda}_{,{\rm inj}}-200$                                       & $\tilde{\Lambda}_{,{\rm inj}}+200$                      \\
$\delta \tilde{\Lambda}$ & -                               &  {equation of state \citep{refId0} }                                               & Uniform & $\delta \tilde{\Lambda}_{,{\rm inj}}$-5000                                      & $\delta \tilde{\Lambda}_{,{\rm inj}}$+5000                     \\
RA                                                                                                                                                                               & rad                            & {48 points uniformly \\distributed in the sky}        & Uniform & 0                                          & 2$\pi$  \\
dec                                                                                                                                                                              & rad                            &                                                       & Cosine  & $-\pi/2$                  & $\pi/2$ \\
$d_{\rm L}$                                                                                                                                   & Mpc                             & {Calculated such that SNR is 1000}                    & Quadratic  & 1 & $d_{\rm L,inj}+$2000                     \\
$\theta_{\rm JN}$                                                                                                                             & rad                            & \{0.2,1.0\}                                       & Sine    & 0                                          & $\pi$   \\
$t_{\rm CE}-t_{\rm CE,inj}$                                                                                                                & sec                               & 0                                                     & Uniform & -0.01                                      & 0.01                     \\
$\psi$                                                                                                                                                          & rad                            & $\pi/2$                              & Uniform & 0~                                         & $\pi$   \\
$\phi_c$                                                                                                                                                       & rad                            & $\pi/2$                              & Uniform & 0                                          & 2$\pi$  
\end{tblr}
\captionsetup{format=hang,justification=centerlast, indention=-1.5cm}
\caption{The injection values, priors, and prior ranges on different parameters are summarized.  We denote by Cosine and Sine uniform priors on each respectively. The subscript $_{\rm ,inj}$ refers to the injected value of the parameter. The injected source frame chirp mass is 1.24. The points in the sky have been chosen following the \textsc{HEALPIX} convention with an $N_{\rm{side}}=2$ corresponding to 48 distinct sky locations. All injections end at a geocentric time GMST = 0. For sampling, we use a uniform prior on $d_{\rm L}$ with the same boundaries however, the posteriors are then reweighted with a quadratic prior as described in the text. \phantom{Hi, how are you}
}
\label{fig:table}
\end{table*}

We sample the 10-dimensional parameter space using \textsc{Bilby} and use \textsc{DYNESTY} \citep{dynesty_paper} as a nested sampler. The parameter space is complicated enough that the standard sampler settings using 1000 live points ($n_{\rm LIVE}$) are inadequate. Convergence of a nested sampler is controlled by $n_{\rm LIVE}$ and the length of the Markov-Chain-Monte-Carlo (MCMC) chain in the unit of its auto-correlation length $n_{\rm ACT}$. We use $n_{\rm LIVE}$ = 4000, $n_{\rm ACT}$ = 5 and fix the maximum MCMC length to 5000. Increasing $n_{\rm LIVE}$ to 5000 does not change the posteriors, confirming we have adequate live points to capture the features of the parameter space. { Since we are using simple non-spinning TaylorF2 waveforms, we use $n_{\rm ACT}$ = 5. Increasing it to 10 does not significantly alter the extrinsic parameter posteriors.} A nested sampler also calculates the evidence at every iteration. The sampling stops when the relative difference of two successive evidences is less than 1 $\times 10^{-5}$. Such a small value is chosen to guarantee convergence. Typically we obtain $10^6$ samples with $10^8$ likelihood evaluations, before meeting the exit criteria with a wallclock runtime of fewer than 3 days using 16 CPUs and n$_{\rm pool}$ = 32. To reduce the cost of generating skymaps, we downsampled posterior samples to 1000.

The constant phase $\phi_c$ of a signal is not important and is analytically marginalized \citep{thrane_talbot_2019}. To ensure the special functions that implement the analytical phase marginalization are accurate enough, we perform PE using nested sampling without phase marginalization and use a large number of live points, that is $n_{\rm LIVE}$ = 6000. However, we noticed no change in posteriors confirming our results are robust. Although we set a uniform prior in $d_{\rm L}$ for sampling, all samples have been reweighted by the standard quadratic prior in $d_{\rm L}$ to obtain the results presented in this paper. The reweighted prior, which is commonly used in the literature, is equivalent to a uniform in comoving volume prior in the local Universe. The antenna pattern is a function of frequency. For PE we update this function at frequency intervals corresponding to a time difference of 4s to save computational resources. Changing this to 1s has no effect on the posterior as the error due to this interpolation is less than the statistical error in our analysis. 

\section{Results}

To quantify biases of ignoring Earth rotation and finite detector size, we performed a parameter estimation run by ignoring those effects. {{Appendix \ref{appendix B} shows the sky-location and luminosity distance posteriors ignoring these effects for some injection parameters.}} The obtained results are { often} biased because the antenna patterns used for the analysis do not capture the effects due to Earth's rotation and the detector size present in the simulated dataset.  Hence, incorporating the effects mentioned above is essential for unbiased PE in the CE era. 

\begin{figure}[htb]
    \includegraphics[width=0.5\textwidth]{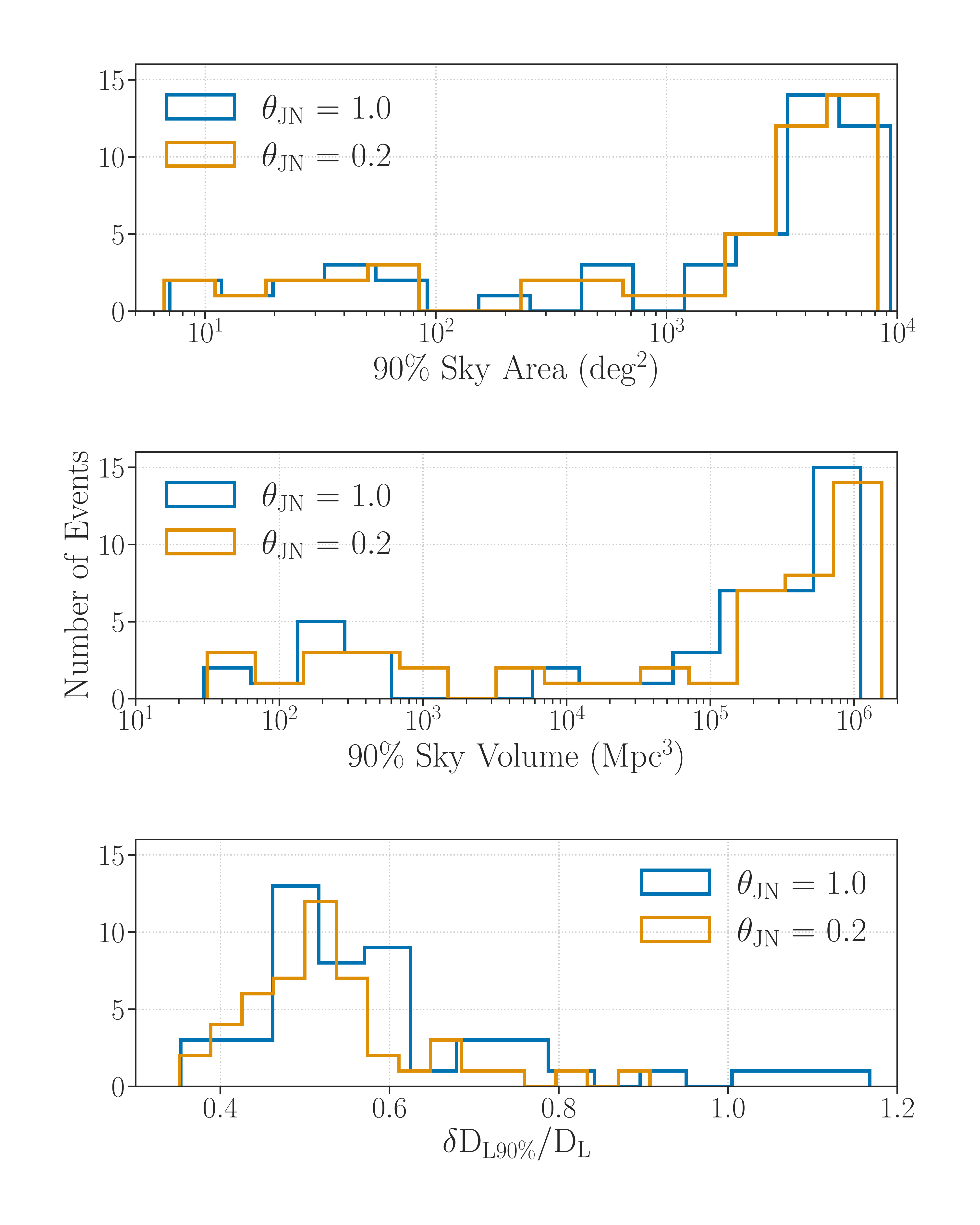}
    \caption{Histograms of 90\% sky areas (top panel), 90\% cosmic volumes (middle panel) and 90\% fractional distance uncertainties (bottom panel) for two sets of 48 injections corresponding to two values of inclination angles $\theta_{\rm{JN}} = 0.2$ shown in orange and $\theta_{\rm{JN}} = 1.0$ shown in blue. \phantom{Alingments are a pain}}
    \label{fig:Histo}
\end{figure}

We find the effects due to the rotation of the Earth and the size of the detector helps in localizing sources in the sky. Figure \ref{fig:Histo} shows the histogram of the fractional distance uncertainties, 2D sky areas, and 3D sky volumes for gravitational-wave events. The fractional distance uncertainties are computed by taking the ratio of the difference of 95 and 5 percentiles of the distance posterior, and the median distance. A typical skymap {} is shown in Figure \ref{fig:snr1000_typ}. The skymap is not unimodal, which is a direct artifact of using a single detector, making Fisher analyses unreliable. The individual skymaps for all the injections are presented in Appendix \ref{appendix A}.

\begin{figure}[htb]
    \includegraphics[width=0.5 \textwidth,height=\textheight,keepaspectratio]{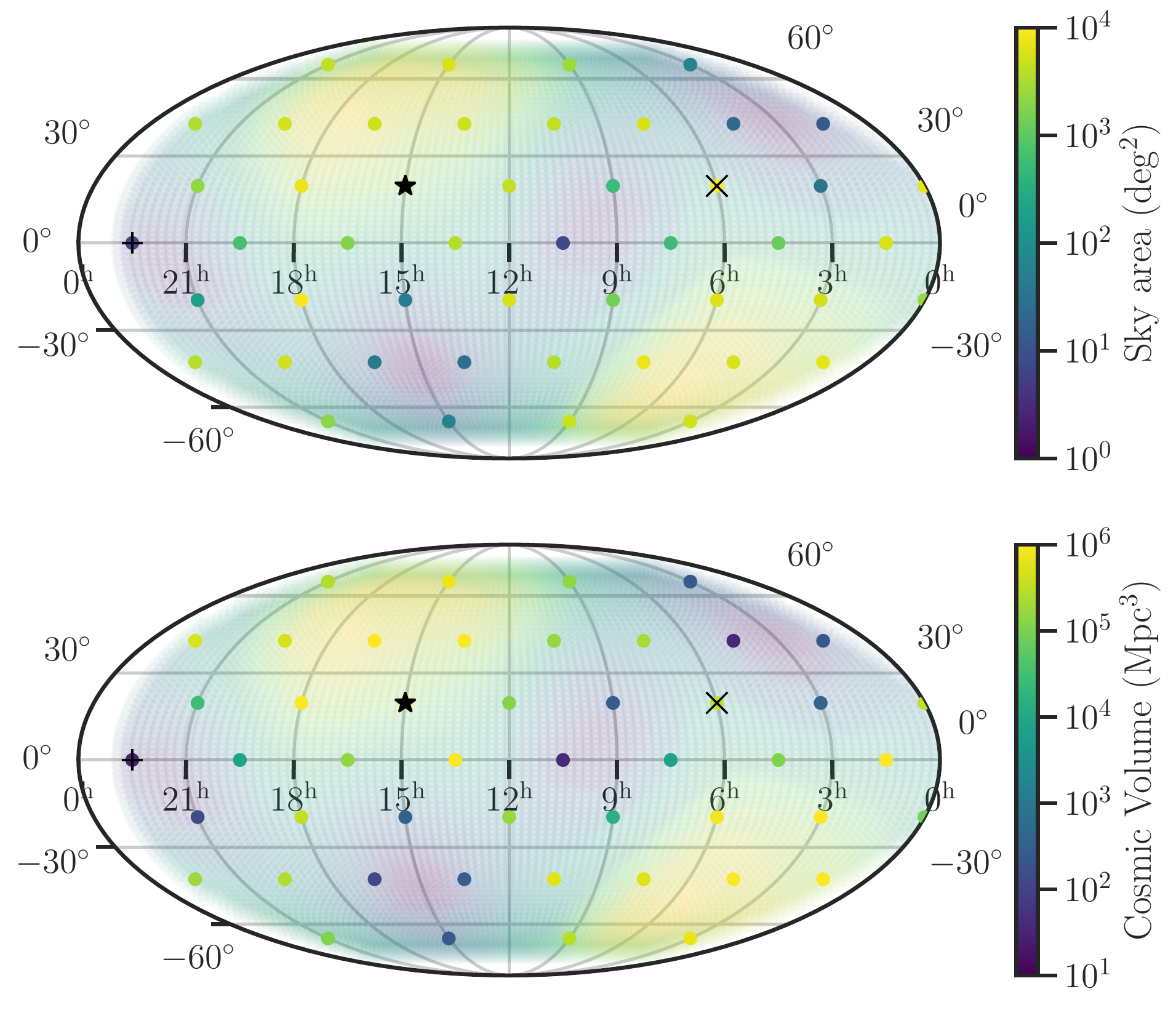}
    \caption{Distribution of 90\% sky areas/volumes (denoted by colored dots at the injected positions) overlayed on the antenna pattern (where bright yellow represents directions of high average sensitivity where darker blue represents directions of low average sensitivity). Note that the color bar applies to the plotted points rather than the overlayed antenna pattern. The + and $\times$ sign denotes the best and worst localization respectively. The skymap of the source marked with a $\star$ is plotted in Figure \ref{fig:snr1000_typ}. \phantom{Hi what's up and how are you doing?}} 
    \label{fig:Sky}
\end{figure}

\begin{figure}[htb]
    \includegraphics[width=0.5\textwidth,height=\textheight,keepaspectratio]{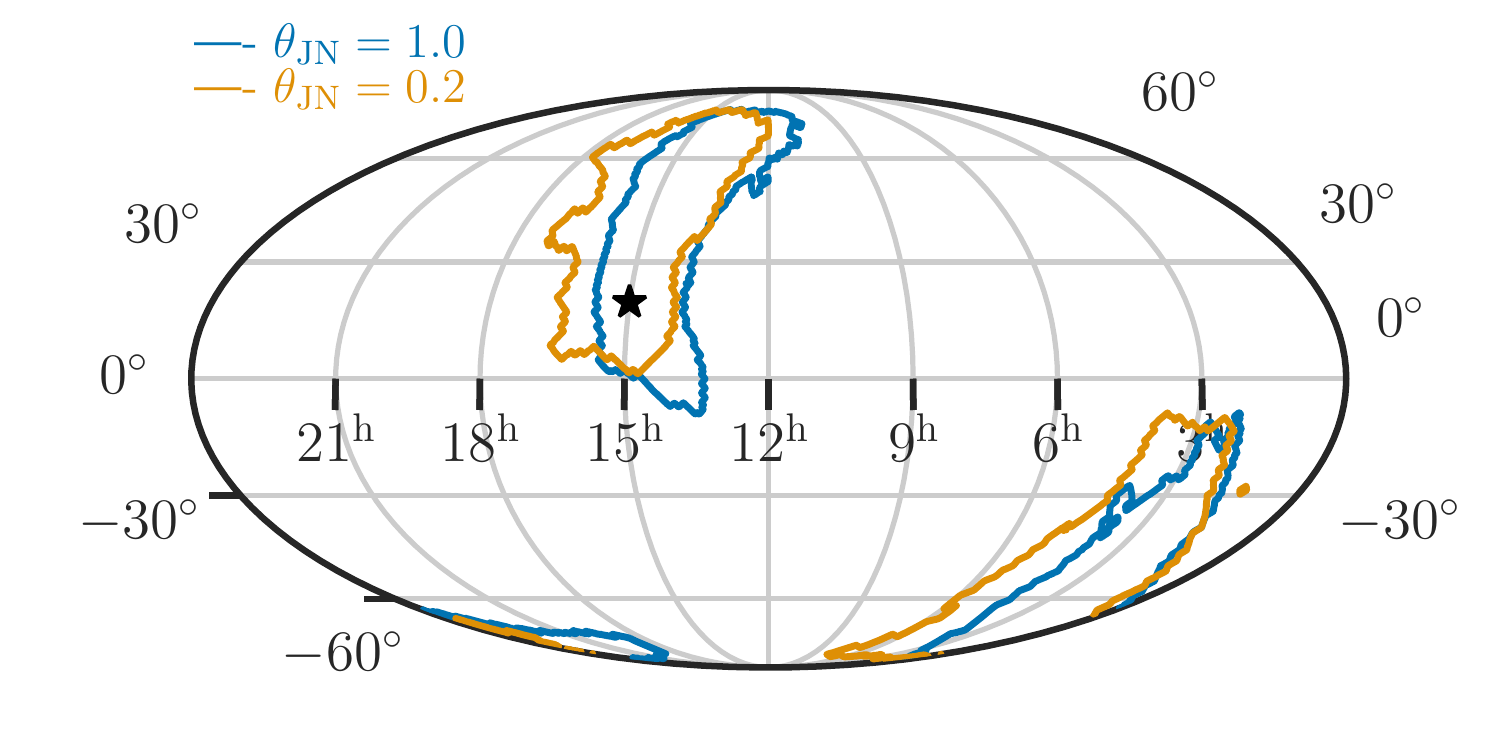}
    \caption{Skymap at the $\star$ position in Figure \ref{fig:Sky}. This is a typical skymap at an SNR of 1000. See Appendix \ref{appendix A} for sky localization posteriors for every injection considered in this study.}
    \label{fig:snr1000_typ}
\end{figure}

Figure \ref{fig:Sky} shows the distribution of the 90 percent credible regions of sky area and sky volume for 48 values of RA and dec with $\theta_{\rm{JN}}$ equal to 1.0. For 3G detectors, as discussed earlier, the antenna pattern changes as a function of frequency. To get a sense of the integrated antenna response we can plot the SNRs fixing $d_{\rm L}$ or vice versa. Both of these plots are exactly the same, as SNR $\times$ $d_{\rm L}$ is a constant for a particular event. We plot this integrated antenna response in the background of Figure \ref{fig:Sky} with yellow denoting the bright spots (directions towards which the detector is more sensitive) and blue denoting the dark spots (directions towards which the detector is less sensitive). The distribution of sky areas and sky volumes over the sky do not vary much with inclination provided the SNR is fixed. 

\begin{figure*}[htb]
    \centering    \includegraphics[width= \textwidth,height=\textheight,keepaspectratio]{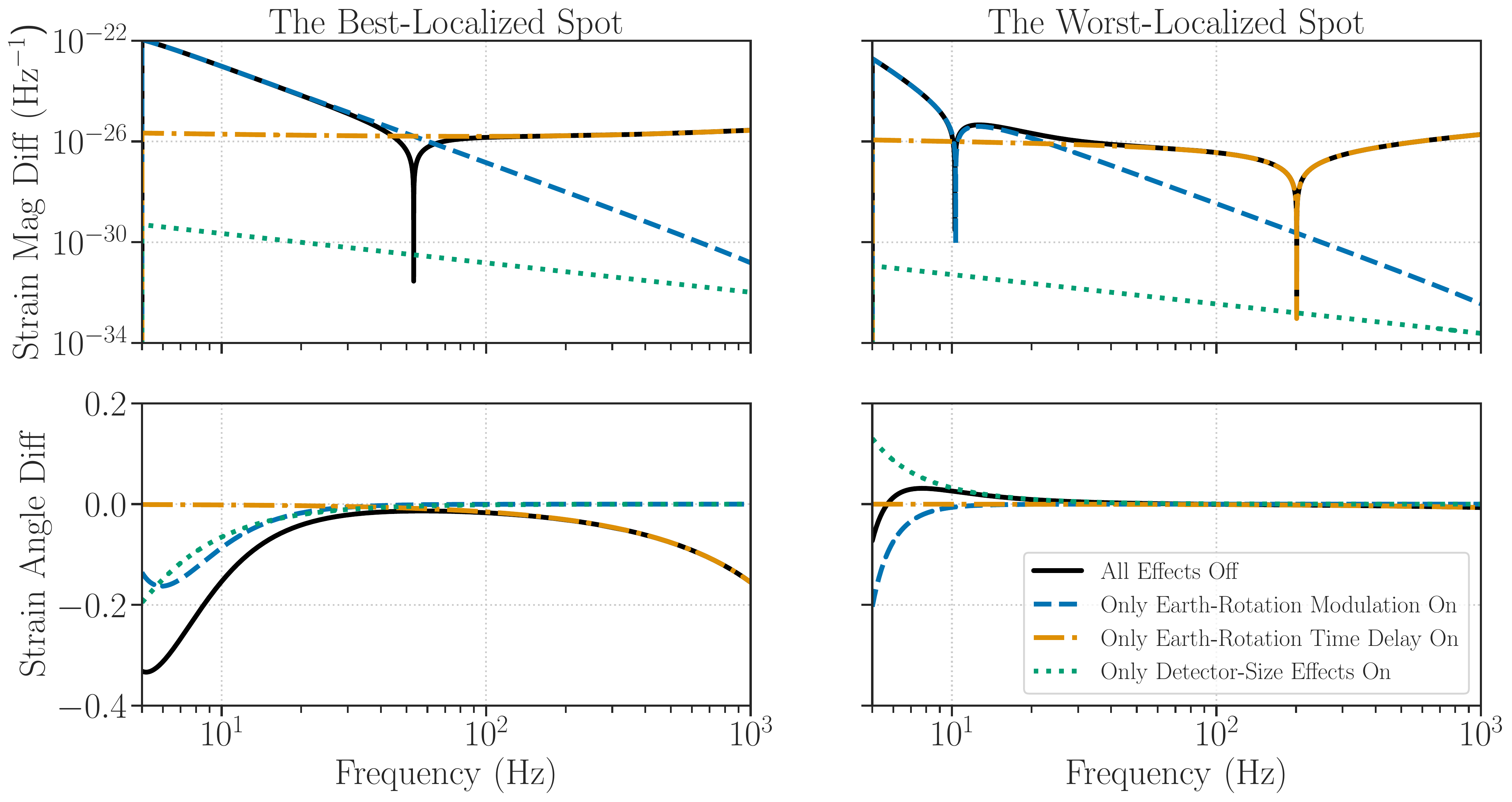}
    \caption{The solid black line represents the absolute value of the difference between the magnitude (top panel: $||h_{\rm CE}|-|h_{\rm ig}||$) and phase (bottom panel: ${\rm angle ~ of}~ h_{\rm CE}/h_{\rm ig}$) of two waveforms, one taking into account all the effects due to the Earth's rotation and detector size ($h_{\rm CE}$) and the other ignoring the same effects ($h_{\rm ig}$). The dashed blue line, dash-dotted orange line, and dotted green line plot the same replacing $h_{\rm CE}$ by a waveform that takes into account only the amplitude modulation due to the rotation of the Earth in CE, the size of CE, and the phase modulation due to rotation of the Earth in CE respectively. The left and right panels correspond to the best and worst sky localizations at $\theta_{\rm{JN}}$ = 1.00. \phantom{Finally, the last figure................Oh wait, appendix is still there.}
    }
    \label{fig:waveform_compare}
\end{figure*}

A degenerate mode may exist in the posterior distributions of the sky positions ($\theta, \phi$), as in Figure \ref{fig:snr1000_typ}, at  $\phi \to \phi - \pi$ and $\theta \to -\theta$, i.e. under the reflection of the wave vector about the CE plane. This is an exact symmetry of the beam pattern function in the absence of the rotation of Earth and detector size effects.  The addition of the above-mentioned effects tries to choose the 'correct'  mode out of all the degenerate modes. This symmetry is also present in the distribution of sky areas and volumes in Figure \ref{fig:Sky}, a direct consequence of the symmetries in the underlying beam pattern function. 

As seen in Figure \ref{fig:Sky}, localization on dark spots is better than bright spots mainly because the frequency dependence of the beam pattern function is more pronounced at dark spots \citep{PhysRevD.96.084004}. Note that the distance to the source is adjusted so that the SNR is always 1000, thereby the effect of SNR on localization is kept constant. Signals from the dark spots are also longer, as the sources are nearer resulting in a lower detector frame mass. The length of the signals varies for about 2.5\% at $\theta_{JN}$ = 1.0. The Earth rotation effect is thus slightly more influential at the dark spot. Moreover, the { quadratic prior on luminosity distance prefers sources that are further, and thus sources on the dark spot are penalized by the prior. This shrinks the 3D sky volume, which in turn reduces the sky area.} All of these three reasons combine to give better localization at the dark spot. 

The four dark spots are not identical due to the orientation of the detector. The dark spot lying in between the arms of the detector is equivalent to the spot diametrically opposite to it, but different from the remaining two. This is primarily due to the frequency-dependent beam patterns. So rotating the detector by $\pi/2$ radians shall change the pattern of distribution of sky areas and volumes among the dark spots and a rotation of $\pi$ radians shall keep the distribution of sky areas and volume invariant. This has been confirmed by PE runs. 

The distribution of 3D sky volumes as shown in the bottom panel of Figure \ref{fig:Sky} follows the same distribution of sky areas. Sources in the dark spots are nearer as SNRs are held constant and have lower inferred $d_{\rm L}$ uncertainty. Fractional uncertainty in inferred $d_{\rm L}$ is about 50\% as shown in the bottom panel of Figure \ref{fig:Histo}. Since these sources also have better 2D sky localization, they cover a lower 3D sky volume than their bright spot counterparts.  

The extent of localization depends on the interplay between the extrinsic parameters. We plot the absolute value of the difference between the waveform amplitude taking into account Earth-rotation and detector-size effects ($h_{\rm CE}$) and the amplitude ignoring these effects ($h_{\rm ig}$) in the top panel of Figure \ref{fig:waveform_compare}. The bottom panel does the same for the phase.  At very low frequencies (5\,Hz - 40\,Hz) the dominant contribution comes from the amplitude modulation due to the rotation of the Earth. This is expected as the signal spends the most time in this band and so the rotation of the Earth causes noticeable changes. Ignoring this effect leads to biases in estimated parameters which has been confirmed by PE runs. At higher frequencies, the dominant contribution comes from effects due to the size of the detector, which is consistent with \citep{PhysRevD.96.084004}.

A greater strain magnitude difference implies better localization. The left panel of Figure \ref{fig:waveform_compare} is the source with 90\% sky area of 7 square degrees (denoted by `+' in Figure \ref{fig:Sky}) and the left panel is the worst localized source at $\theta_{\rm JN}$ = 1.0 with a 90\% sky area of 9300 square degrees (denoted by `$\times$' in Figure \ref{fig:Sky}).

To get a quantitative estimate of the deviation of a waveform taking into account the Earth-rotation and finite-size effects ($h_{\rm CE}$) from a waveform ignoring these effects ($h_{\rm ig}$) we define a ratio $m$, which we refer to as the match. It is given by,
 \begin{equation}
    m = \frac{\text{max}\Big\vert\int_0^{\infty} \frac{h^*_{\textrm{\rm{CE}}} (f)h_{\textrm{ig}}(f)|_{\Phi = t = 0} \exp(2\pi i f t)}{S_n(f)}df \Big\vert}{\left( \int_0^{\infty} \frac{h^*_{\textrm{\rm{CE}}} (f)h_{\textrm{\rm{CE}}}(f)}{S_n(f)}df \right)}
\end{equation}
 Here $S_n(f)$ is the projected CE power spectral density. In other words, the above-defined quantity is the phase and time marginalized SNR \citep{PhysRevD.85.122006} collected by a waveform ignoring the Earth-rotation and detector-size effects divided by the optimal SNR of the injected waveform. For the parameters with best localization at SNR of 1000 and $\theta_{\rm{JN}} = 1.0$ (denoted by + in Figure \ref{fig:Sky}), the match (m) comes out to be 95.4\% while for the worst case (denoted by $\times$ in the top panel of Figure \ref{fig:Sky}) we have a match of 99.9\%. A lesser match ($m$) implies the effects due to detector size and Earth-rotation on the waveform are more pronounced, resulting in better localization. 

 To summarize we have three key results:-
 \begin{itemize}
     \item Ignoring effects due to the rotation of Earth and due to detector size leads to biased inference.
     \item Unlike present-generation detectors, it is possible to extract information about the source location even using 1 CE, for high SNRs. The skymaps may be multimodal which is not ideal for electromagnetic followup.
     \item For a fixed SNR, sources on the dark spot of the detector have a better localization compared to sources on the bright spot.
 \end{itemize}
 
\section{Summary and Discussions}
In this work we perform full Bayesian parameter estimation for typical BNS systems, neglecting the spin degrees of freedom and taking into account the effects due to the rotation of the Earth and the size of Cosmic Explorer. We study the localization of sources at various positions in the sky with a fixed SNR of 1000. The goal of the study is to understand the localization capabilities of a single CE. We find that it is indeed possible to localize the source precisely ($\sim$ 10 sq deg) using one CE for some set of parameters at SNR 1000. Around 10\% of the injections have 2D sky localizations comparable to GW170817. The parameter space might however be multimodal, making Fisher estimates unrealistic. We choose an SNR of 1000 as 99\% of events have a sky area less than 10000 square degrees at this SNR. Lower SNRs will have poorer localization, especially at bright spots. An SNR 1000 event at CE is equivalent to an SNR of 21 in LIGO Hanford design sensitivity.

We fix the spins to zero in our analysis as BNS are expected to slow down due to rotational energy loss via magnetically driven plasma winds.\citep{Contopoulos_1999, 1969ApJ...157..869G, Spitkovsky_2006}. A dimensionless aligned spin parameter of 0.05 changes the length of the signal by 1s which would not affect localization. Even for a spin of 0.99 the length of a signal changes only by 0.34\%. So it is reasonable to assume that including spins would have no significant effect on source localization. In the absence of precession in an aligned spin case, we expect the intrinsic and the extrinsic parameters to be weakly correlated so that our results on localization capabilities will not change in presence of a weakly spinning prior for sampling \citep{PhysRevD.103.083011}. However, degeneracies between spins and mass ratios are well known in the literature \citep{PhysRevD.52.848, PhysRevD.90.044074, PhysRevD.88.042002, PhysRevD.98.083007, Tiwari_2018, PhysRevD.87.024035} and thus we might underestimate the errors on intrinsic parameters in our study. For this reason, we make no comments on the intrinsic parameters, which shall be studied in future work.

We use \textsc{TaylorF2} as our waveform model, which might be inaccurate at such high SNRs, but will be representative of the effects of interest. Additionally, more physical effects need to be incorporated into the waveform models for such strong signals. Orbital precession, eccentricities, and higher-order multipoles may play vital roles and need to be consistently and efficiently implemented. Higher-order multipoles (HM) of emission, in addition to the dominant quadruple (2,2) modes, need special mention. During the inspiral phase, the frequency of a mode $(l,m)$ is given by $m\Omega$, where $\Omega$ is the orbital frequency. The frequency of each multipole $m$ corresponds to a (2,2) frequency of $2f/m$ \citep{PhysRevLett.120.161102}. This means that the time to the merger of a pure (3,3) mode from 6 Hz is equal to the time to the merger of a pure (2,2) mode from 4 Hz. Thus modes with higher multipoles last longer in-band and make effects due to the rotation of Earth more pronounced, which might improve sky localization. In addition, the amplitudes of various modes have a different dependence on the inclination of the source's orbit as viewed from the Earth, and this additional information should allow a better determination of the source distance \citep{Calderon} However, we do not expect HMs to resolve the multimodalities in the sky localization \citep{PhysRevD.103.083011}. Also, we are uncertain if HMs are detectable for nearly equal mass binaries like those studied in this work.

For simplicity and to speed up convergence we generate zero noise injections. The addition of colored noise will eventually be required for developing a PE pipeline in the CE era. However, any drastic changes in the degeneracies or the width of the posterior are not expected in the analysis of noisy data. Given the rates of detection in CE we expect most signals will overlap \citep{2021PhRvD.104d4003S, PhysRevD.105.104016, PhysRevD.106.104045, 10.1093/mnras/stab2358, 10.1093/mnras/stad1542}. Our analysis cannot handle overlapping signals. However, it is unlikely two high SNR signals would overlap. Moreover, chunks of the data containing high SNR signals can be easily identified. So our method is reasonable at high SNRs.

The likelihood evaluations are accelerated using multibanding, allowing us to do PE on signals for about an hour. Each likelihood evaluation takes around 20 ms, approximately 500 times faster than a standard evaluation. This makes standard Bayesian PE feasible, which is implemented using \textsc{DYNESTY} as a nested sampler in the framework of \textsc{Bilby}. The relative error in evidence falls below $10^{-5}$ in about 3 days using 16 cores collecting around $10^6$ samples. The samplers and techniques used are highly optimized for PE in ground-based detectors of the present era. We were able to tweak the settings to sample the parameter space robustly for next-generation ground-based detectors. It is possible to come up with a much better sampling algorithm requiring fewer likelihood evaluations. Our work is a first step in that direction, giving a better understanding of the parameter space and the problem at hand.

Although we can localize some sources to 10 square degrees, most of them have sky areas of a few hundred to a few thousand square degrees. We expect only a handful of sources to be well localized for a year-long observation using a single CE. Localization will { also} become poorer for lower SNRs as these effects start to lose significance. To come up with a reliable number for well-localized events we need injections from an expected astrophysical distribution which will be done as a part of follow-up studies. For purposes of electromagnetic follow-up, it would be extremely beneficial to have another identical CE or one with a shorter arm-length of 20 km, to form a baseline and improve localization. A sole 20 km long detector is expected to have similar Earth-rotation effects other than the fact it will collect a slightly lower SNR due to differences in the PSDs. However, the effects due to the size of the detector will become less pronounced and so the sky areas and sky volumes for sources at the dark spot is expected to increase. We used only one CE in our study to understand if the effects becoming relevant in the future detectors, will have any influence on source localization or not. We find the Earth-rotation and detector-size effects play vital roles in localizing sources, but not enough to meet the demands of electromagnetic follow-up. However, the codes developed are general and can handle multiple detectors with minor modifications. 

$Software:$ Analysis in this paper made use of \textsc{Bilby}v1.1.5 \citep{Bilby_paper, Smith2020, RomeroShaw2020}, \textsc{LALSuite}v7.2.4 \citep{lalsuite}, \textsc{NumPy}v1.23.1 \citep{Harris2020}, \textsc{SciPy}v1.8.1 \citep{Virtanen2020}, \textsc{Astropy}v5.1 \citep{2013, 2018} and \textsc{Healpy}v1.16.1 \citep{Zonca2019, 2005ApJ...622..759G}. Plots were produced using \textsc{ligo.skymap}v1.0.3 \citep{singer_2019} and \textsc{Matplotlib}v3.5.2 \citep{4160265}.

$Data~and~ code~ availability :$ The codes used for this project are hosted in \url{https://git.ligo.org/pratyusava.baral/3g_pe}.

\section{Acknowledgements}
We thank the anonymous referee for constructive suggestions and comments that helped in the considerable improvement of the manuscript. the We thank Noah Wolfe for reviewing the manuscript and providing valuable comments.
The authors are thankful to Neil Cornish, Matthew Evans, Kevin Kuns, Alex Nitz, Anarya Ray, Caitlin Rose, Soumendra Kishore Roy, and Bangalore Sathyaprakash for useful comments regarding this work and the manuscript. {\text{blue} We also thank the anonymous referee for useful suggestions which improved the overall quality of the work.} This work was supported by the National Science Foundation awards PHY-2207728 and PHY-2110576. The authors are grateful for computational resources provided by the LIGO Laboratory and supported by National Science Foundation Grants PHY-0757058 and PHY-0823459, PHY-0823459, PHY-1626190 and PHY-1700765 and those provided by Cardiff University, and funded by an STFC grant supporting UK Involvement in the Operation of Advanced
LIGO. This material is based upon work supported by NSF’s LIGO Laboratory which is a major facility fully funded by the National Science Foundation. This work carries LIGO Document No. LIGO-P2300057.

\appendix 
\section{Skymaps for all injections}\label{appendix A}
\begin{figure*}[h!]
     \centering
     \begin{subfigure}[b]{0.22\textwidth}
         \centering
        \includegraphics[width=\textwidth]{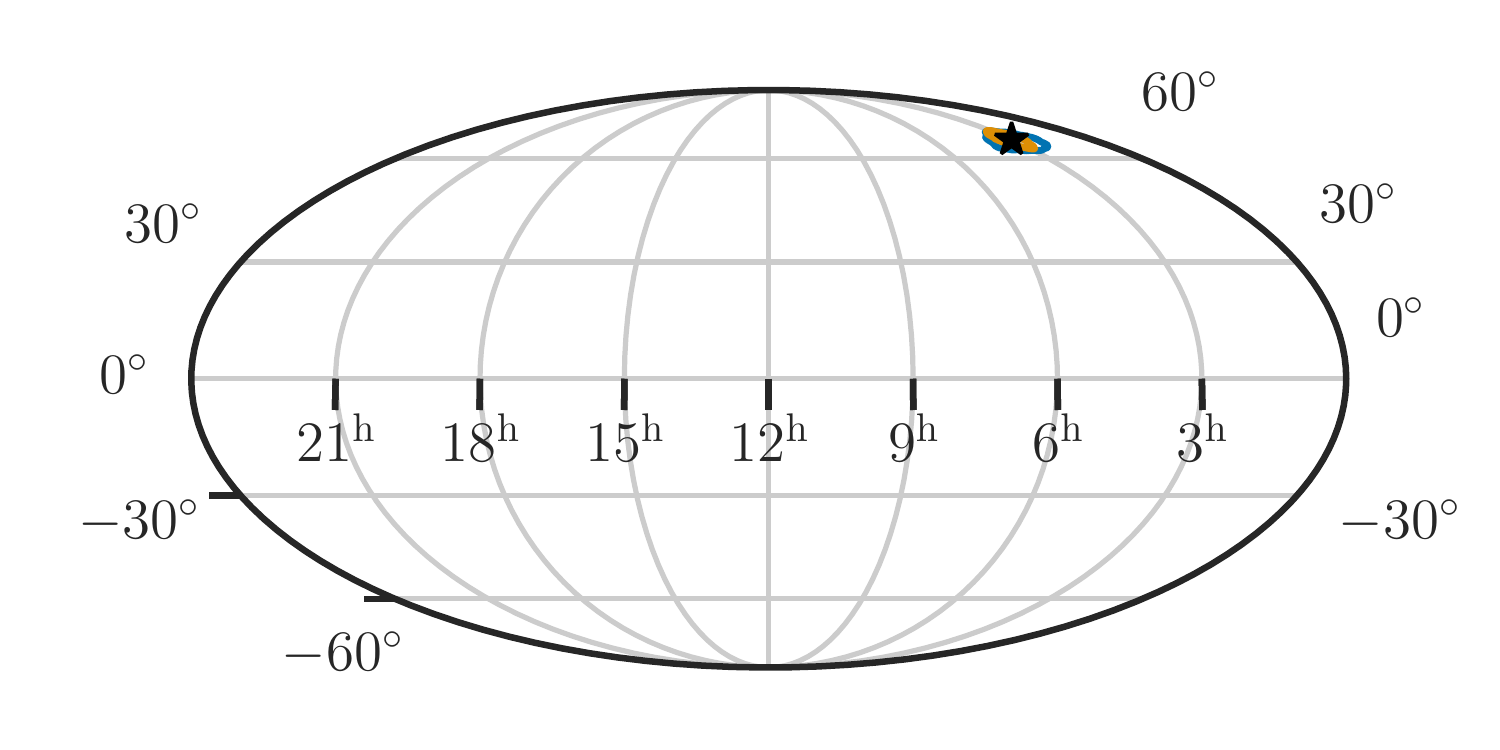}
    \end{subfigure}
     \hfill
     \begin{subfigure}[b]{0.22\textwidth}
         \centering
        \includegraphics[width=\textwidth]{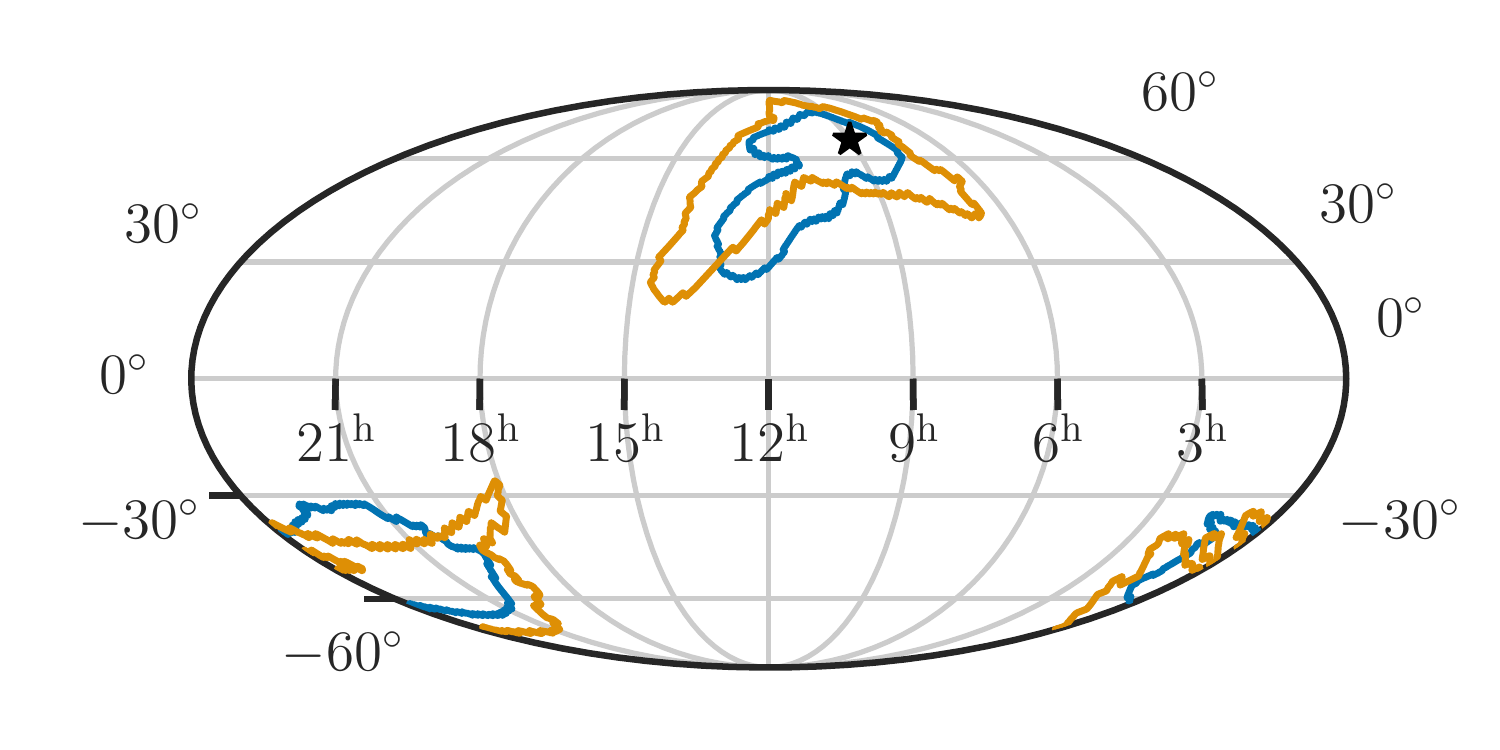}
     \end{subfigure}
     \hfill
     \begin{subfigure}[b]{0.22\textwidth}
         \centering
        \includegraphics[width=\textwidth]{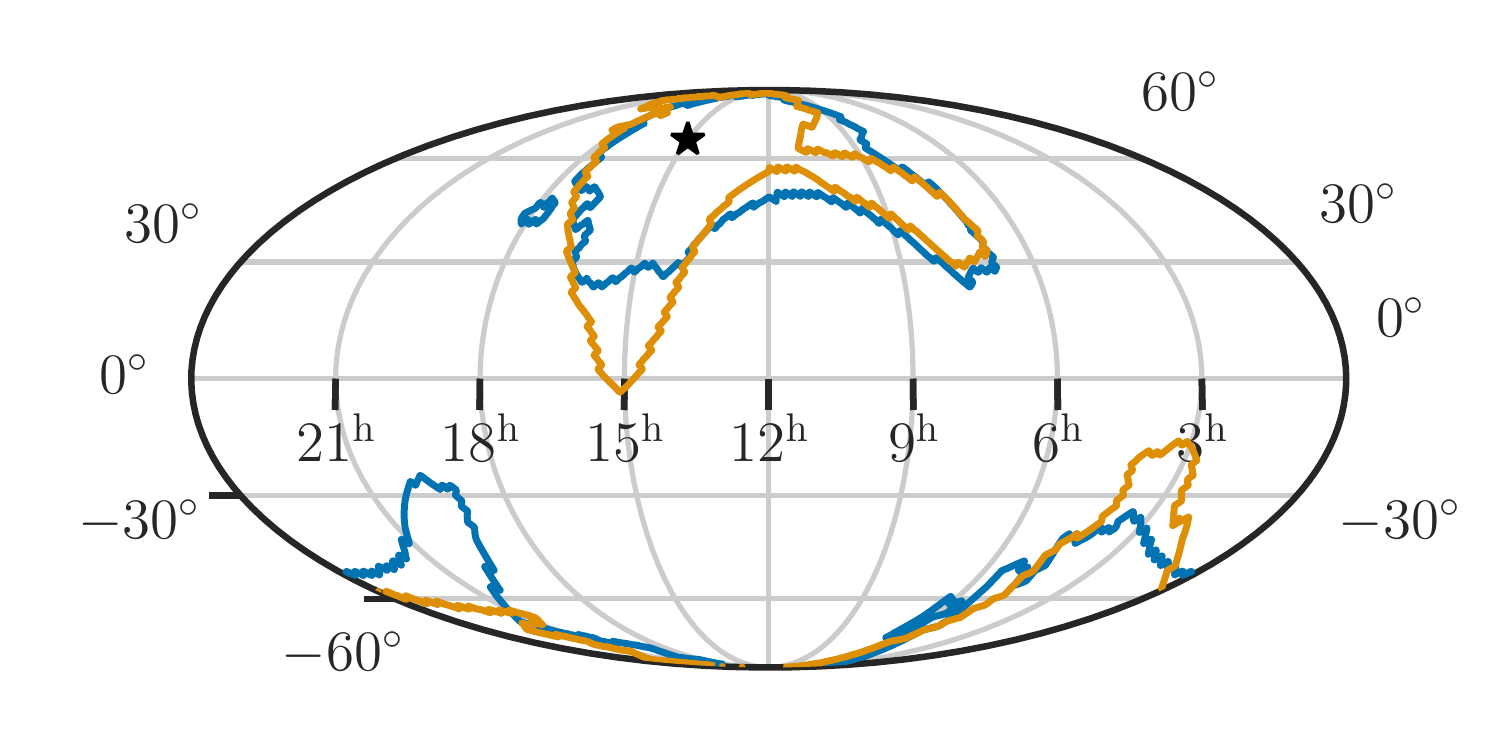}
    
     \end{subfigure}
     \hfill
     \begin{subfigure}[b]{0.22\textwidth}
         \centering
\includegraphics[width=\textwidth]{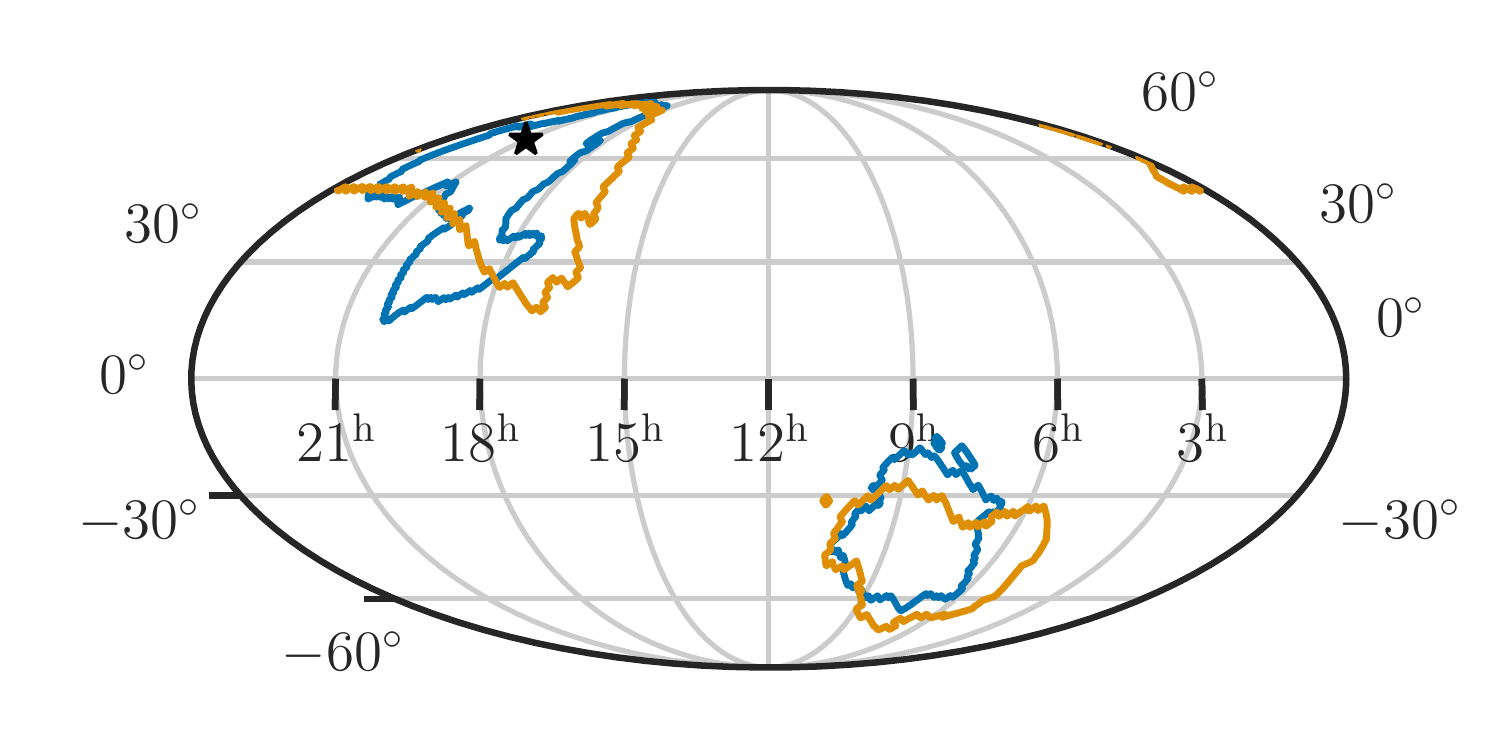}
     \end{subfigure}
     \hfill
     \begin{subfigure}[b]{0.22\textwidth}
         \centering
\includegraphics[width=\textwidth]{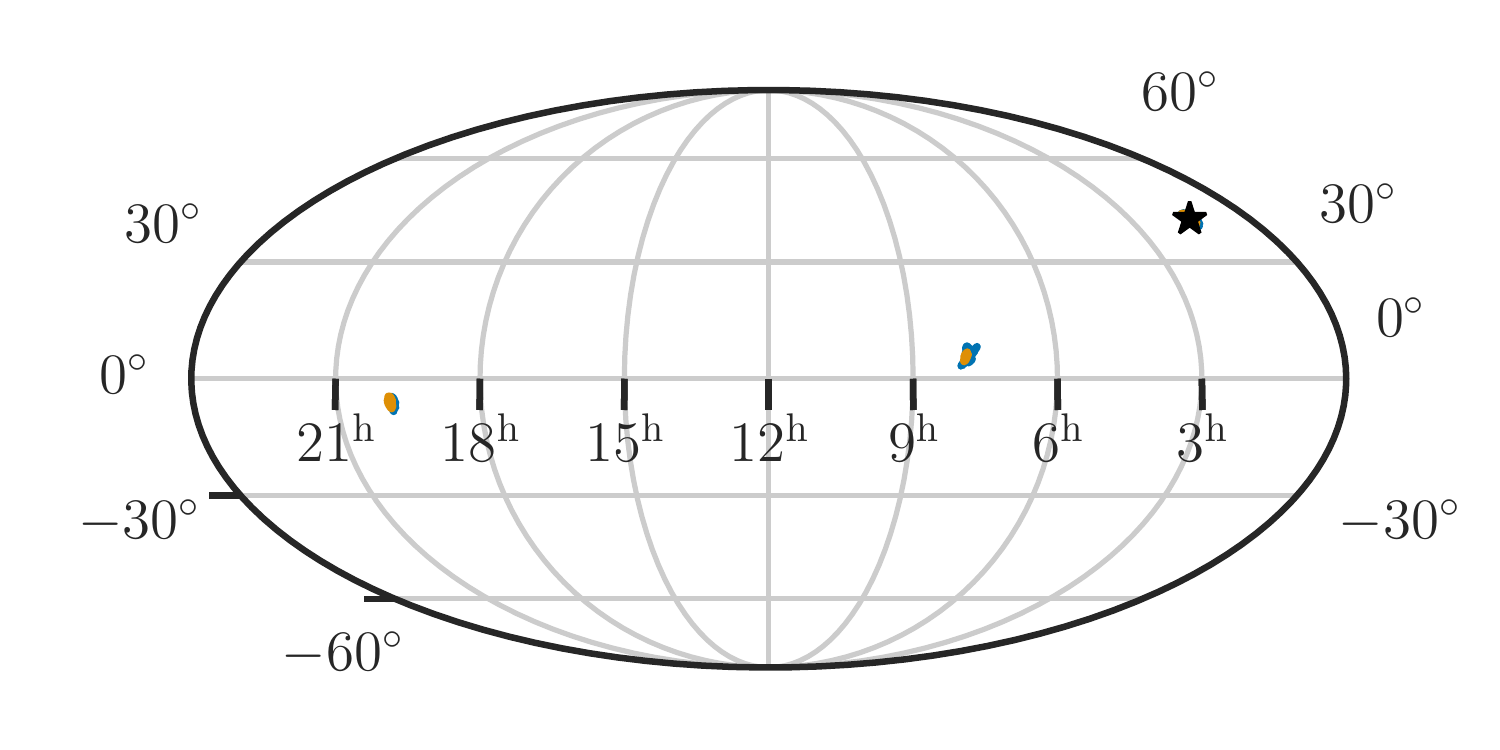}
     \end{subfigure}
     \hfill
     \begin{subfigure}[b]{0.22\textwidth}
         \centering
        \includegraphics[width=\textwidth]{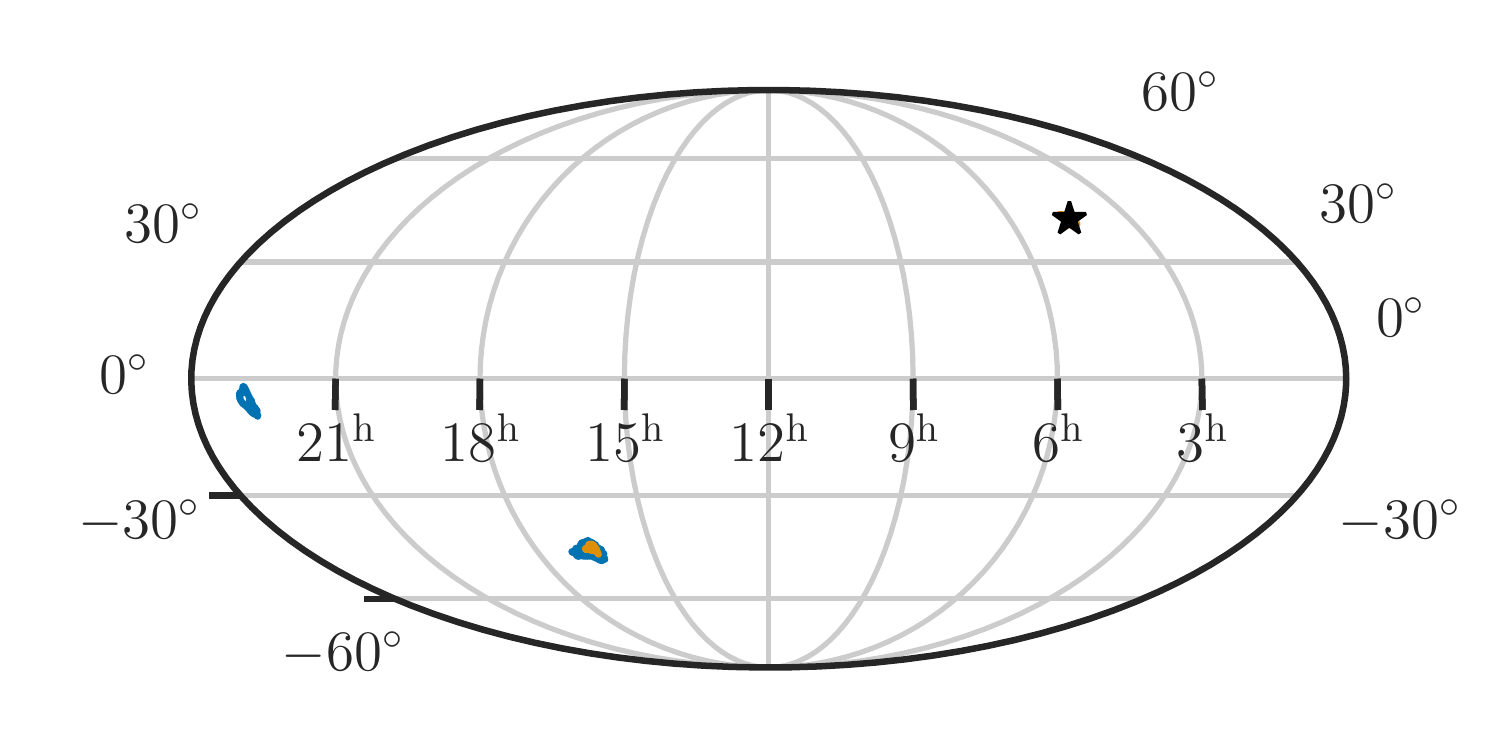}
     \end{subfigure}
     \hfill
     \begin{subfigure}[b]{0.22\textwidth}
         \centering
        \includegraphics[width=\textwidth]{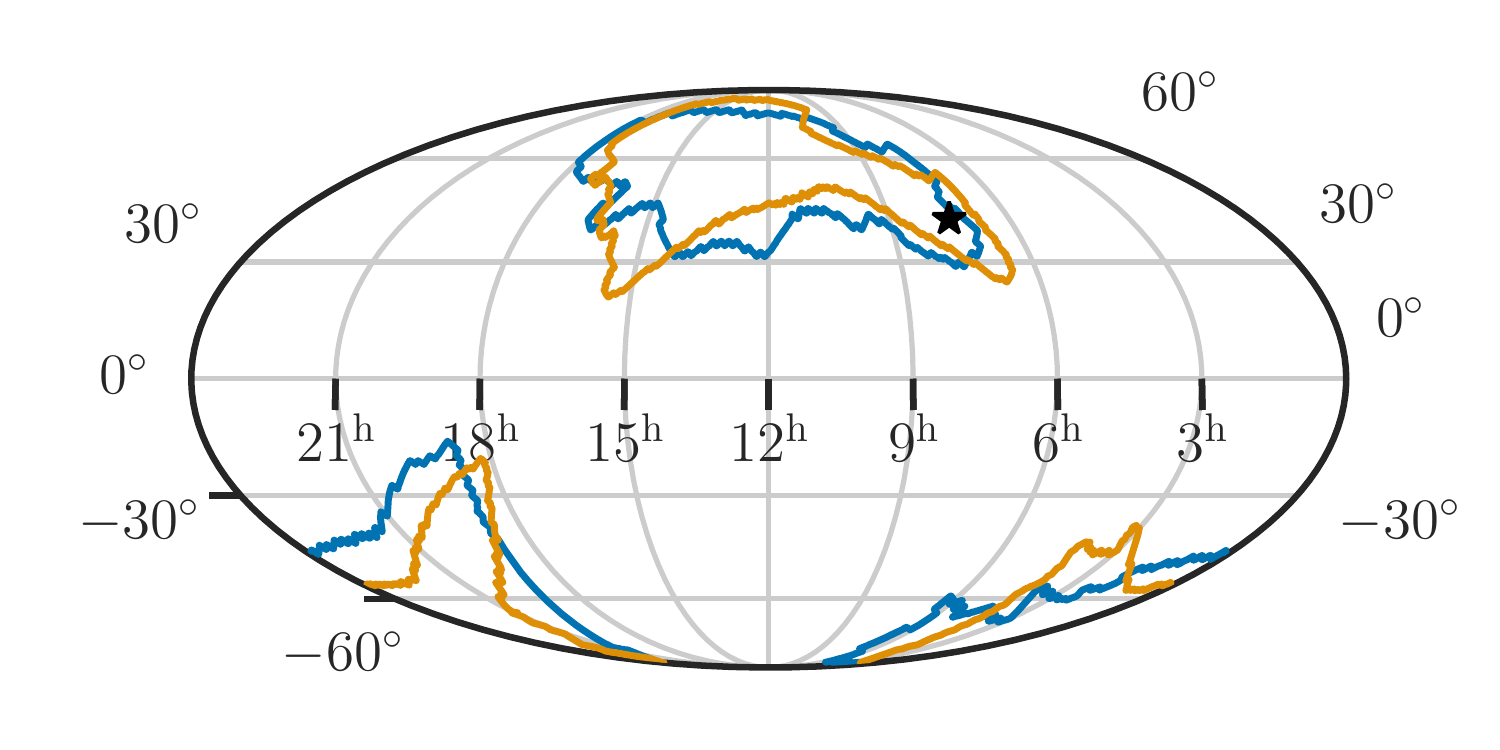}
    
     \end{subfigure}
     \hfill
     \begin{subfigure}[b]{0.22\textwidth}
         \centering
\includegraphics[width=\textwidth]{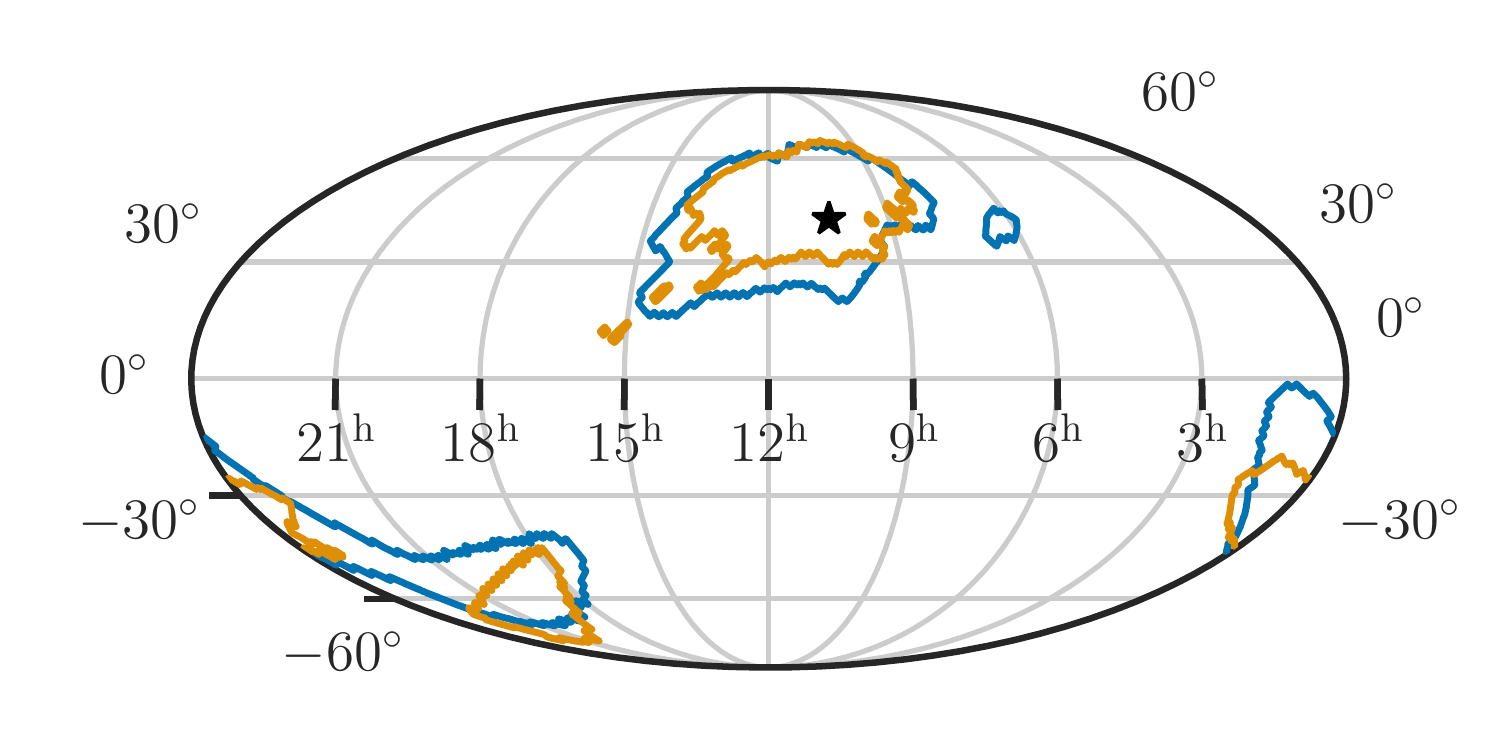}
     \end{subfigure}
     \hfill
     \begin{subfigure}[b]{0.22\textwidth}
         \centering
        \includegraphics[width=\textwidth]{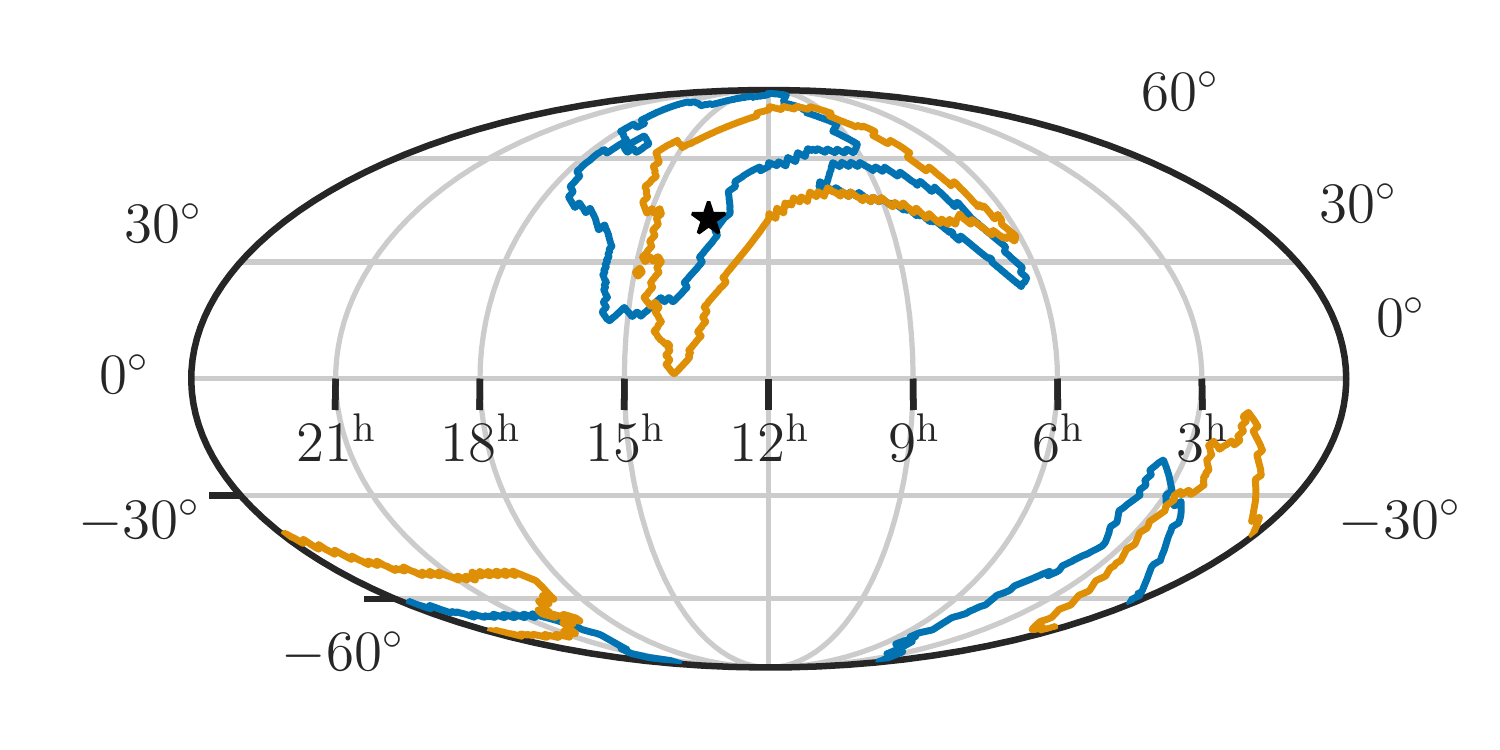}
    \end{subfigure}
     \hfill
     \begin{subfigure}[b]{0.22\textwidth}
         \centering
        \includegraphics[width=\textwidth]{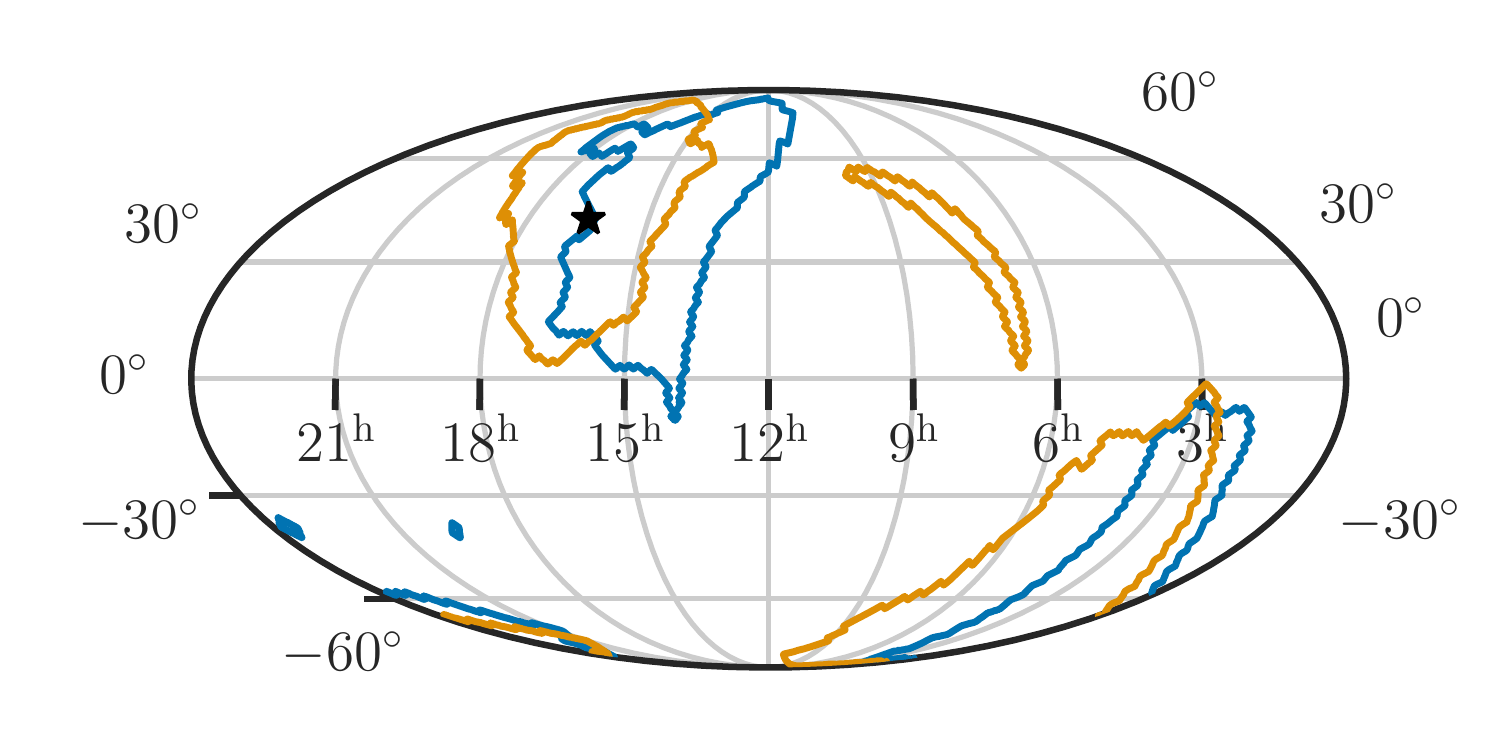}
     \end{subfigure}
     \hfill
     \begin{subfigure}[b]{0.22\textwidth}
         \centering
        \includegraphics[width=\textwidth]{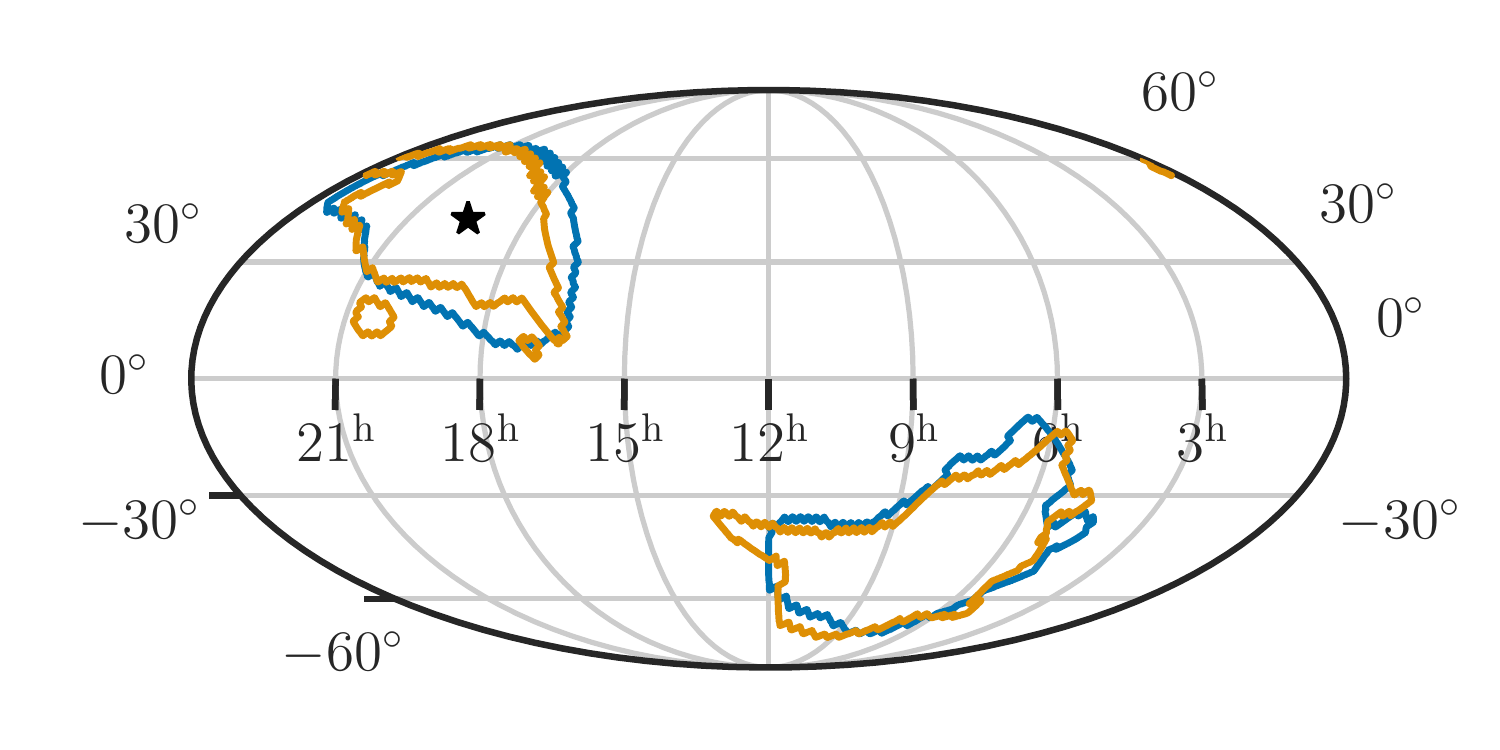}
    
     \end{subfigure}
     \hfill
     \begin{subfigure}[b]{0.22\textwidth}
         \centering
\includegraphics[width=\textwidth]{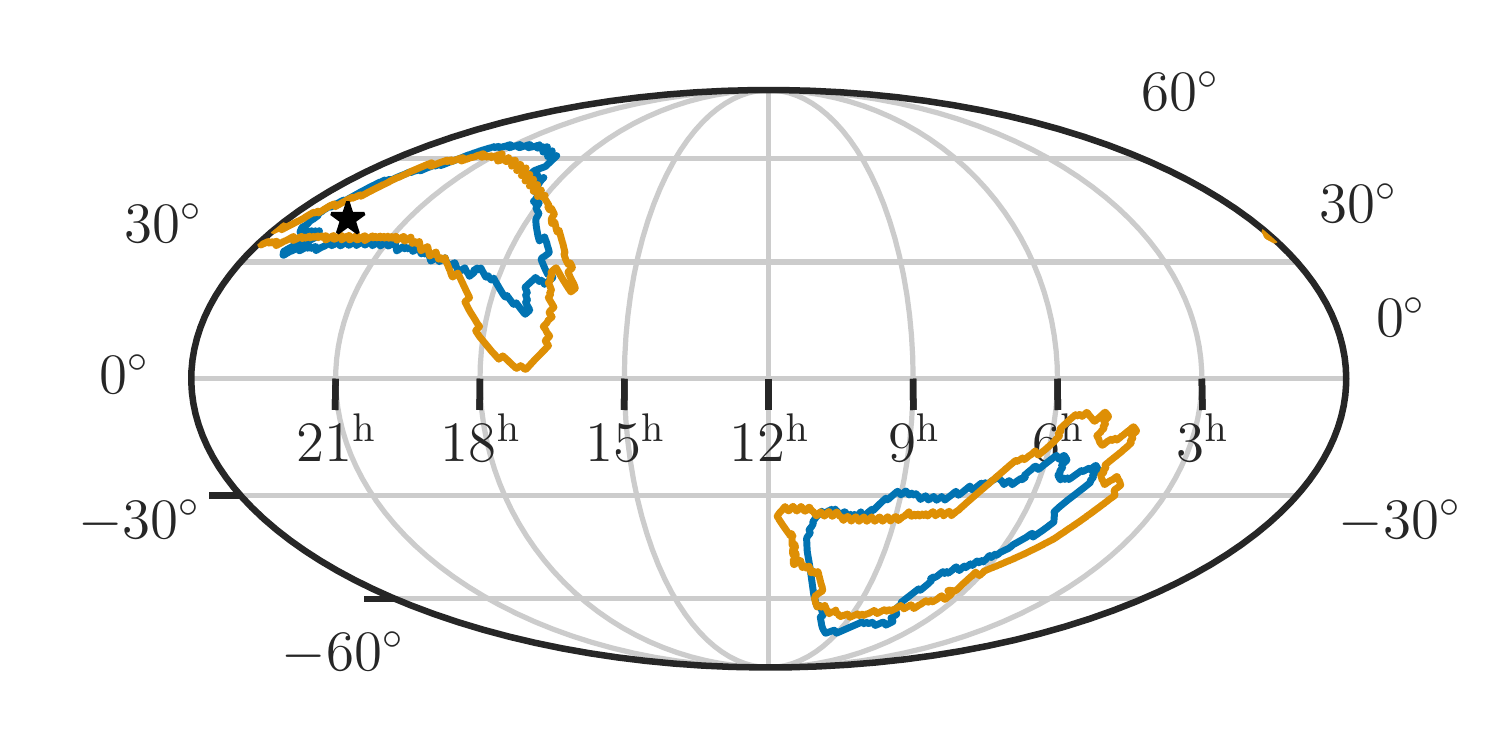}
     \end{subfigure}
     \hfill
     \begin{subfigure}[b]{0.22\textwidth}
         \centering
        \includegraphics[width=\textwidth]{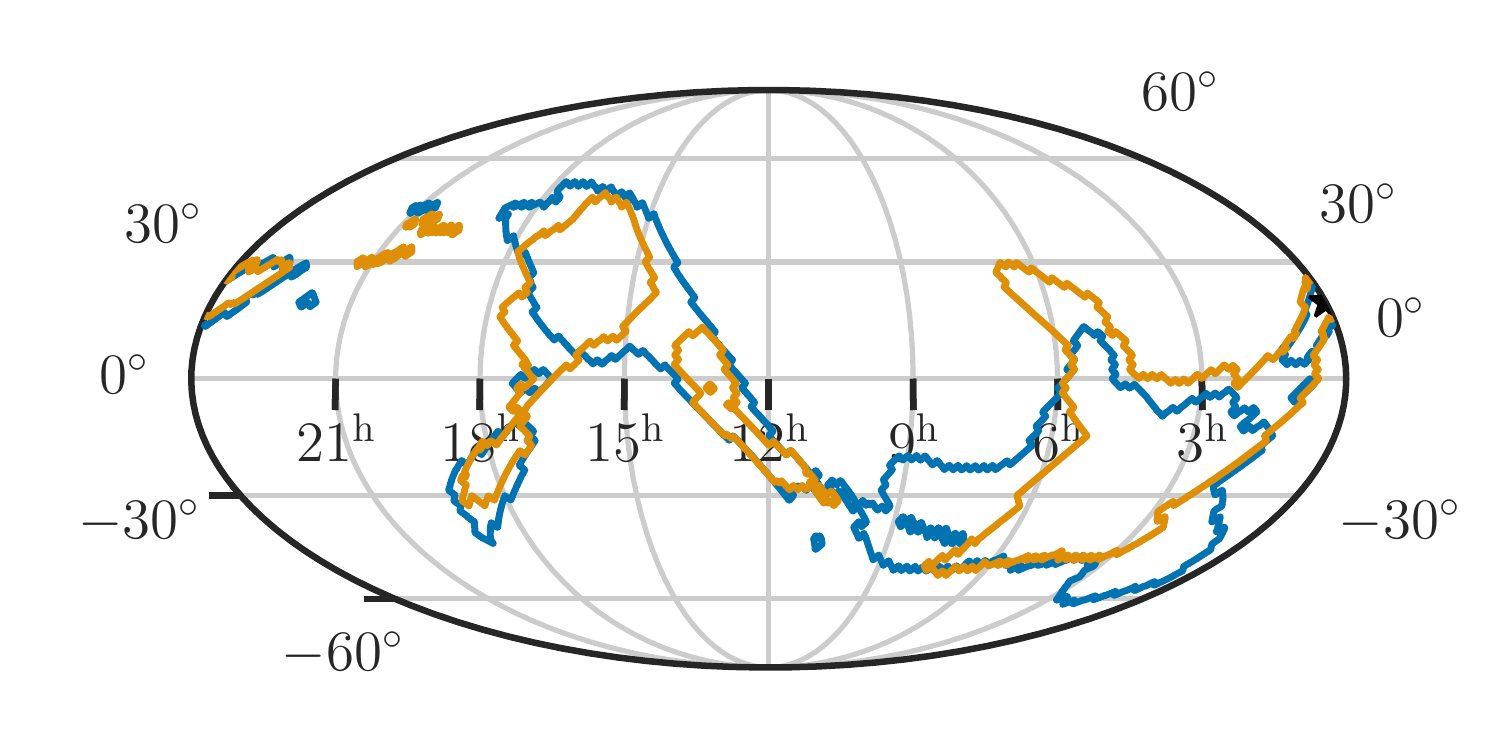}
    \end{subfigure}
     \hfill
     \begin{subfigure}[b]{0.22\textwidth}
         \centering
        \includegraphics[width=\textwidth]{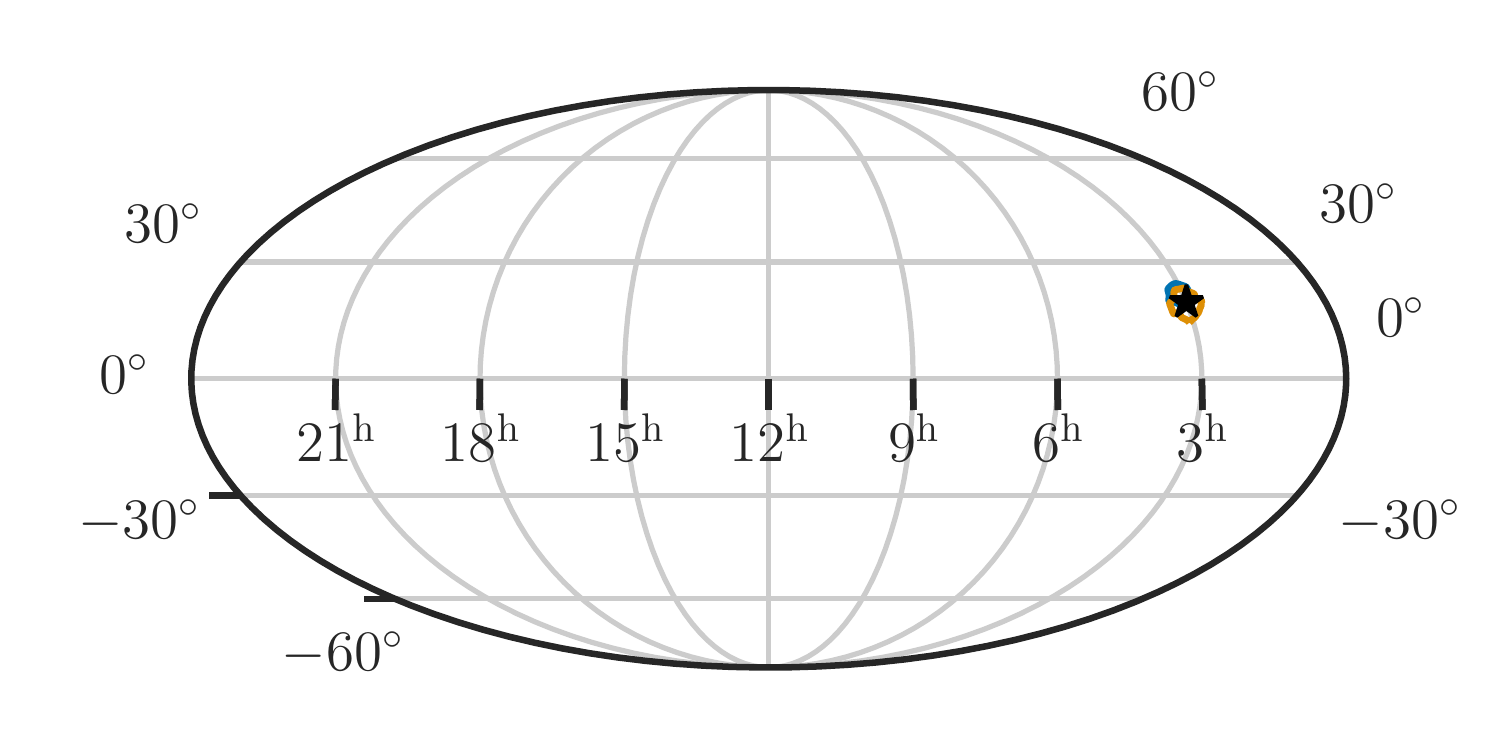}
     \end{subfigure}
     \hfill
     \begin{subfigure}[b]{0.22\textwidth}
         \centering
        \includegraphics[width=\textwidth]{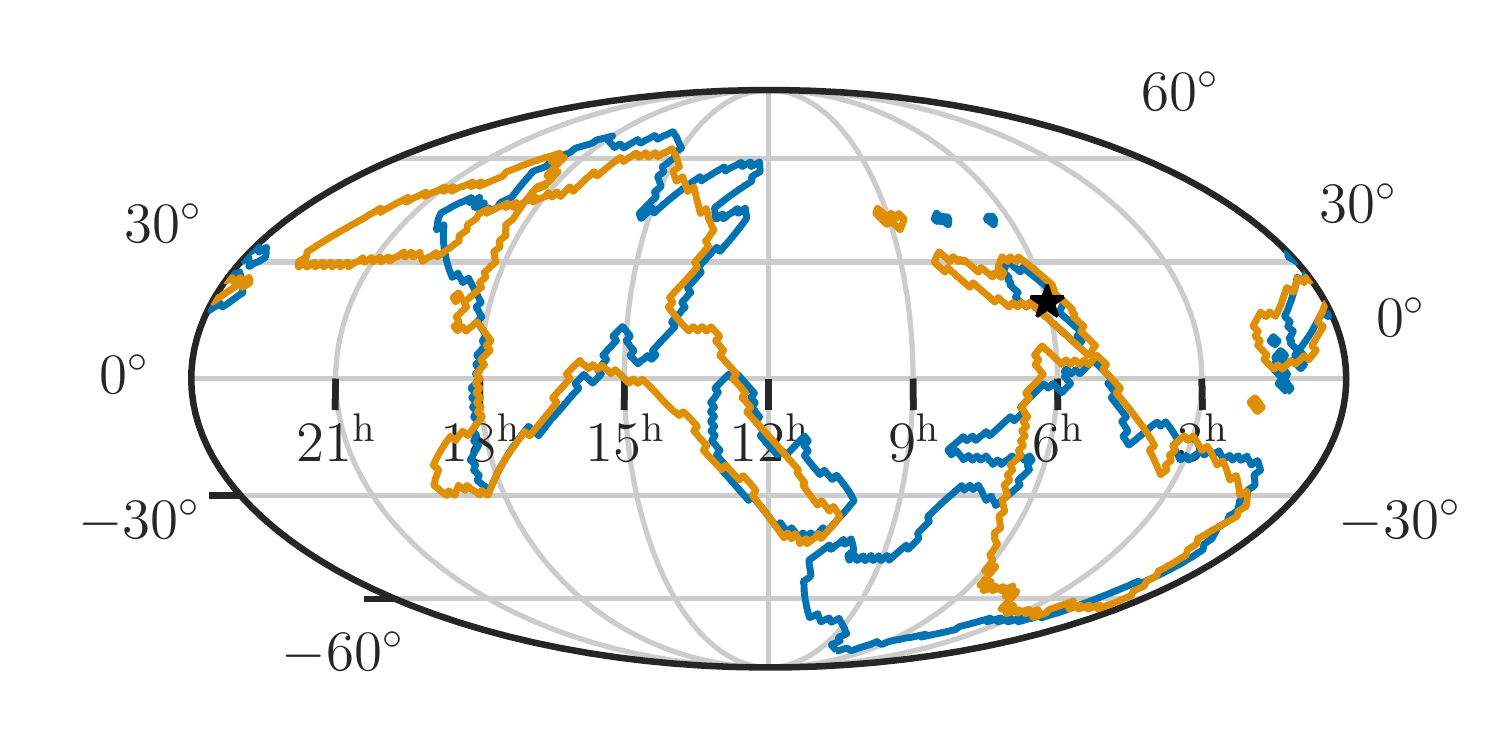}
    
     \end{subfigure}
     \hfill
     \begin{subfigure}[b]{0.22\textwidth}
         \centering
\includegraphics[width=\textwidth]{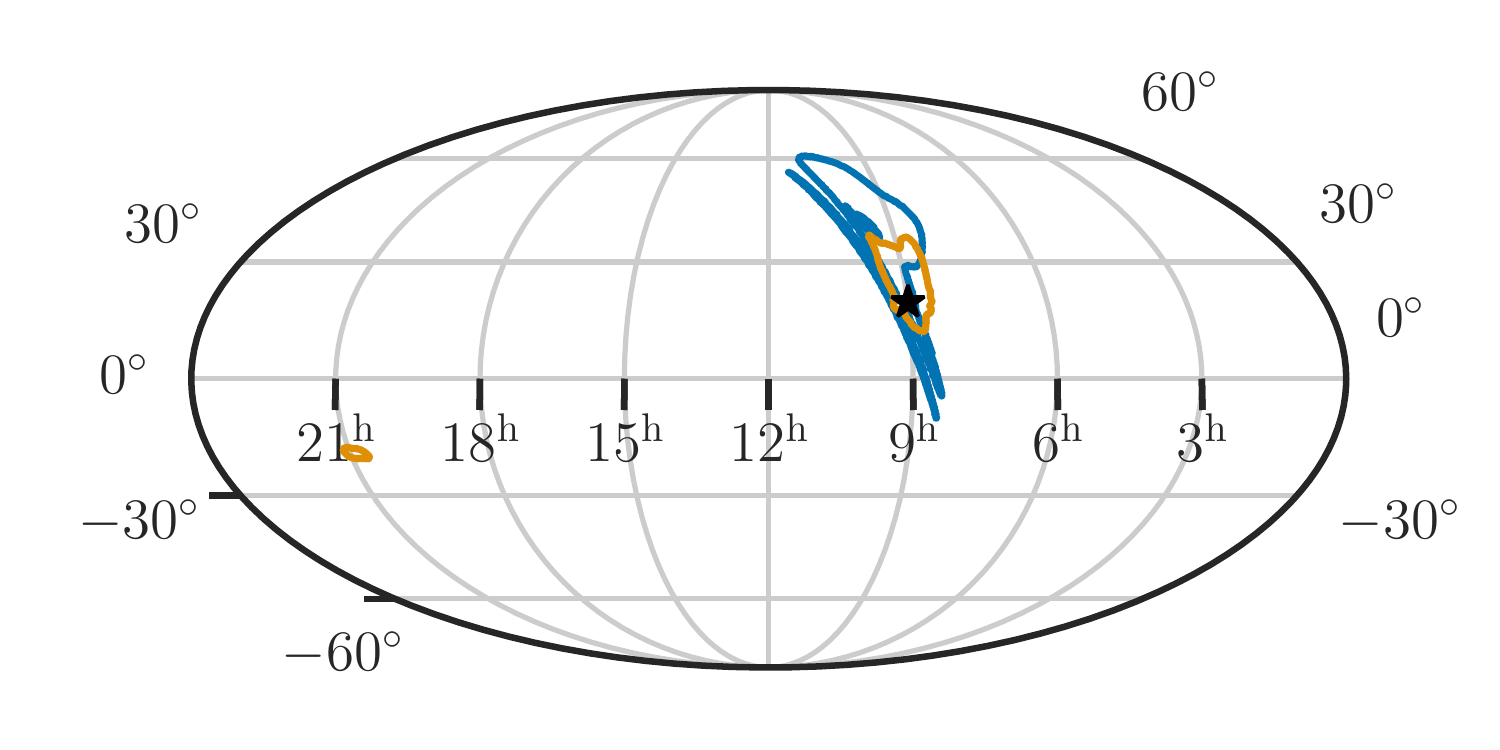}
     \end{subfigure}
     \hfill
     \begin{subfigure}[b]{0.22\textwidth}
         \centering
        \includegraphics[width=\textwidth]{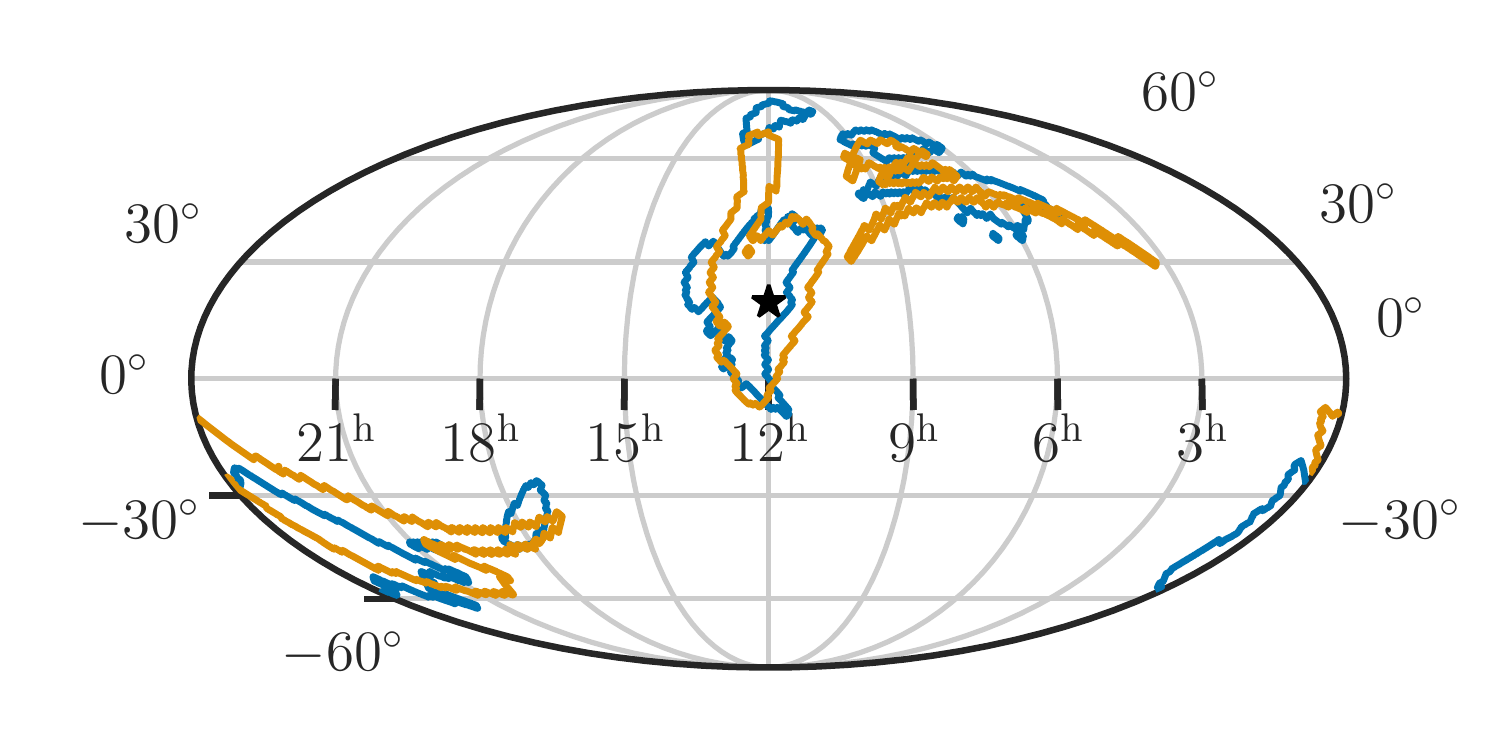}
    \end{subfigure}
     \hfill
     \begin{subfigure}[b]{0.22\textwidth}
         \centering
        \includegraphics[width=\textwidth]{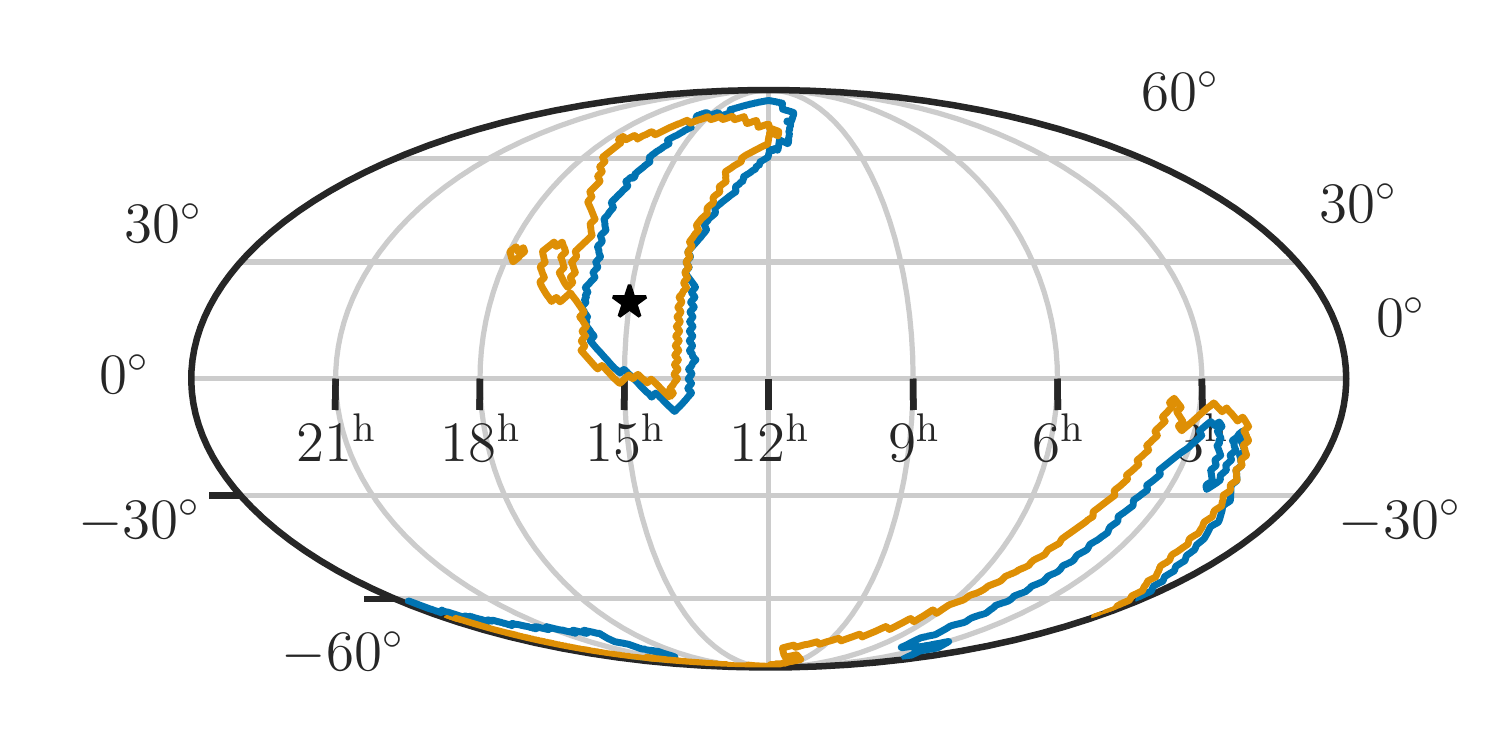}
     \end{subfigure}
     \hfill
     \begin{subfigure}[b]{0.22\textwidth}
         \centering
        \includegraphics[width=\textwidth]{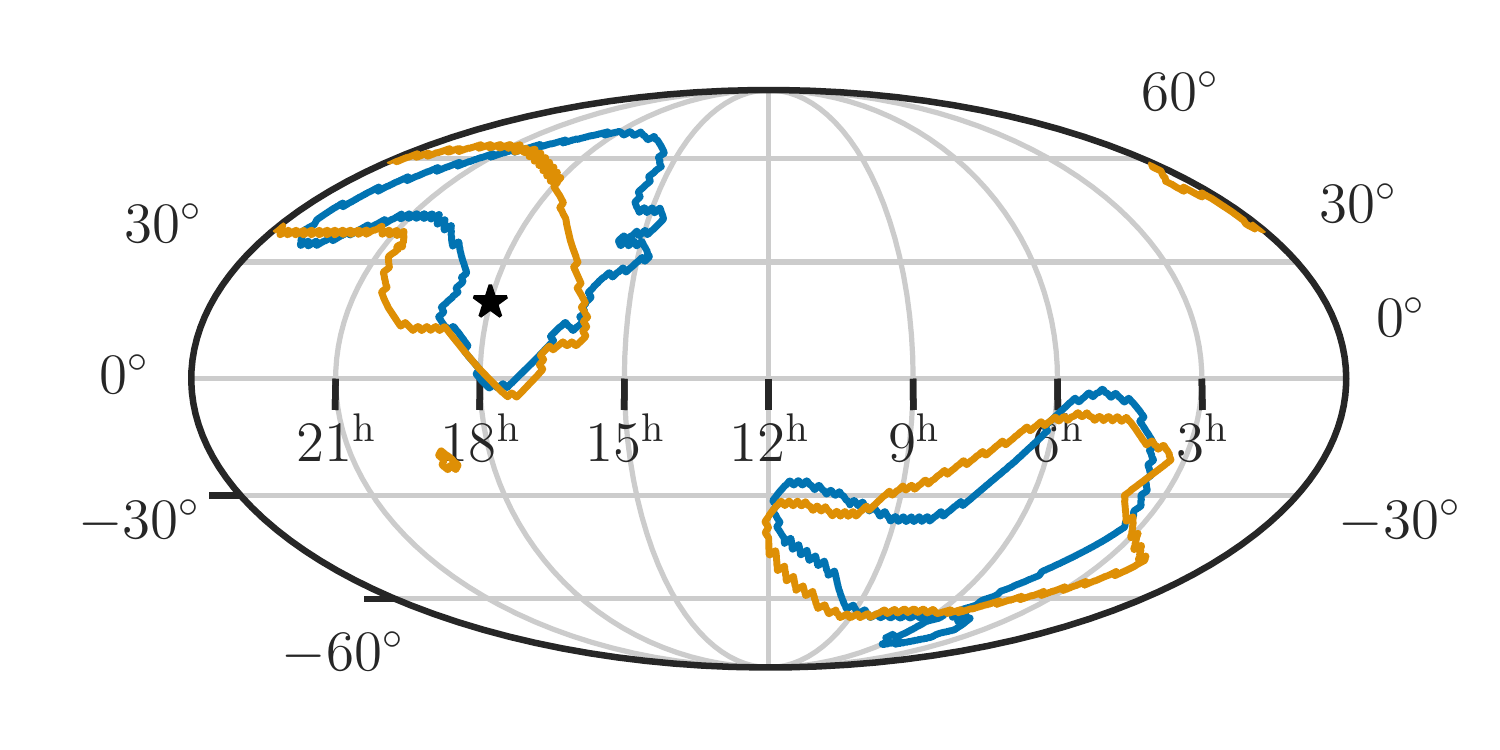}
    
     \end{subfigure}
     \hfill
     \begin{subfigure}[b]{0.22\textwidth}
         \centering
\includegraphics[width=\textwidth]{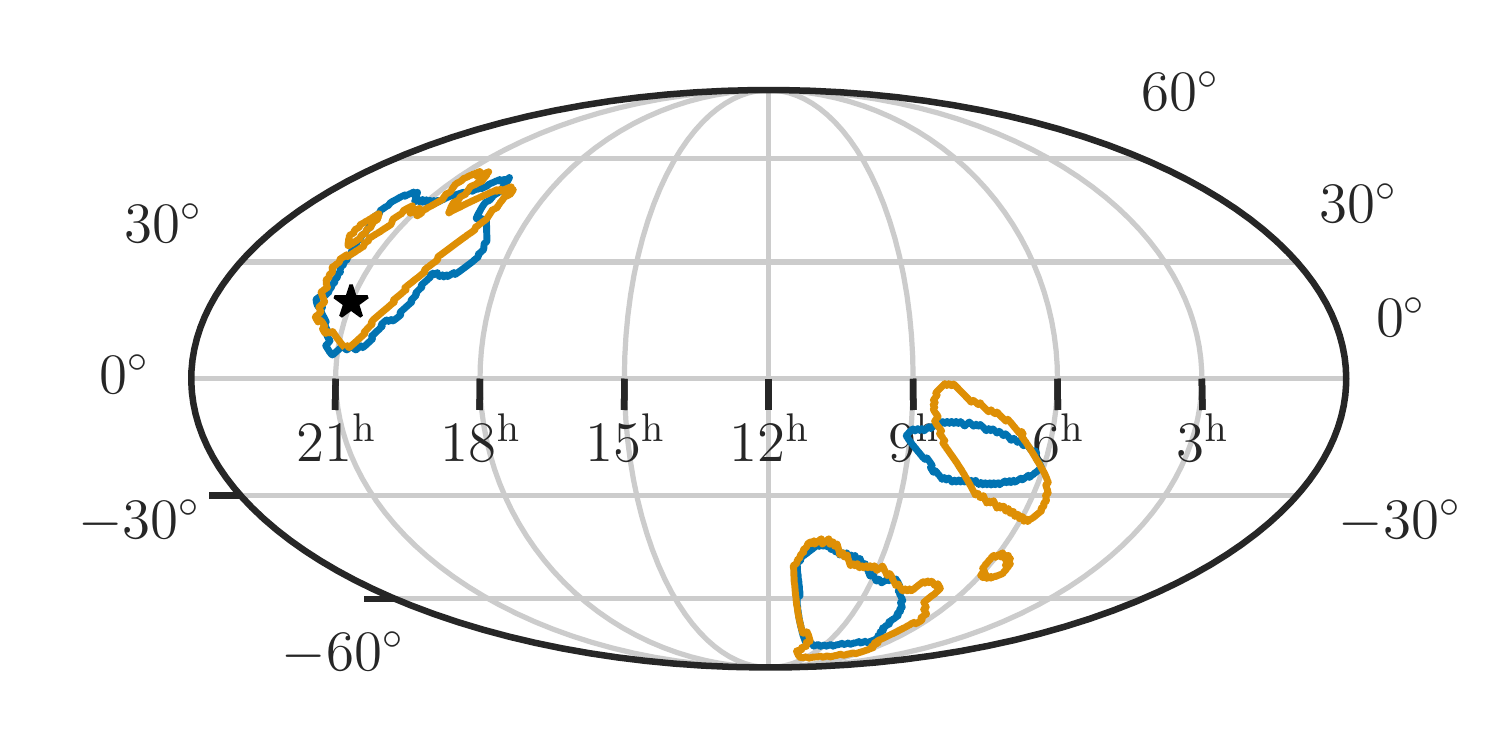}
     \end{subfigure}
     \hfill
     \begin{subfigure}[b]{0.22\textwidth}
         \centering
        \includegraphics[width=\textwidth]{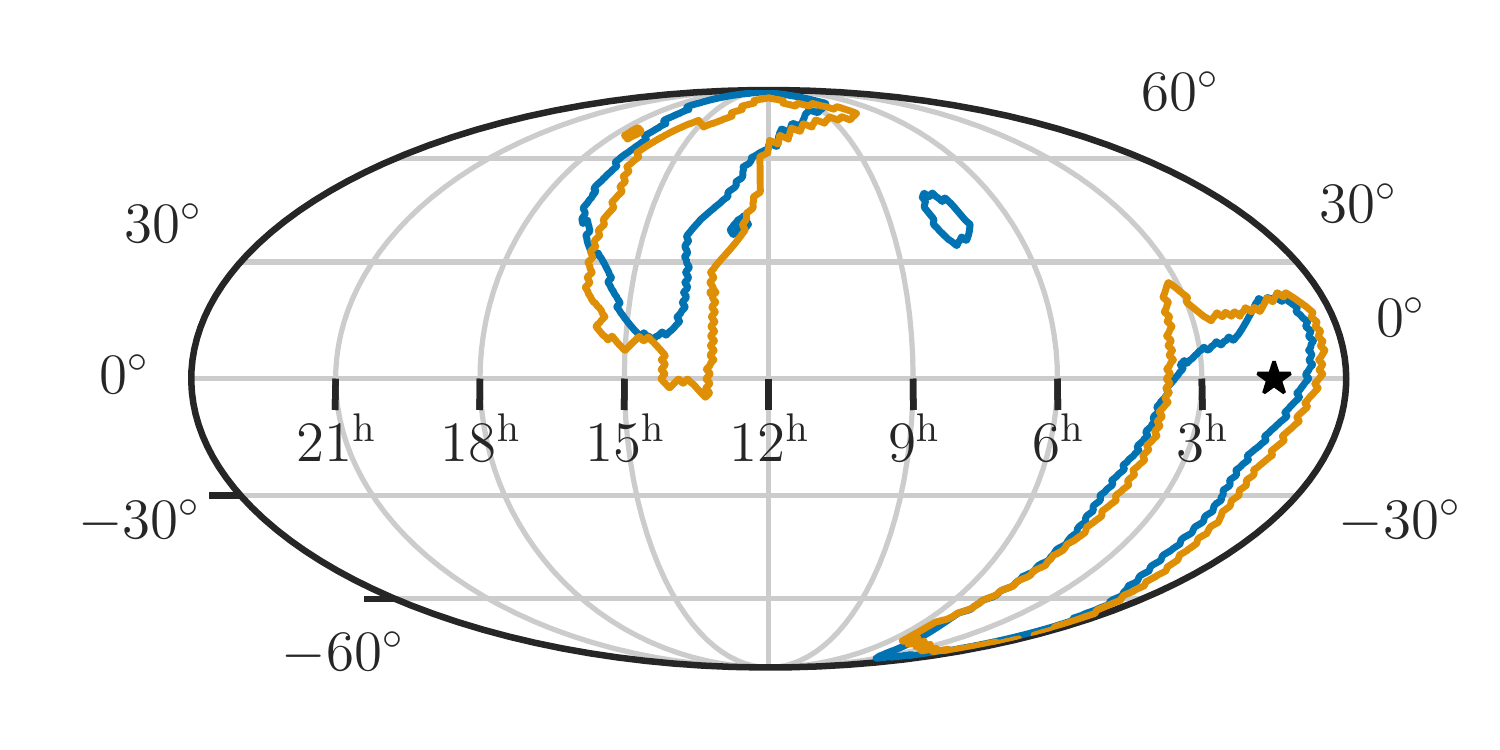}
    \end{subfigure}
     \hfill
     \begin{subfigure}[b]{0.22\textwidth}
         \centering
        \includegraphics[width=\textwidth]{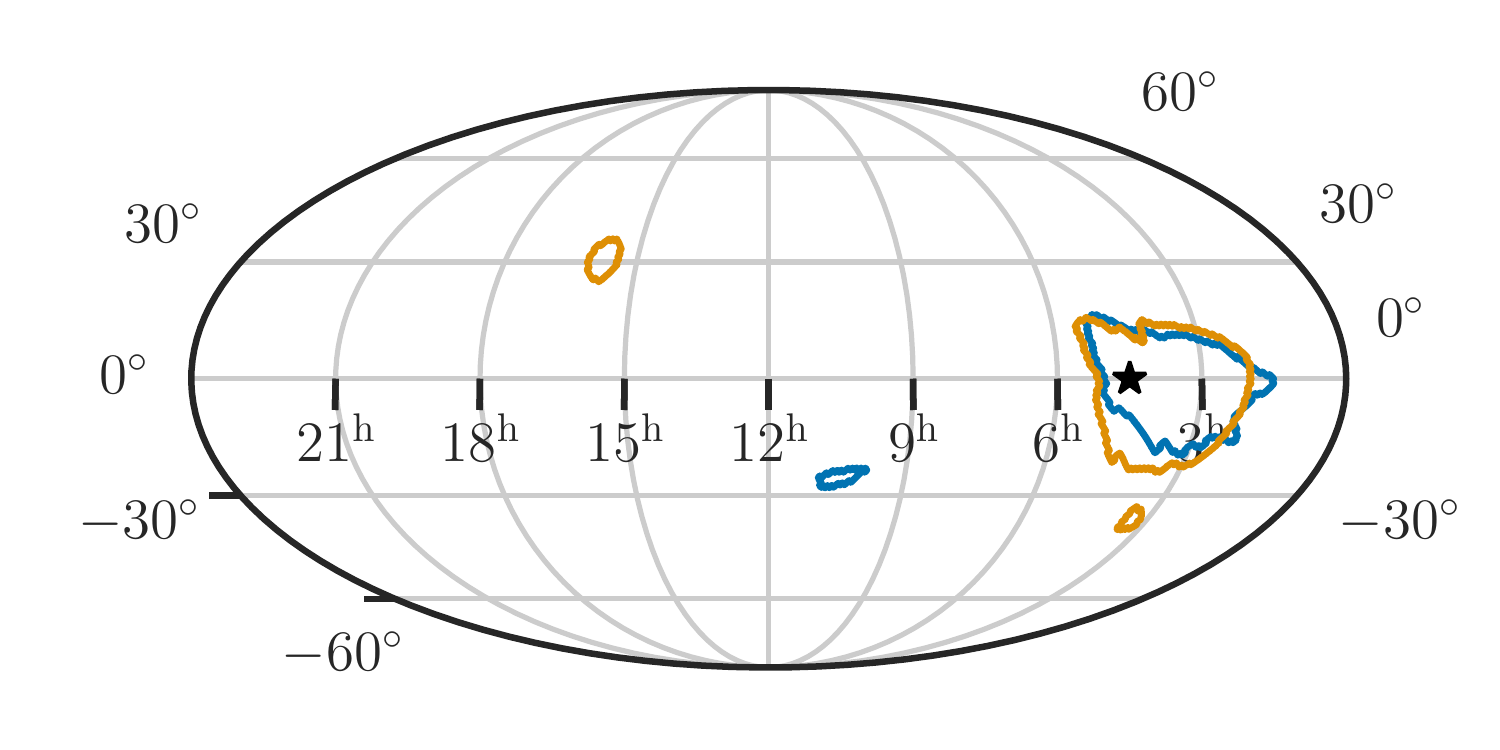}
     \end{subfigure}
     \hfill
     \begin{subfigure}[b]{0.22\textwidth}
         \centering
        \includegraphics[width=\textwidth]{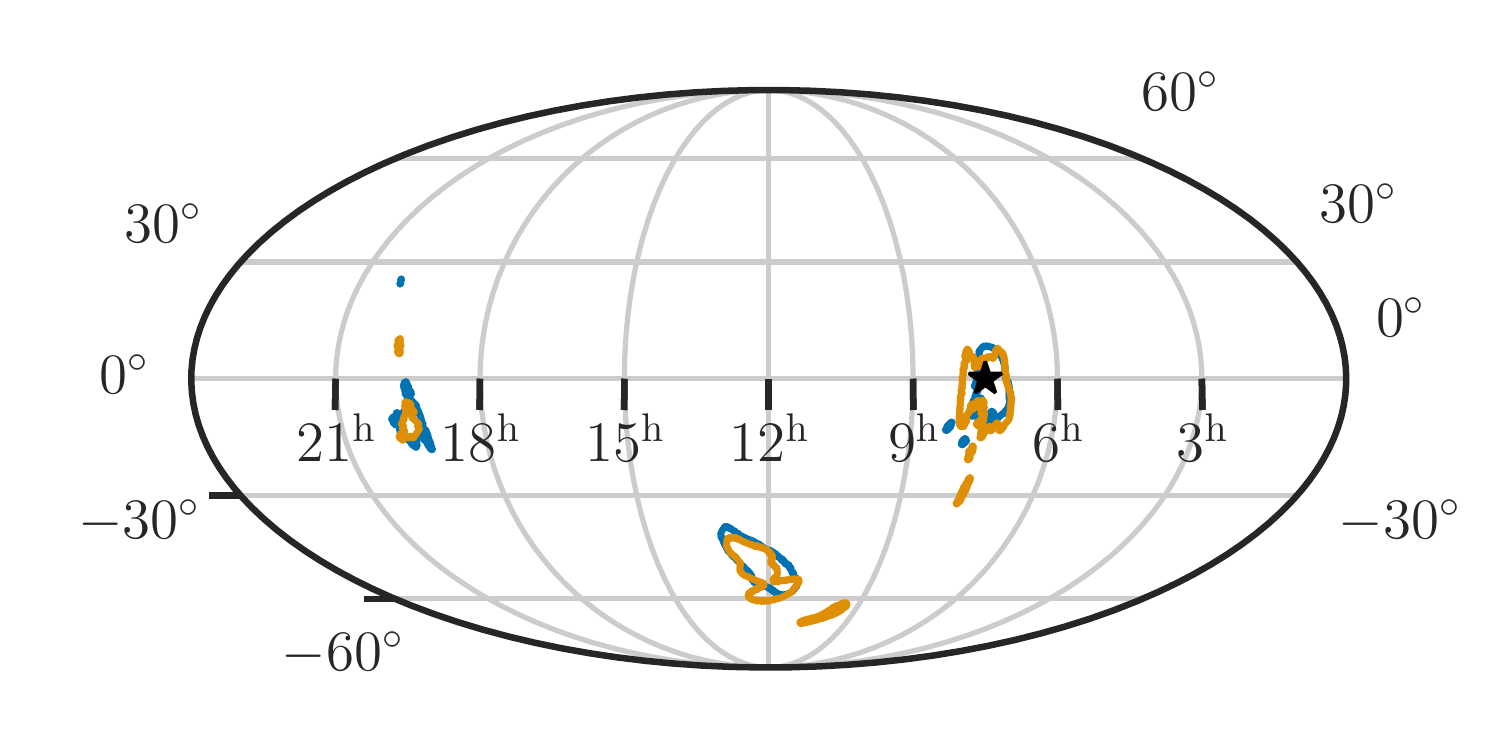}
    
     \end{subfigure}
     \hfill
     \begin{subfigure}[b]{0.22\textwidth}
         \centering
\includegraphics[width=\textwidth]{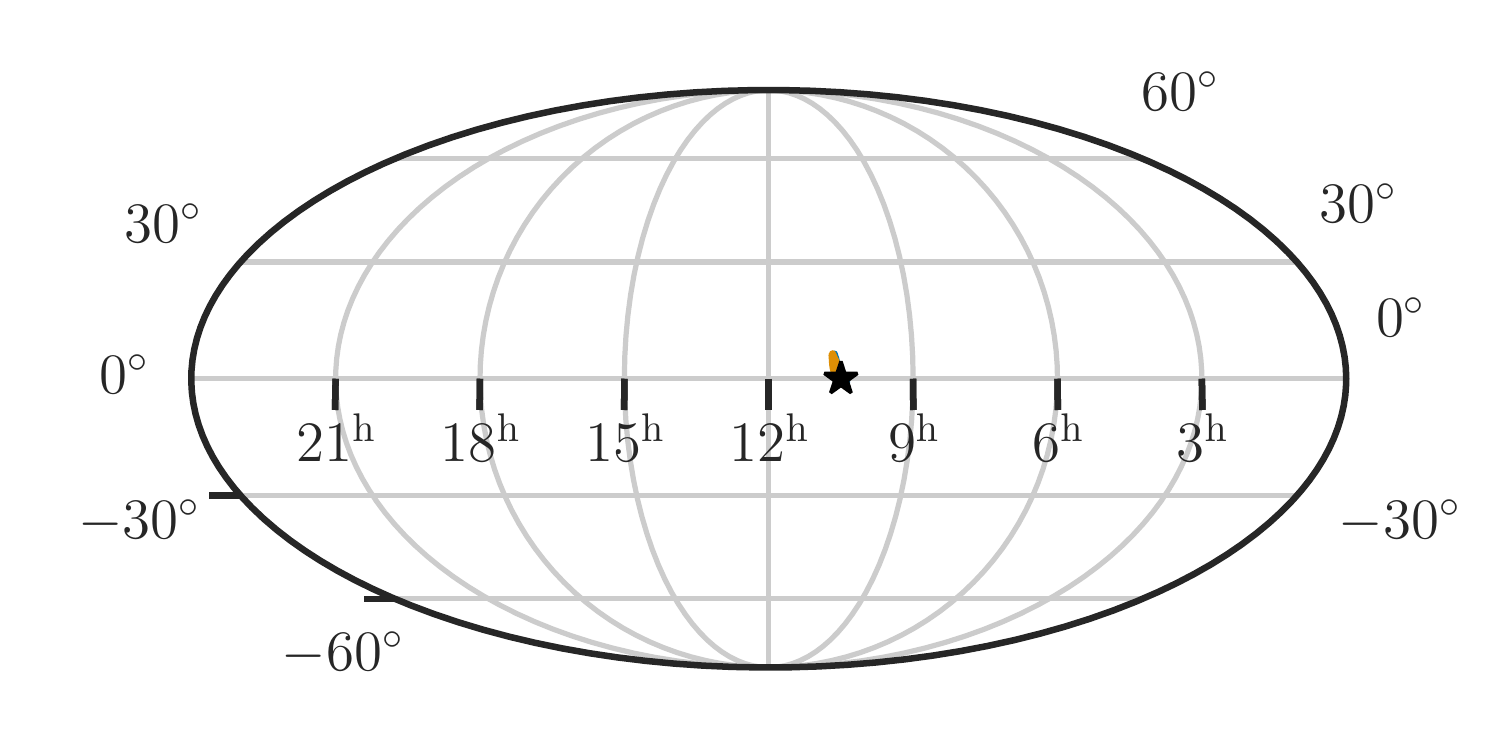}
     \end{subfigure}
     \hfill
     \begin{subfigure}[b]{0.22\textwidth}
         \centering
        \includegraphics[width=\textwidth]{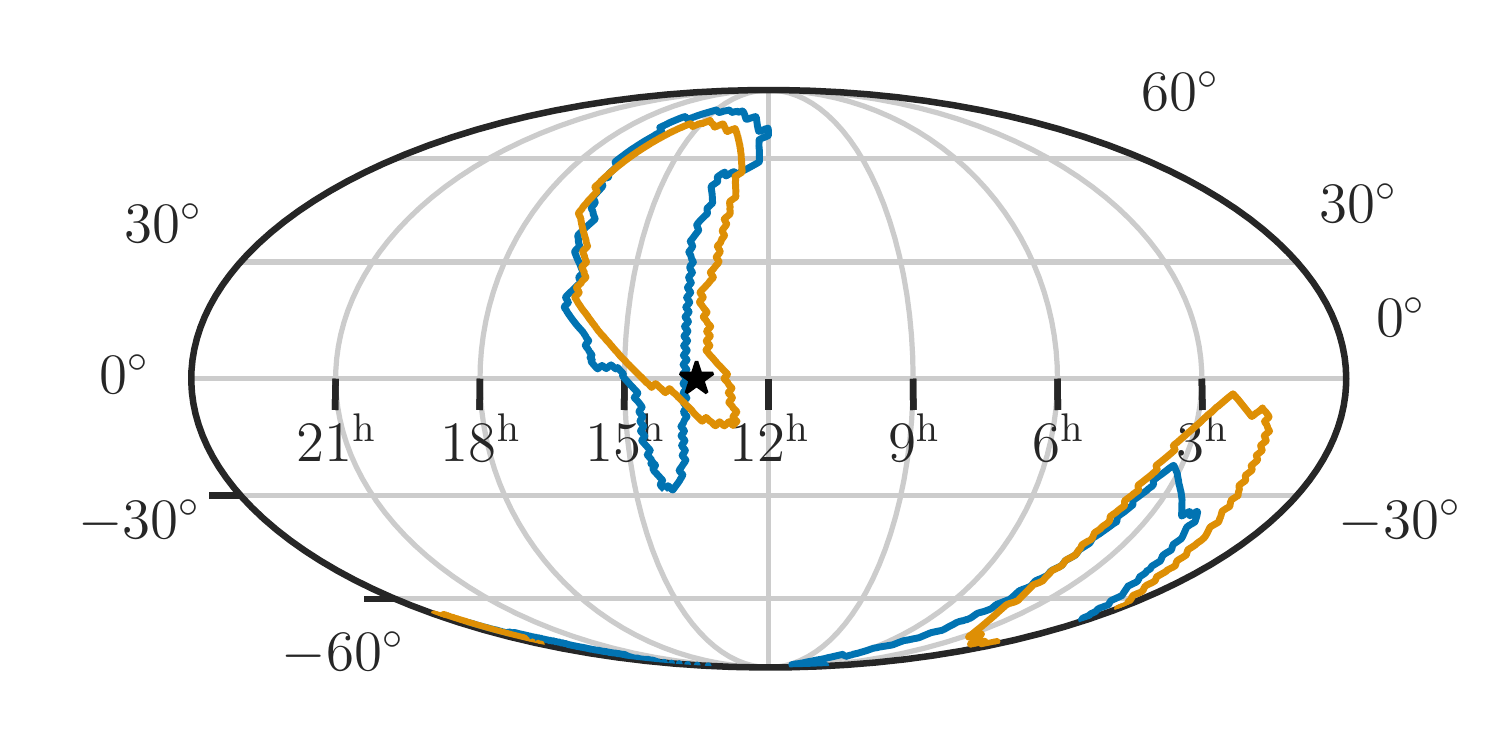}
    \end{subfigure}
     \hfill
     \begin{subfigure}[b]{0.22\textwidth}
         \centering
        \includegraphics[width=\textwidth]{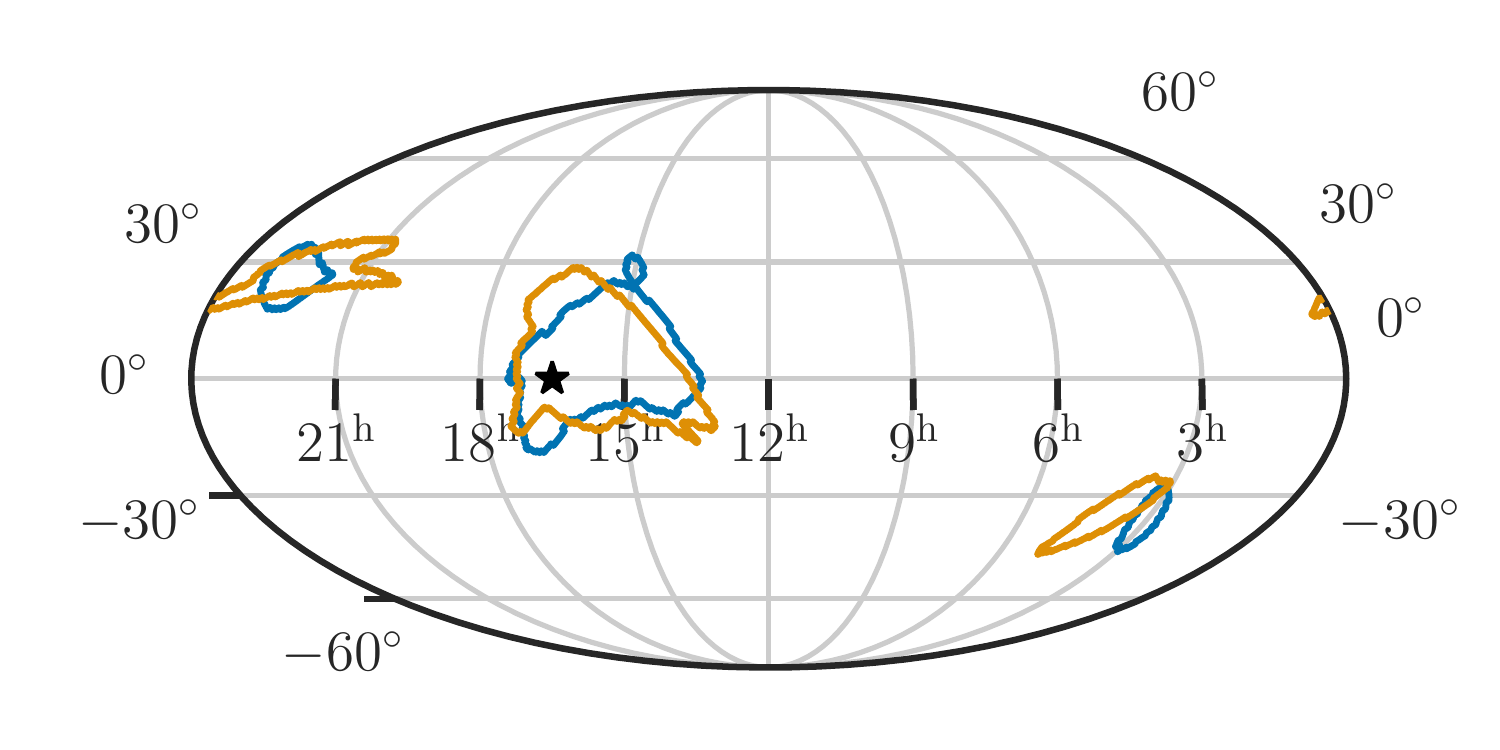}
     \end{subfigure}
     \hfill
     \begin{subfigure}[b]{0.22\textwidth}
         \centering
        \includegraphics[width=\textwidth]{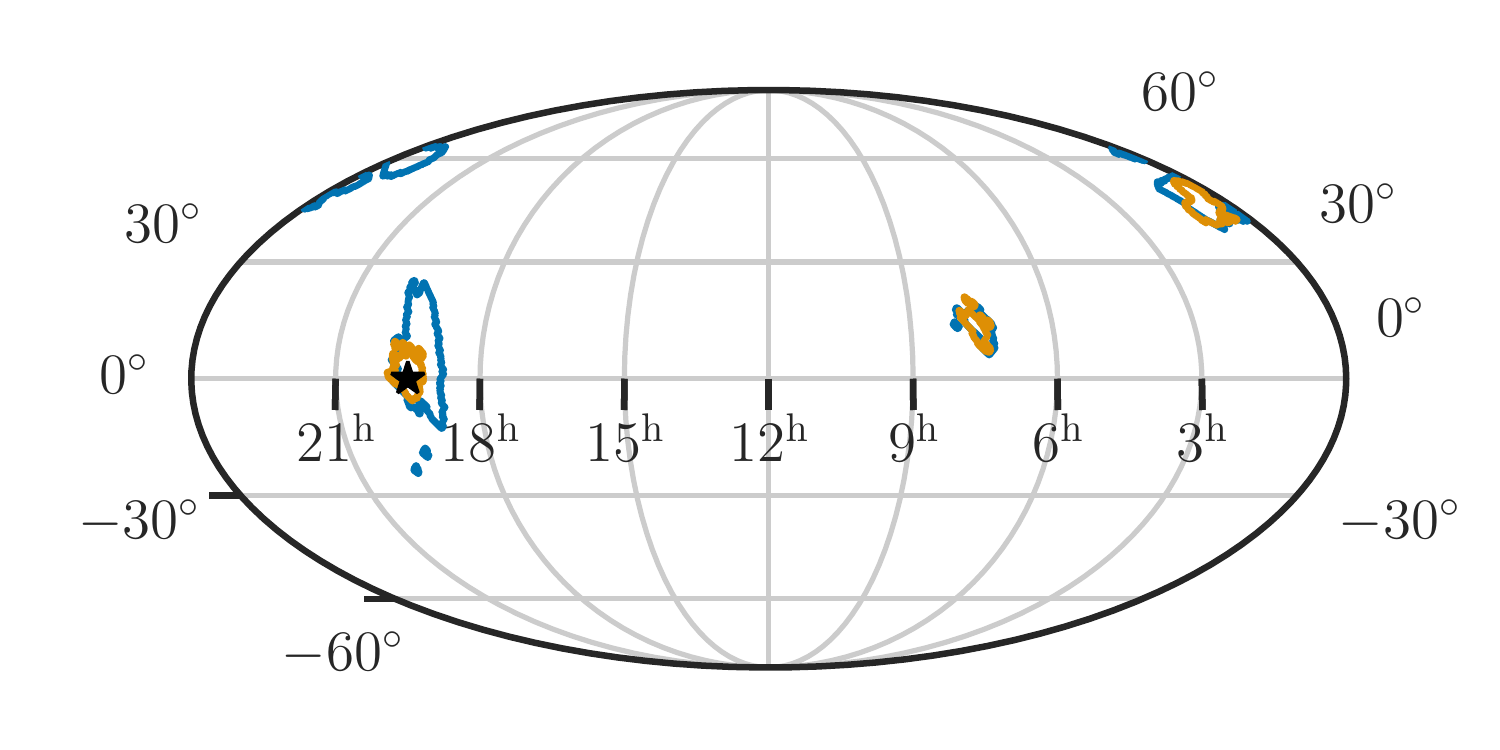}
    
     \end{subfigure}
     \hfill
     \begin{subfigure}[b]{0.22\textwidth}
         \centering
\includegraphics[width=\textwidth]{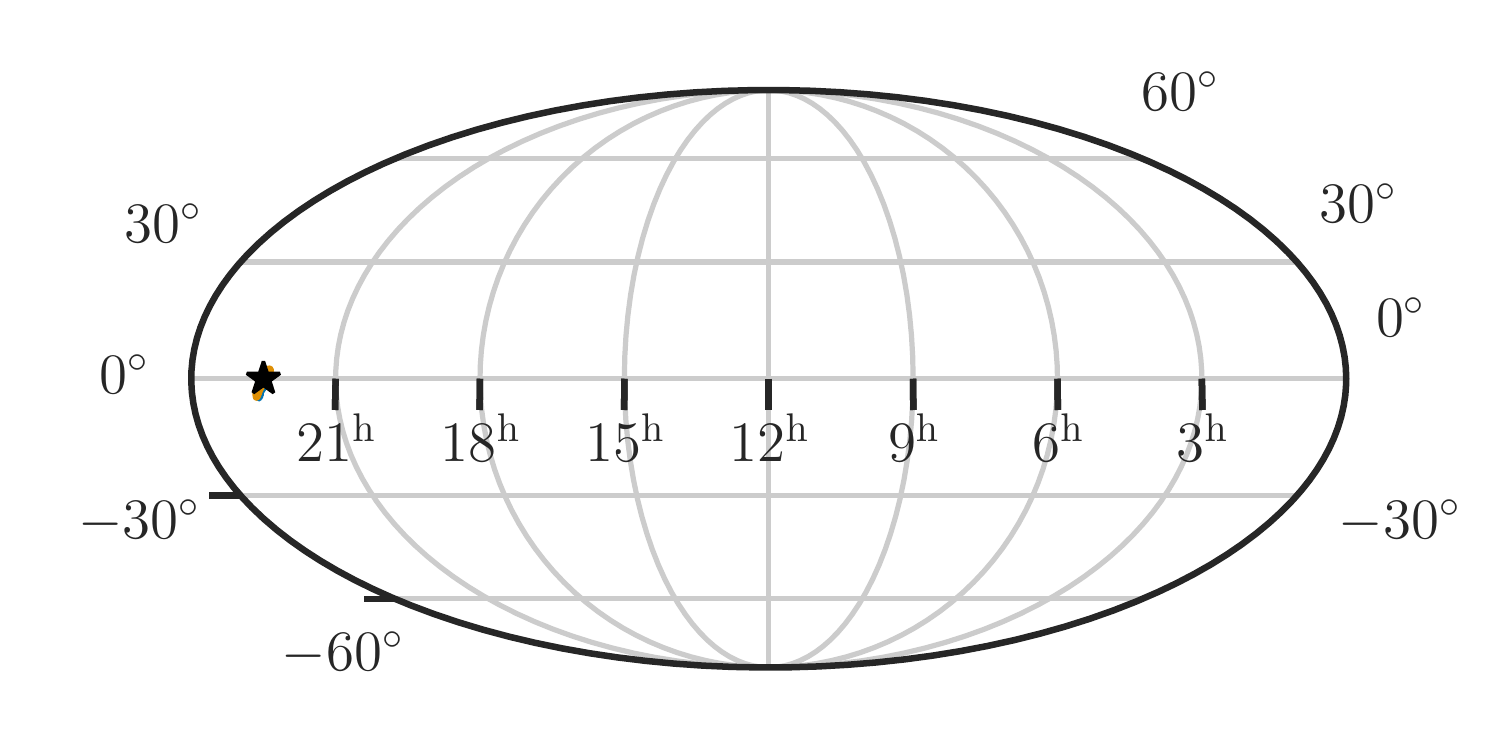}
     \end{subfigure}
     \hfill
     \begin{subfigure}[b]{0.22\textwidth}
         \centering
        \includegraphics[width=\textwidth]{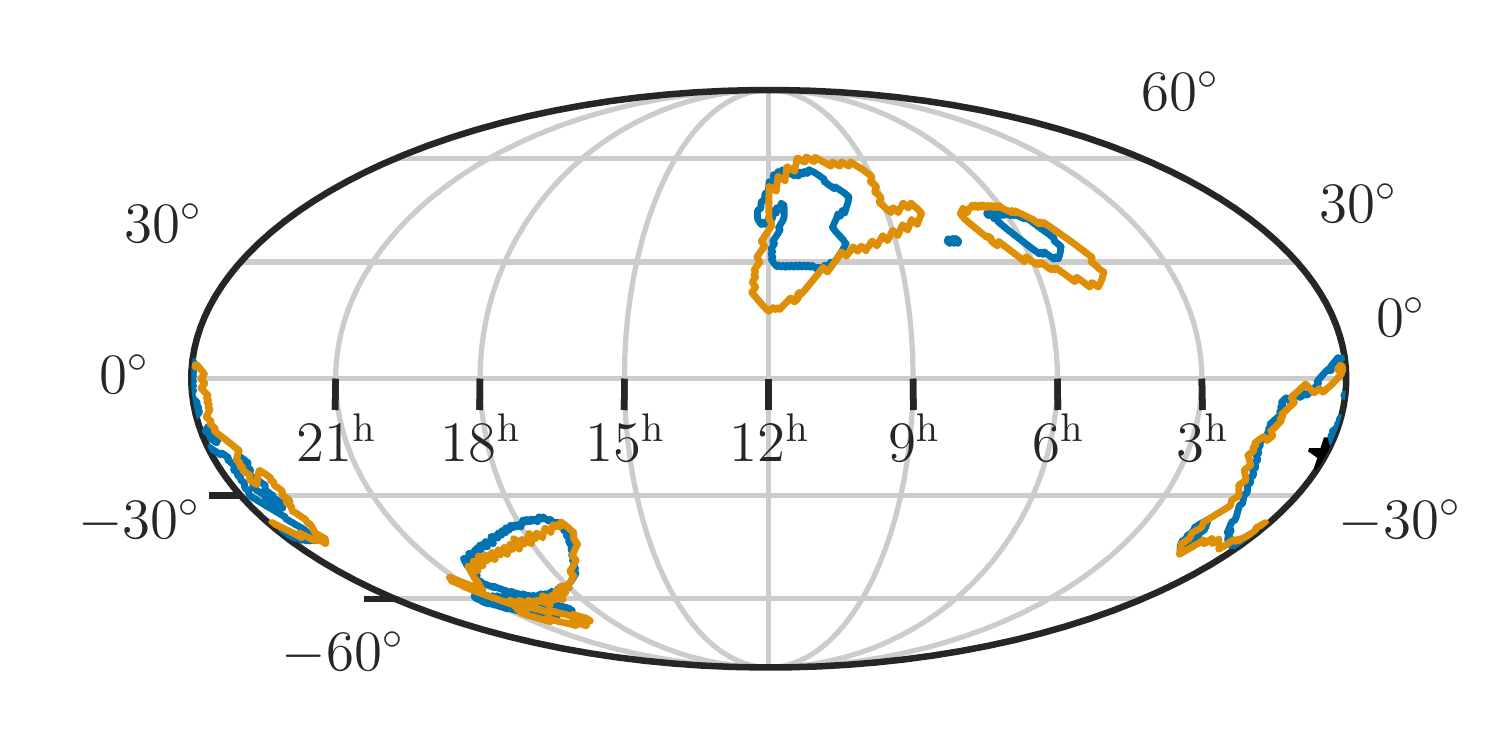}
    \end{subfigure}
     \hfill
     \begin{subfigure}[b]{0.22\textwidth}
         \centering
        \includegraphics[width=\textwidth]{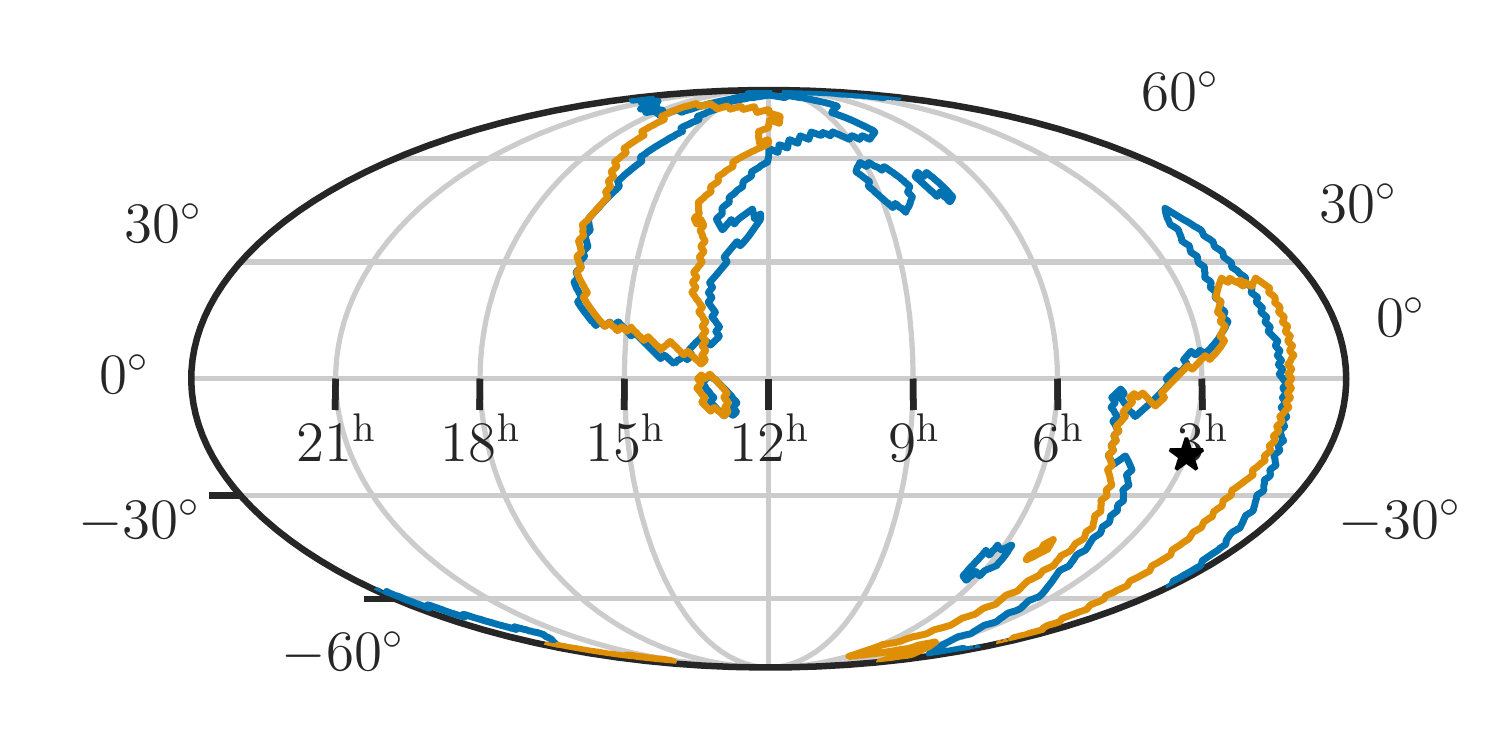}
     \end{subfigure}
     \hfill
     \begin{subfigure}[b]{0.22\textwidth}
         \centering
        \includegraphics[width=\textwidth]{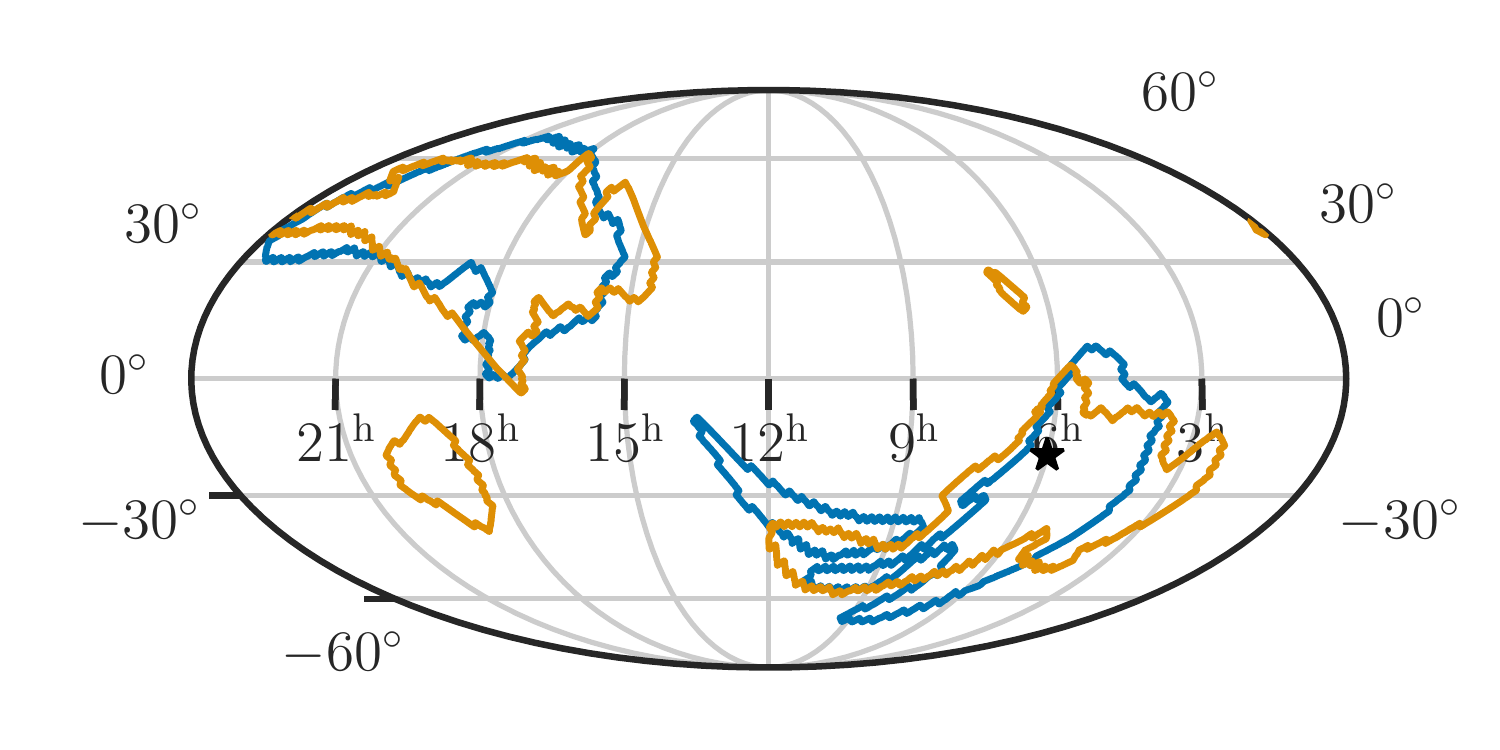}
    
     \end{subfigure}
     \hfill
     \begin{subfigure}[b]{0.22\textwidth}
         \centering
\includegraphics[width=\textwidth]{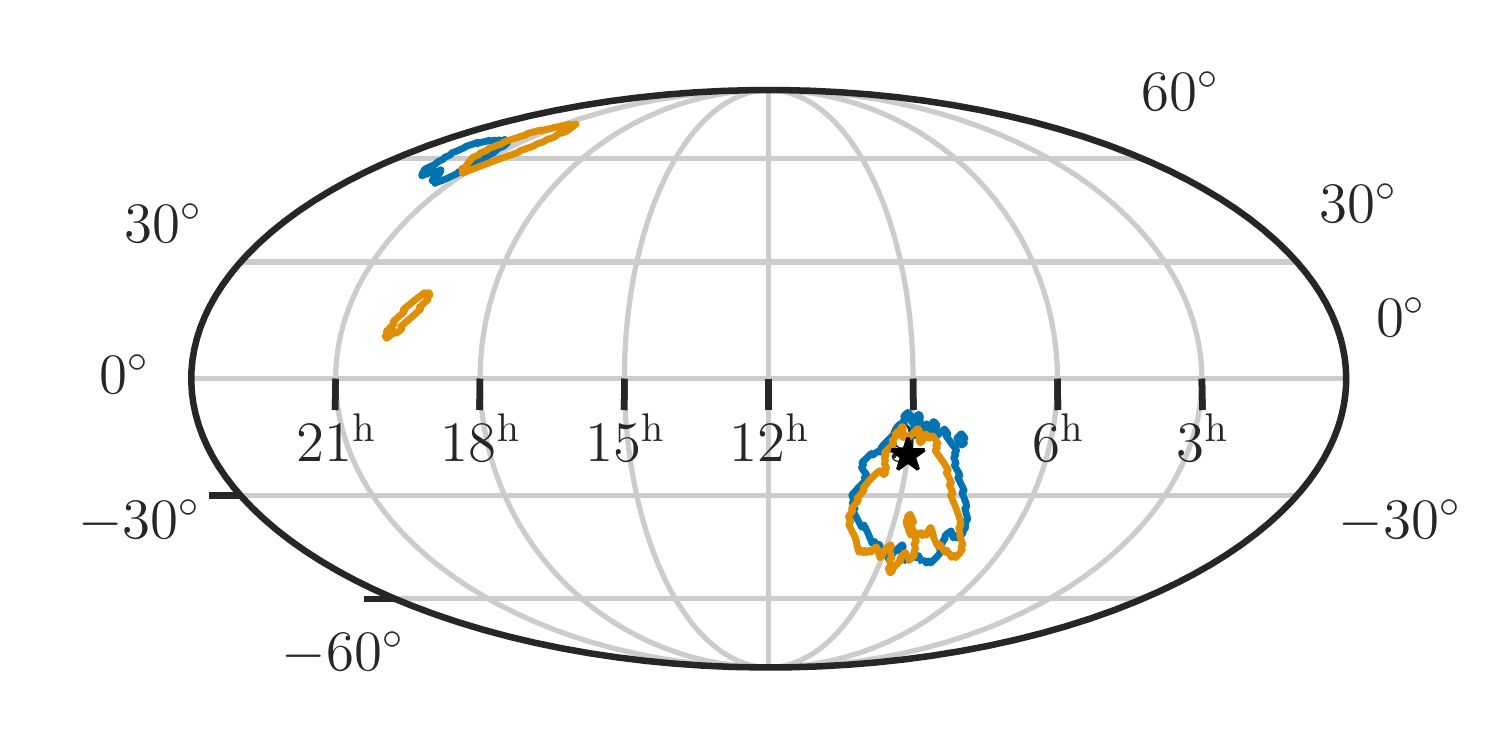}
     \end{subfigure}
     \hfill
     \begin{subfigure}[b]{0.22\textwidth}
         \centering
        \includegraphics[width=\textwidth]{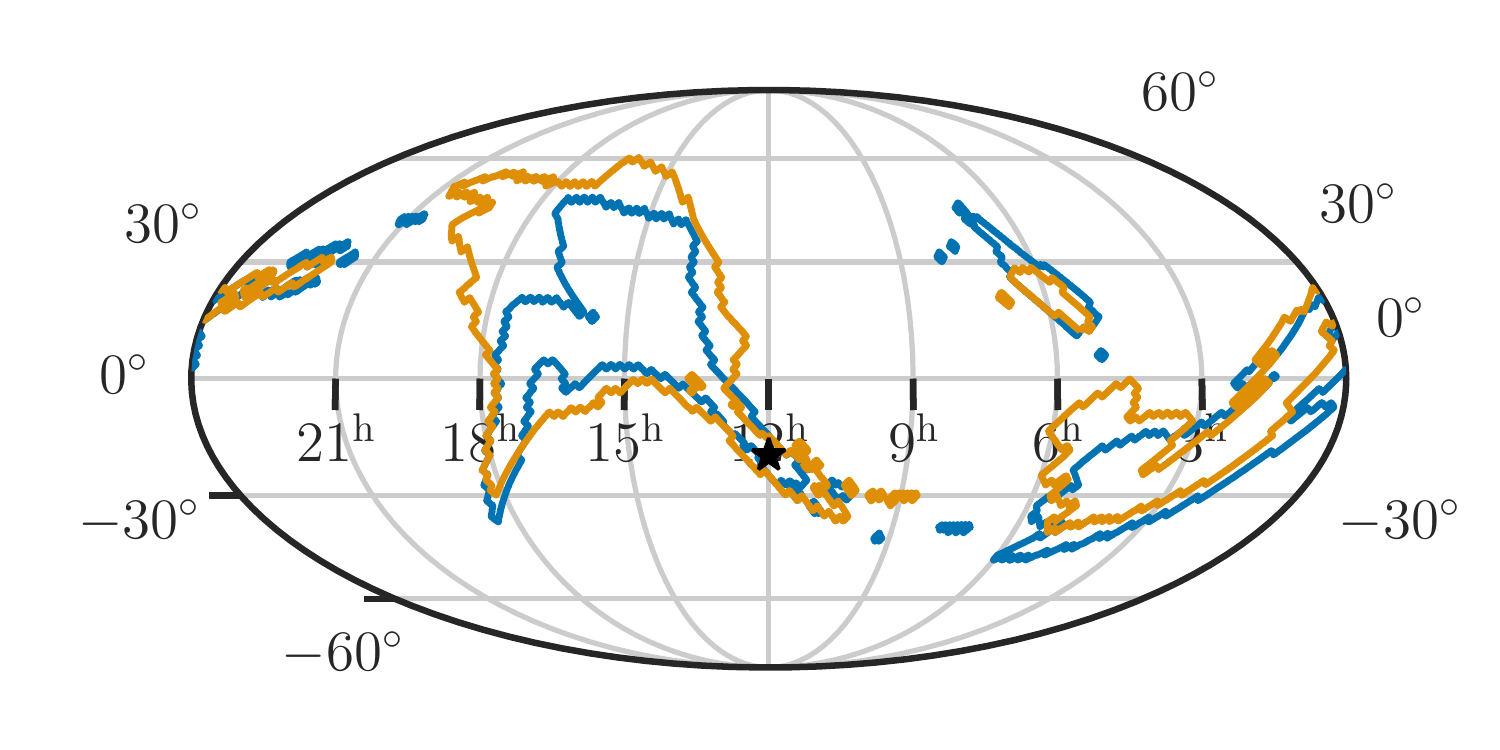}
    \end{subfigure}
     \hfill
     \begin{subfigure}[b]{0.22\textwidth}
         \centering
        \includegraphics[width=\textwidth]{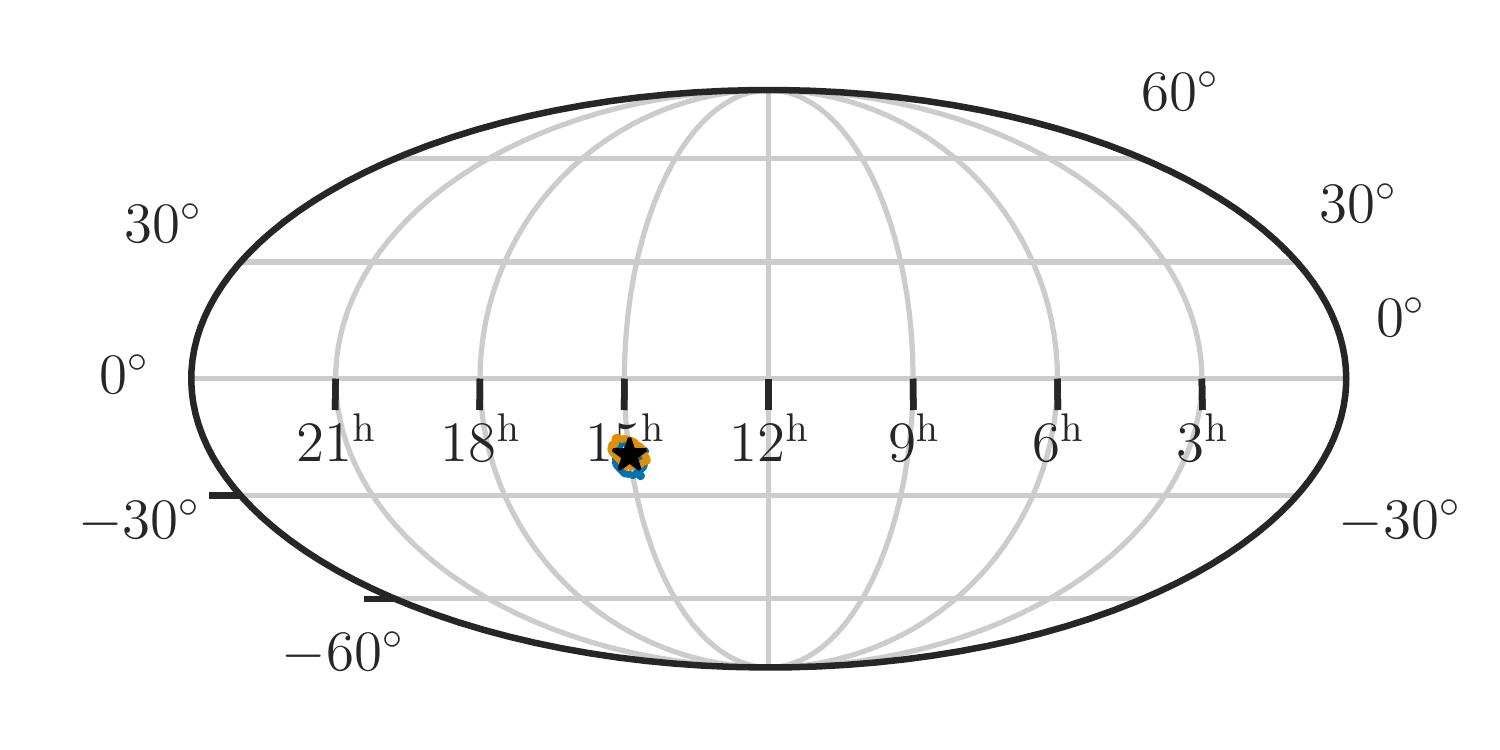}
     \end{subfigure}
     \hfill
     \begin{subfigure}[b]{0.22\textwidth}
         \centering
        \includegraphics[width=\textwidth]{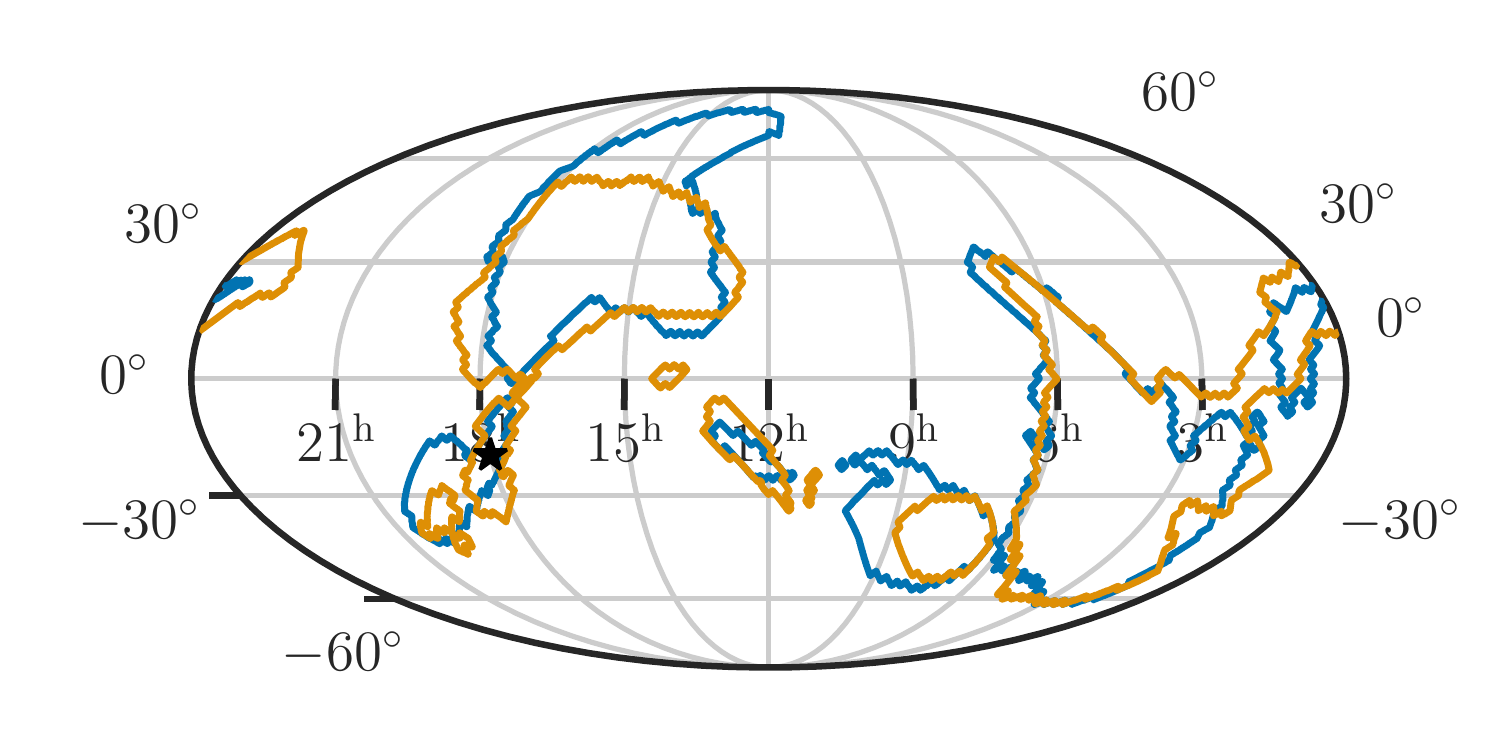}
    
     \end{subfigure}
     \hfill
     \begin{subfigure}[b]{0.22\textwidth}
         \centering
\includegraphics[width=\textwidth]{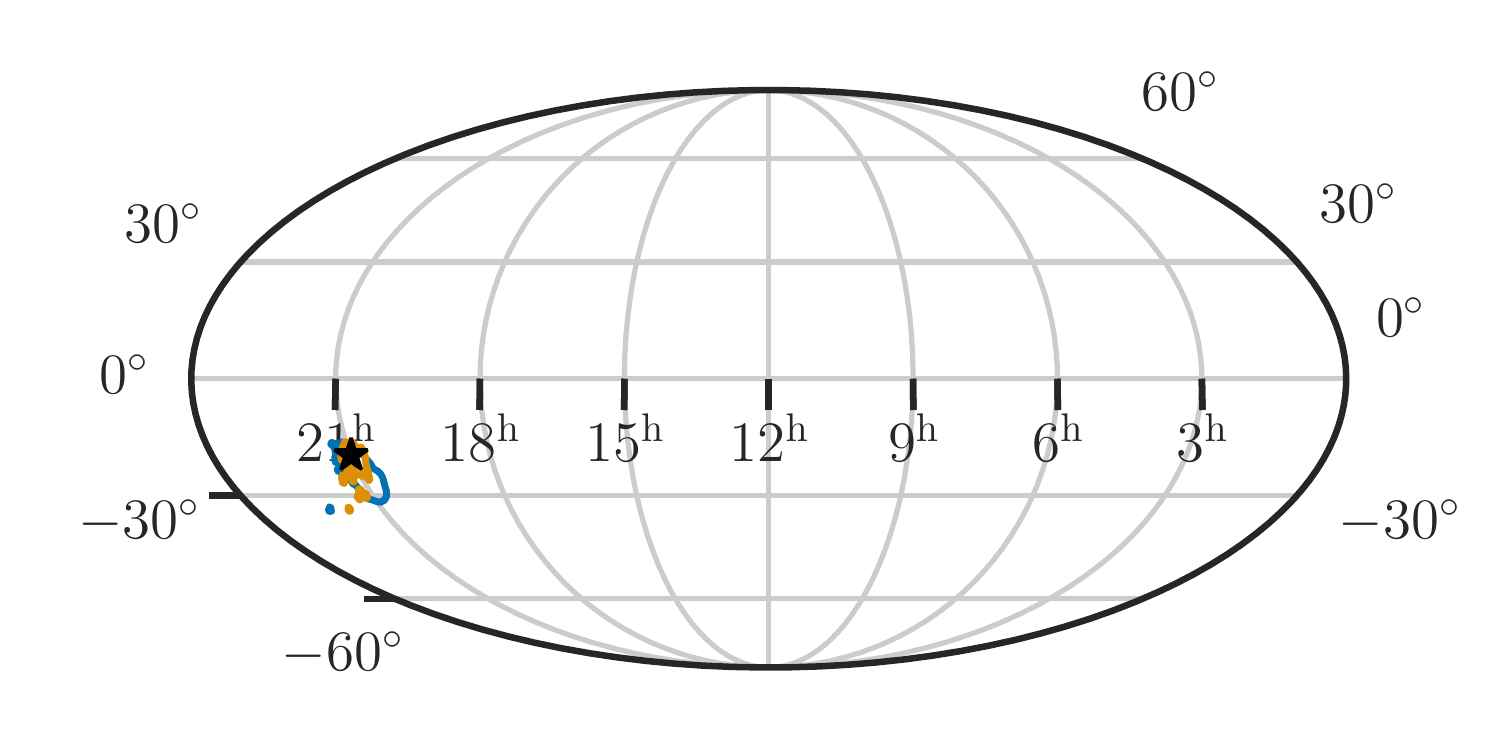}
     \end{subfigure}
     \hfill
     \begin{subfigure}[b]{0.22\textwidth}
         \centering
        \includegraphics[width=\textwidth]{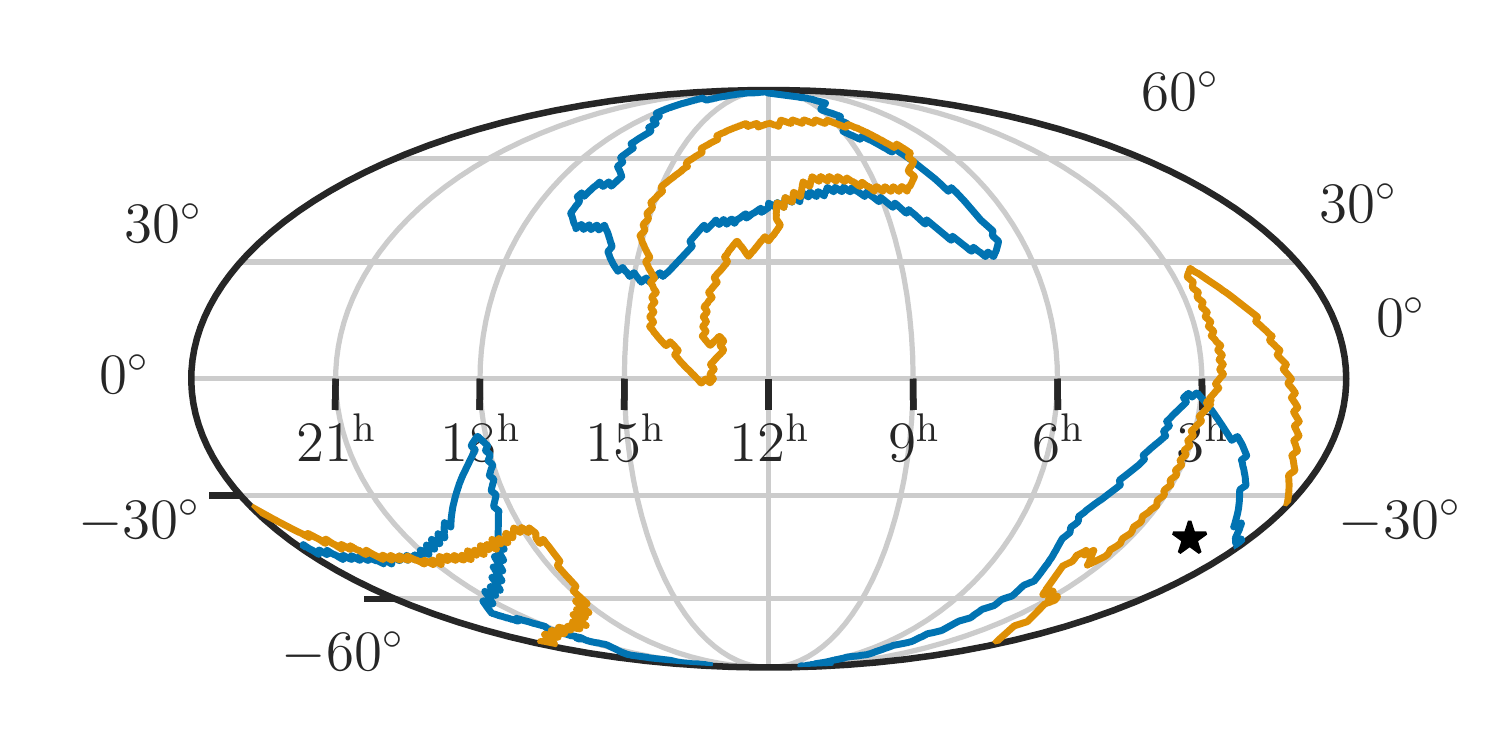}
    \end{subfigure}
     \hfill
     \begin{subfigure}[b]{0.22\textwidth}
         \centering
        \includegraphics[width=\textwidth]{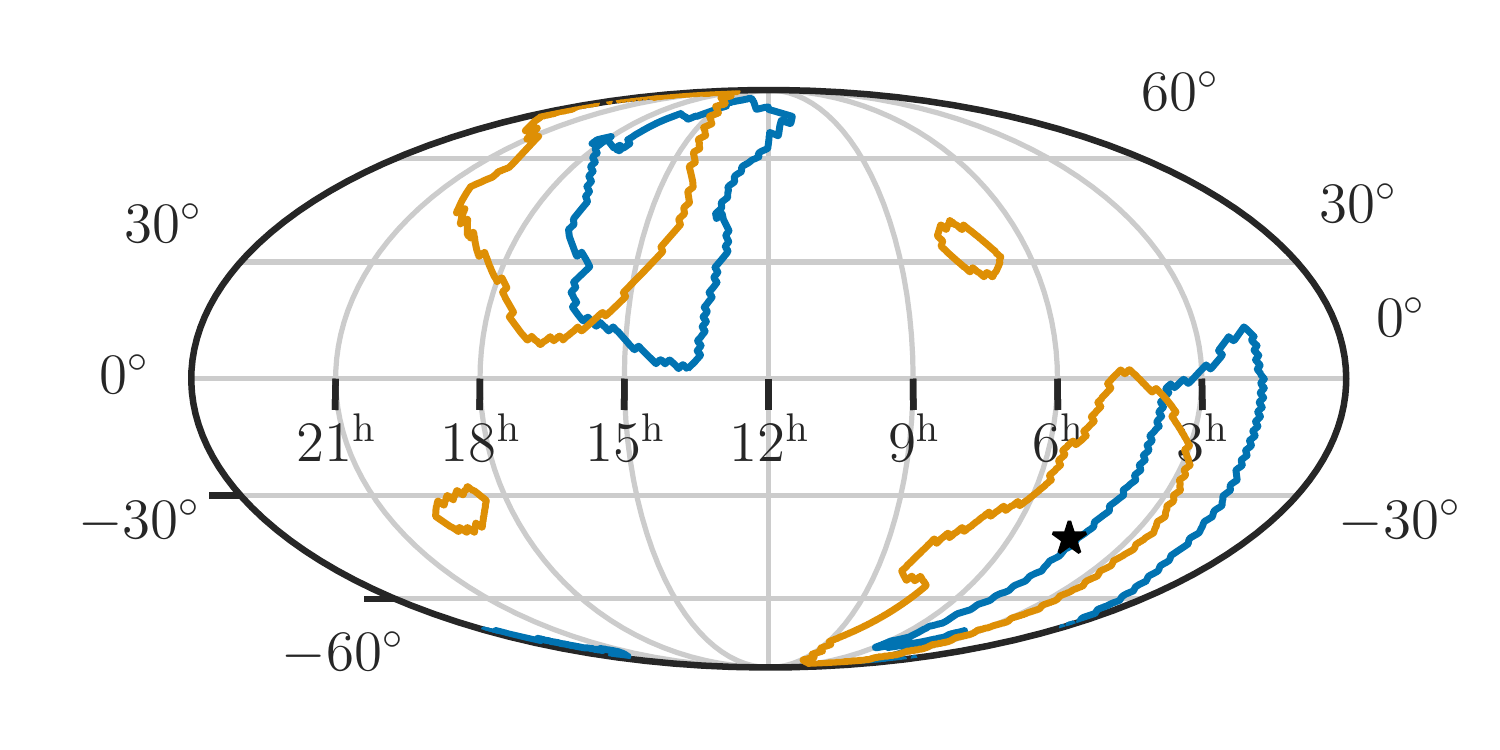}
     \end{subfigure}
     \hfill
     \begin{subfigure}[b]{0.22\textwidth}
         \centering
        \includegraphics[width=\textwidth]{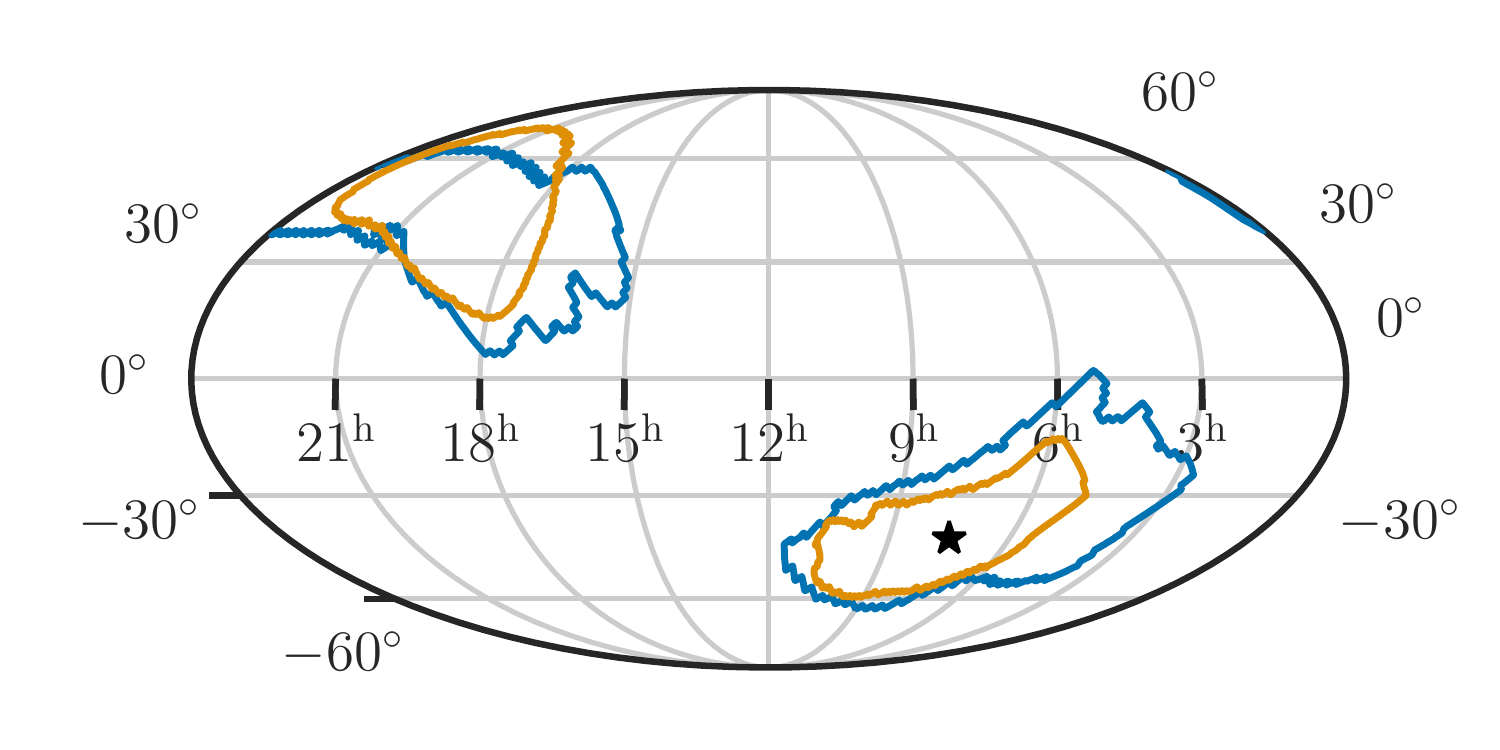}
    
     \end{subfigure}
     \hfill
     \begin{subfigure}[b]{0.22\textwidth}
         \centering
\includegraphics[width=\textwidth]{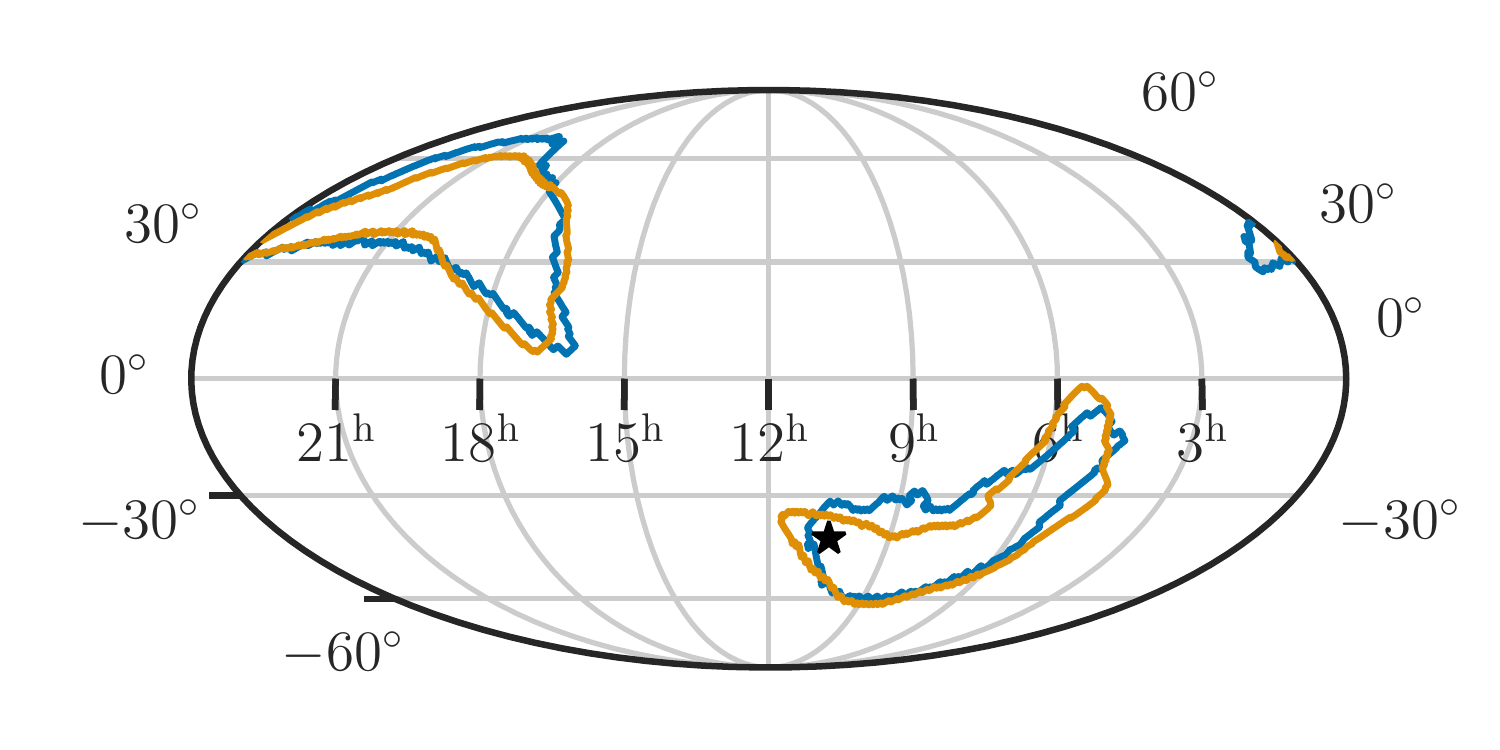}
     \end{subfigure}
     \hfill
     \begin{subfigure}[b]{0.22\textwidth}
         \centering
        \includegraphics[width=\textwidth]{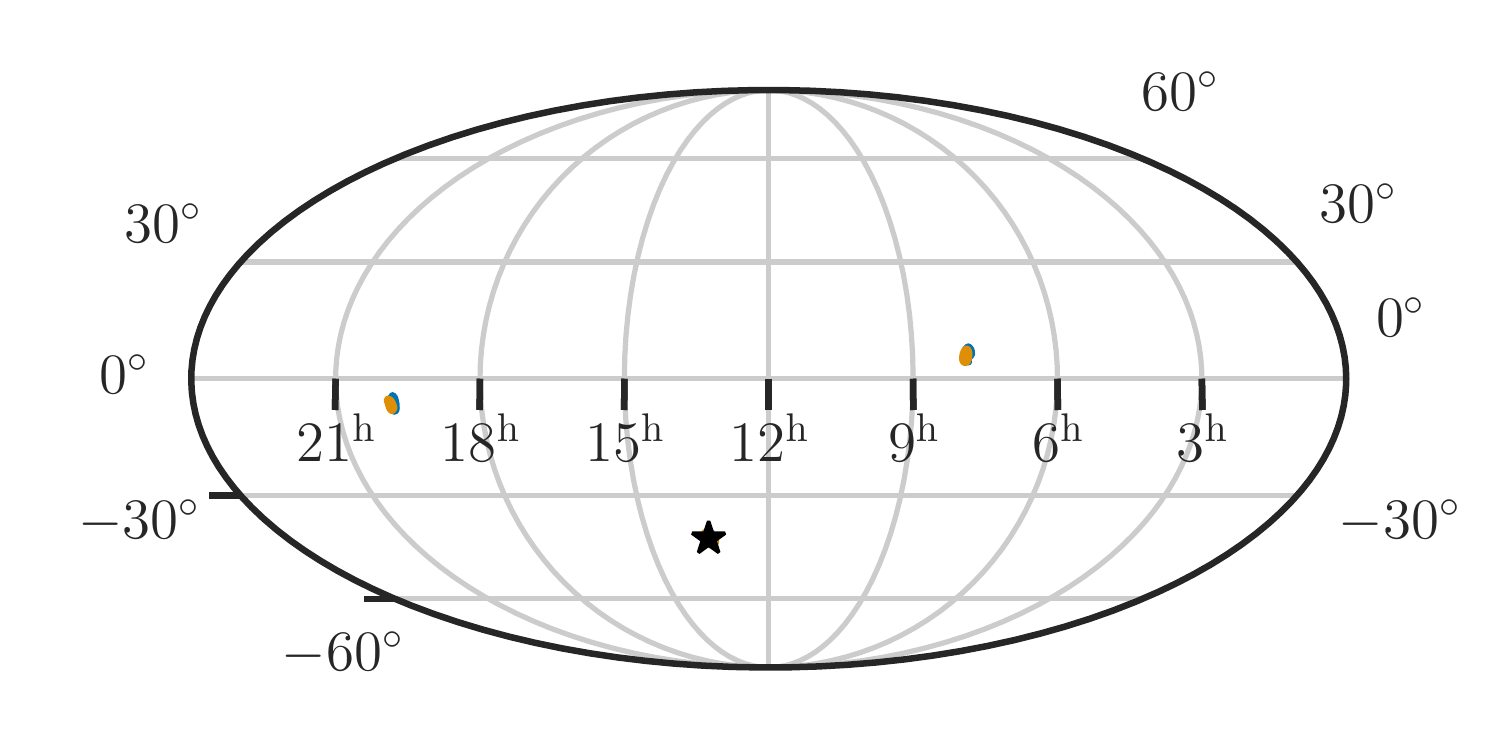}
    \end{subfigure}
     \hfill
     \begin{subfigure}[b]{0.22\textwidth}
         \centering
        \includegraphics[width=\textwidth]{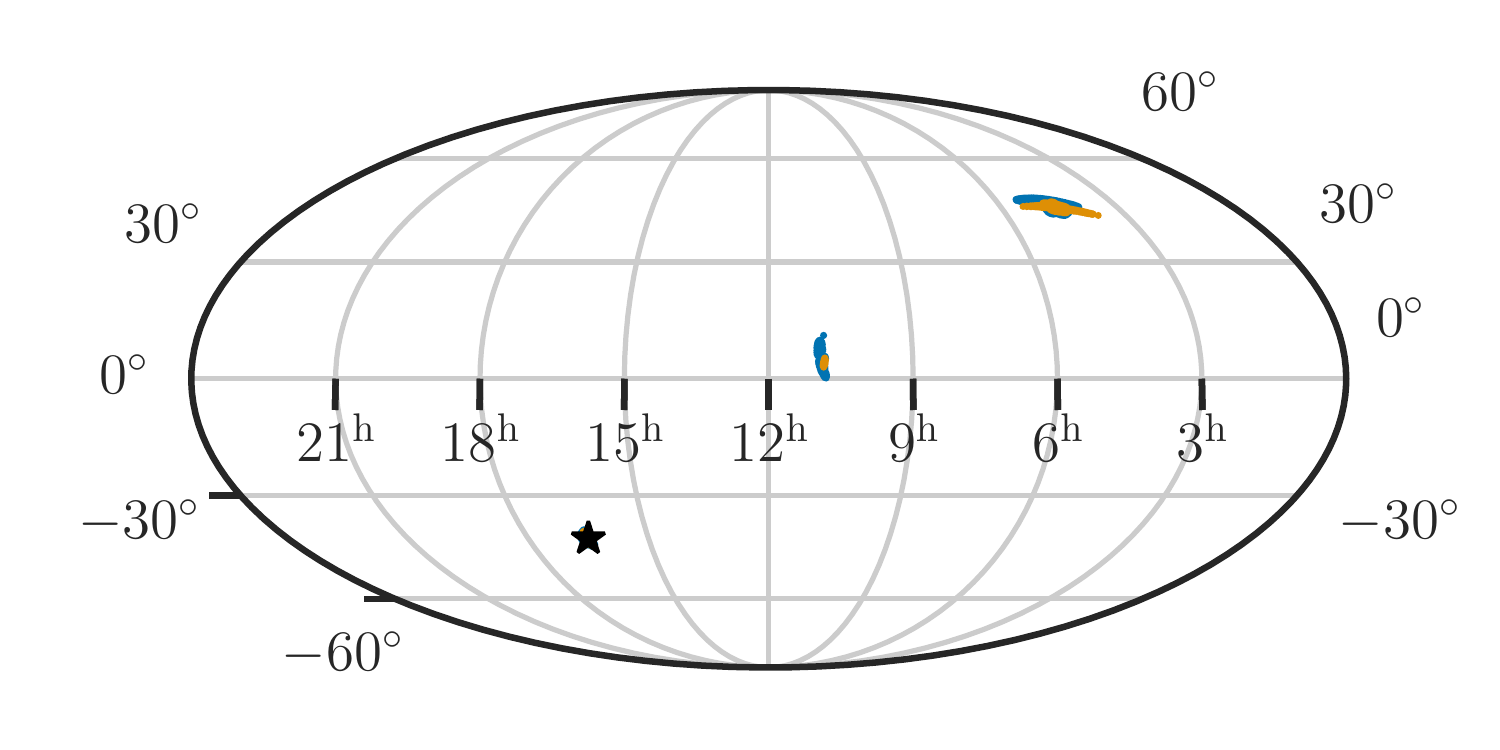}
     \end{subfigure}
     \hfill
     \begin{subfigure}[b]{0.22\textwidth}
         \centering
        \includegraphics[width=\textwidth]{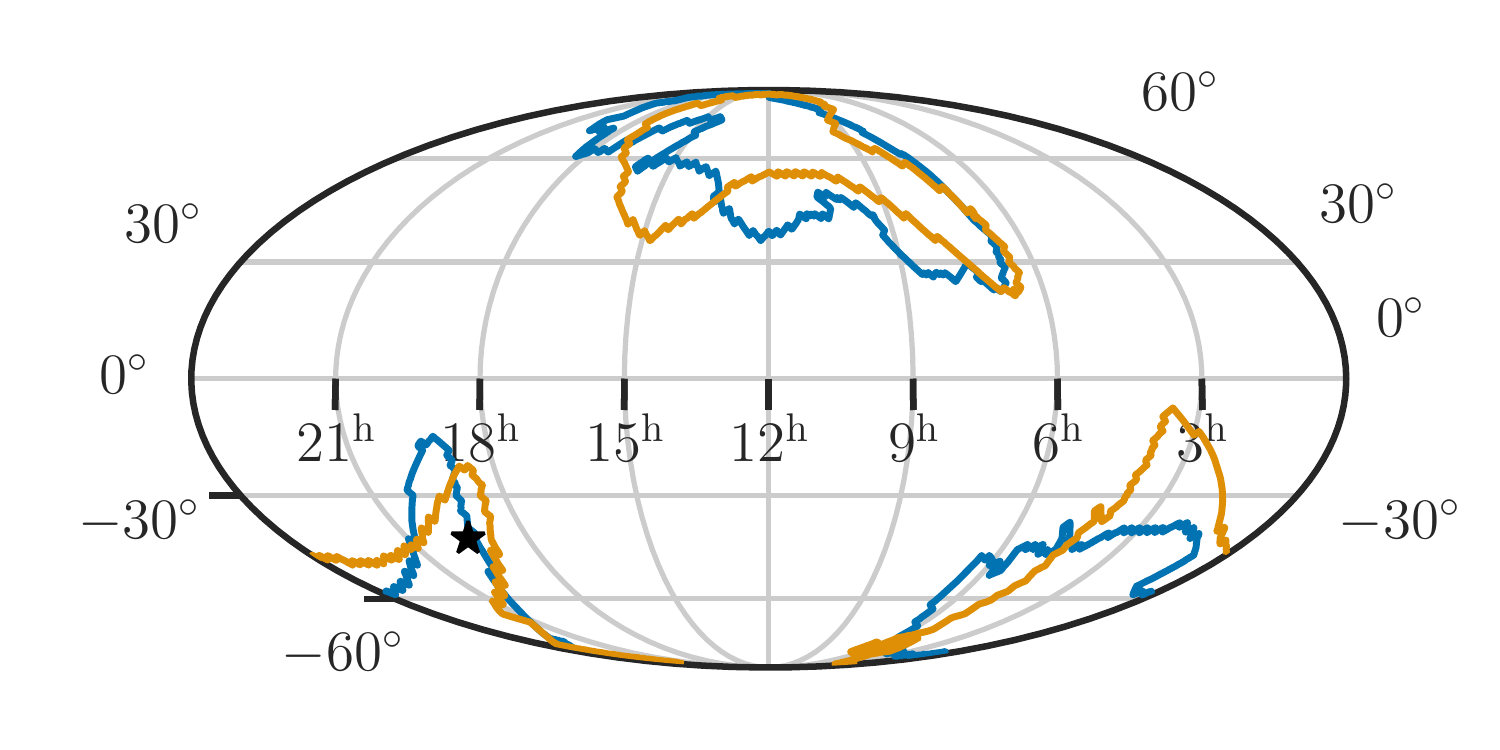}
    
     \end{subfigure}
     \hfill
     \begin{subfigure}[b]{0.22\textwidth}
         \centering
\includegraphics[width=\textwidth]{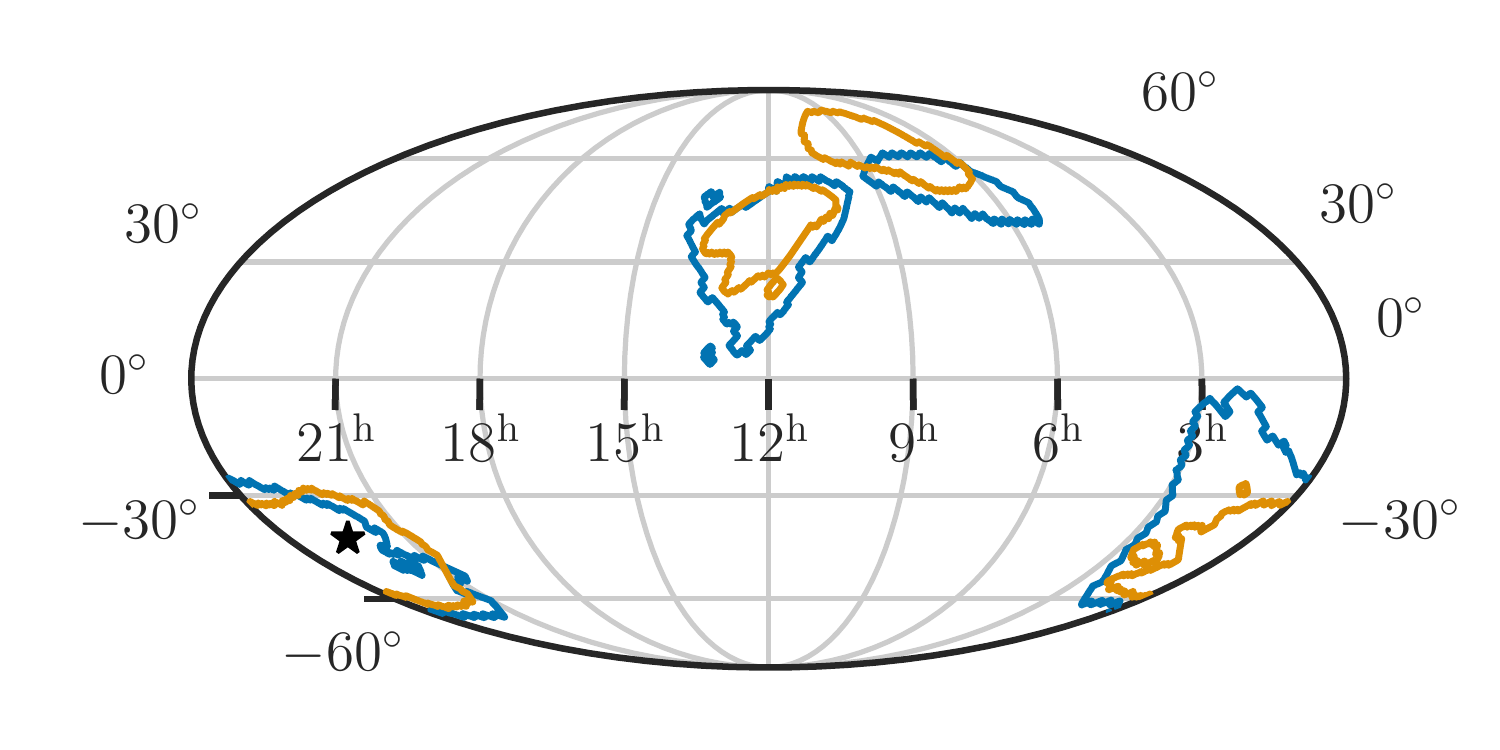}
     \end{subfigure}
     \hfill
     \begin{subfigure}[b]{0.22\textwidth}
         \centering
        \includegraphics[width=\textwidth]{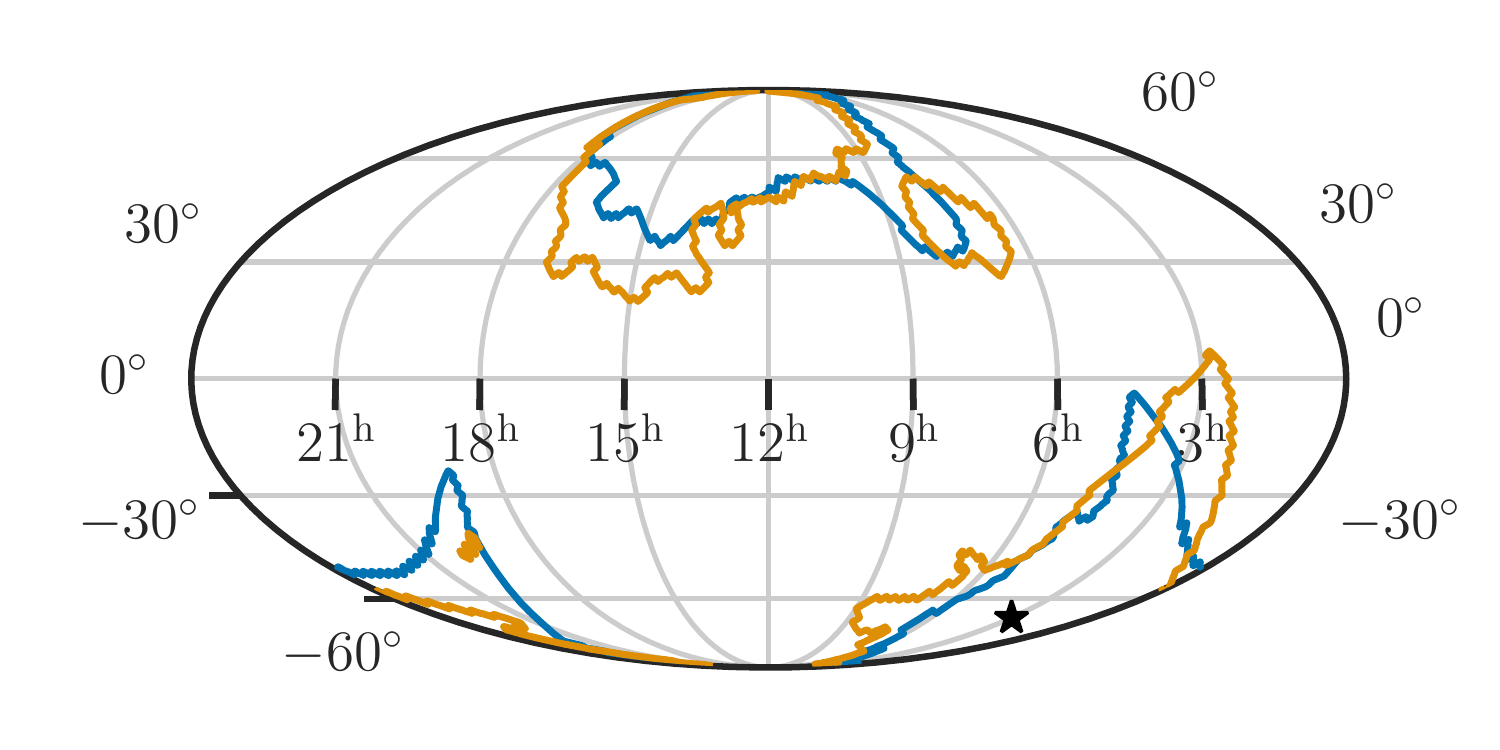}
    \end{subfigure}
     \hfill
     \begin{subfigure}[b]{0.22\textwidth}
         \centering
        \includegraphics[width=\textwidth]{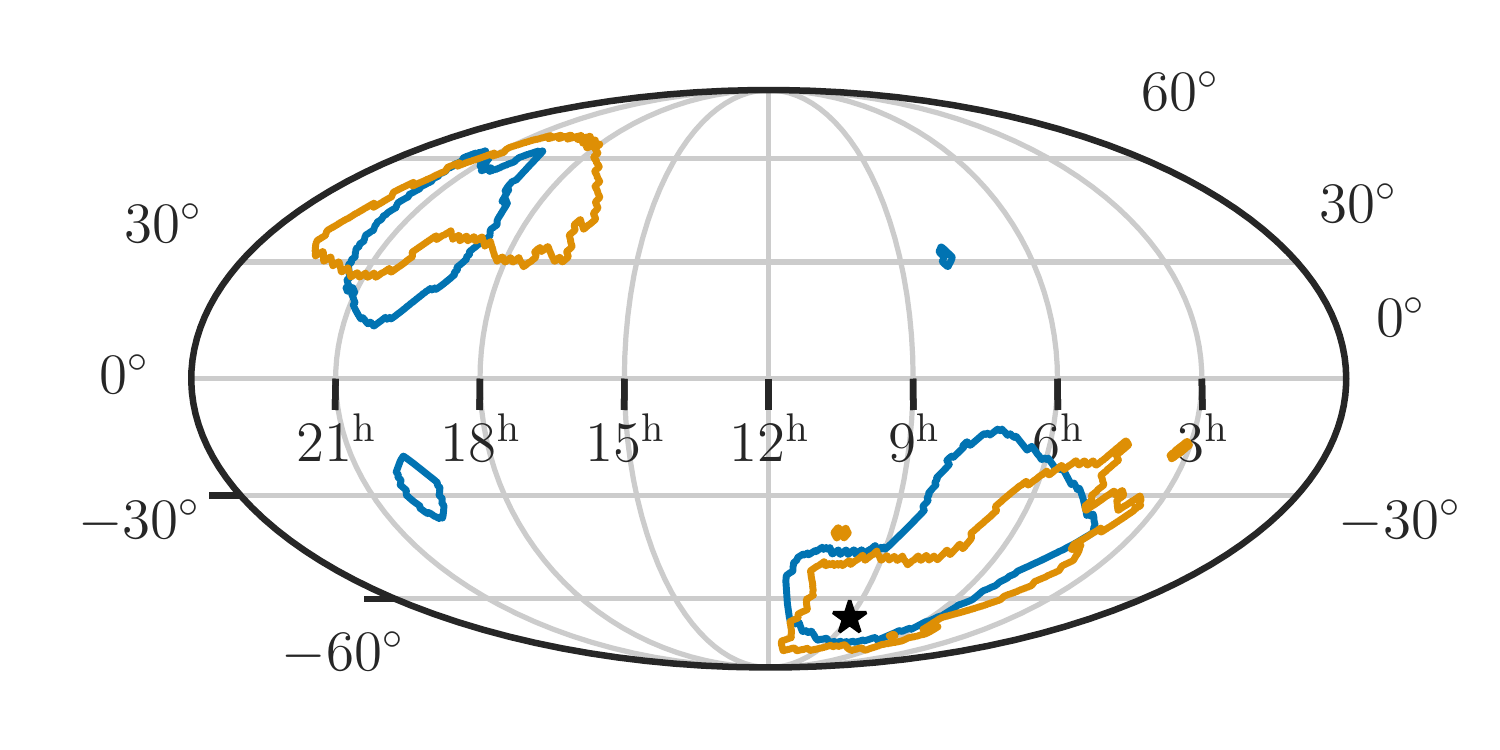}
     \end{subfigure}
     \hfill
     \begin{subfigure}[b]{0.22\textwidth}
         \centering
        \includegraphics[width=\textwidth]{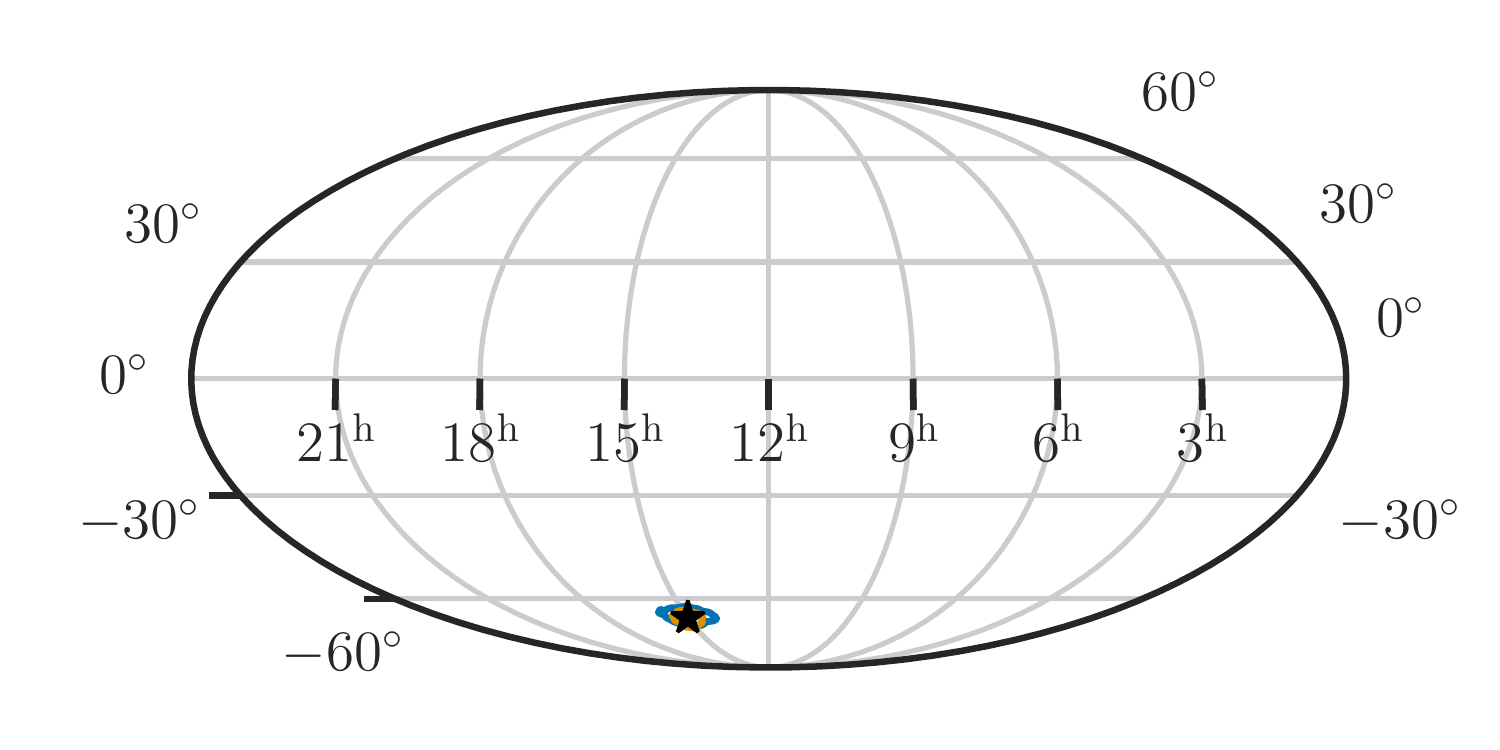}
    
     \end{subfigure}
     \hfill
     \begin{subfigure}[b]{0.22\textwidth}
         \centering
\includegraphics[width=\textwidth]{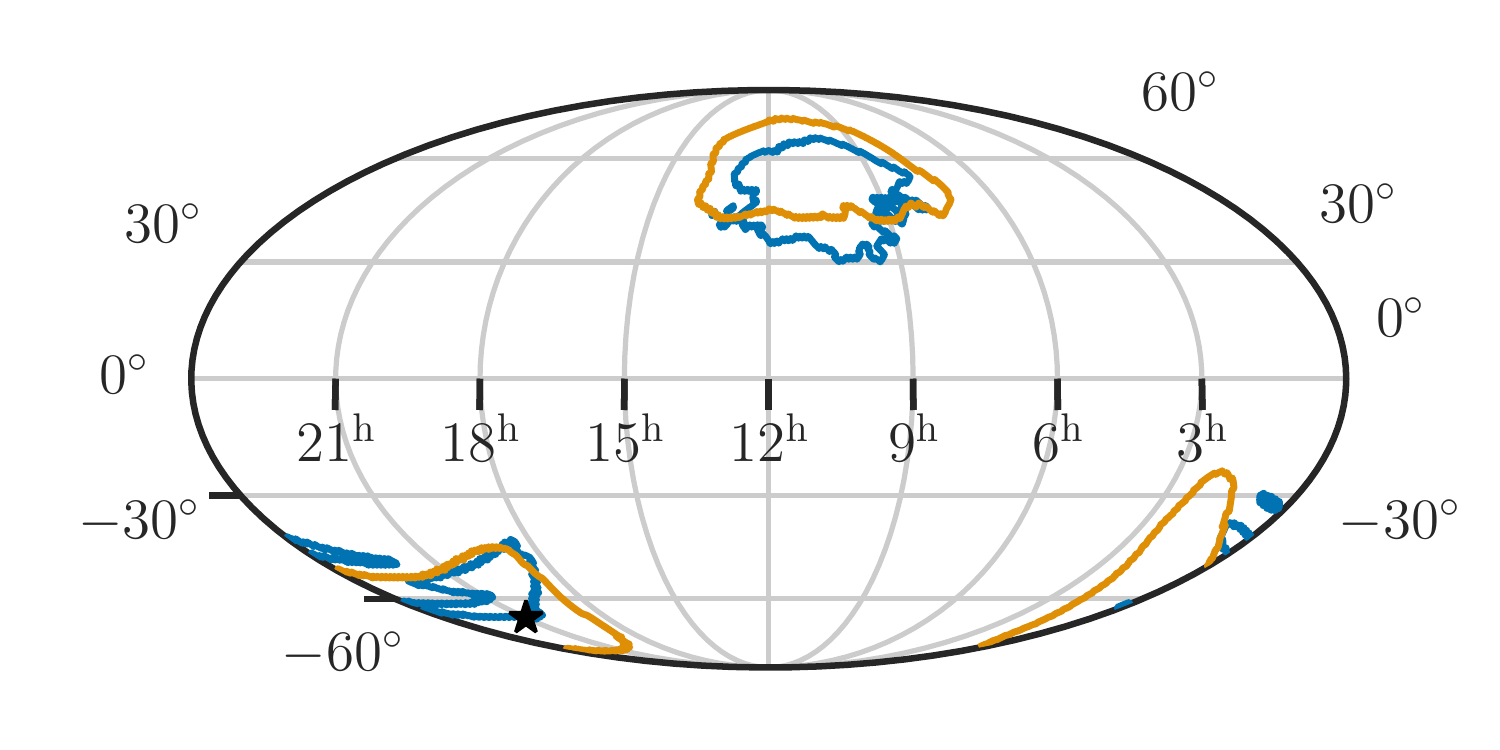}
     \end{subfigure}
     \hfill
        \caption{Posterior distributions showing the 2D sky areas for all the injections used in this paper using a Mollweide projection in an Earth-Centric coordinate system. The blue and orange contours refer to injections with $\theta_{\rm JN}$ values of 1.0 and 0.2 respectively.}
        \label{fig:three graphs}
\end{figure*}

We present the 2-D skymaps for all the injections in Figure \ref{fig:three graphs}. Note that all events have an optimal SNR of 1000. All the posteriors have support at the injected values. Some injections are well-localized but multimodal.
\section{The best localized injection}\label{appendix B}
\begin{figure*}[h!]
     \centering
     \begin{subfigure}[b]{0.49\textwidth}
         \centering
        \includegraphics[width=\textwidth]{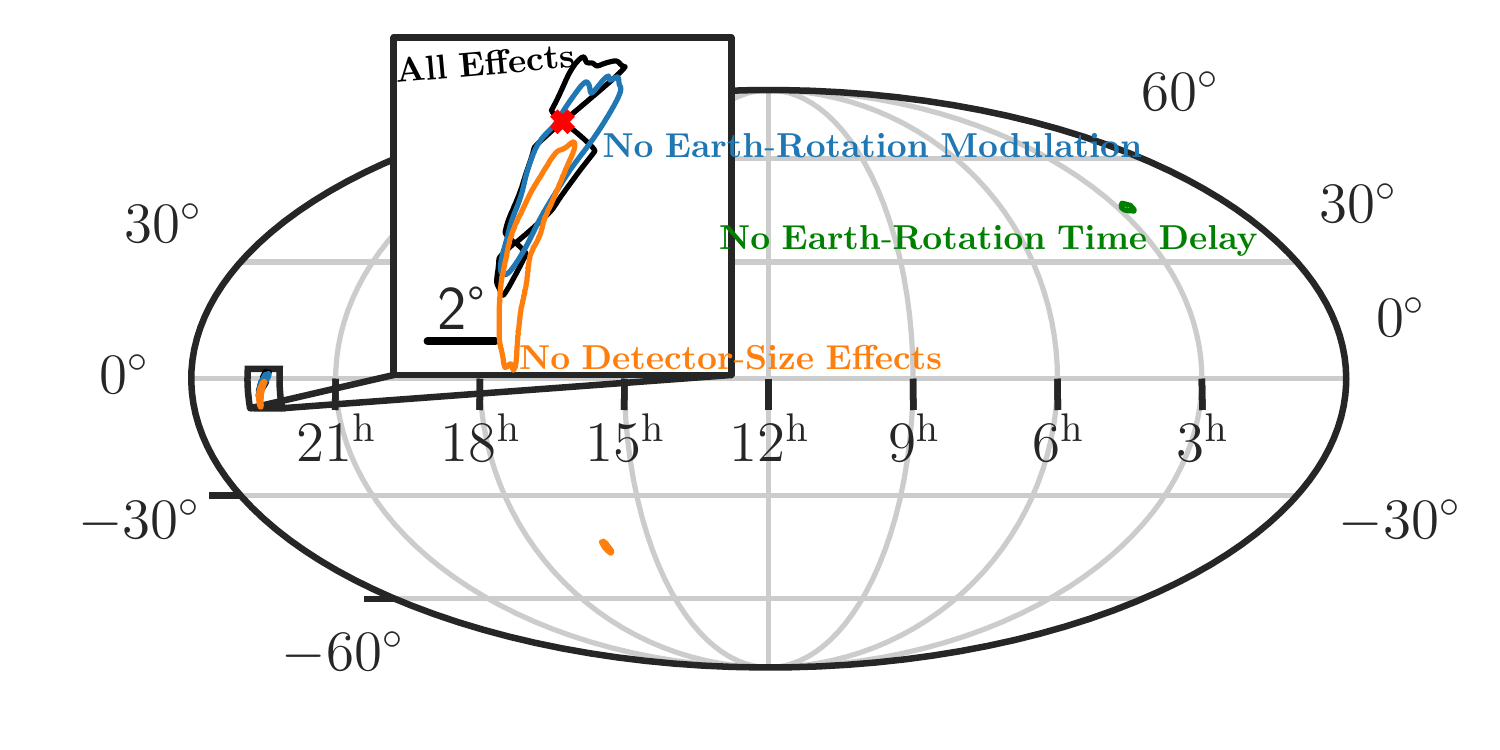}
    \end{subfigure}
     \hfill
     \begin{subfigure}[b]{0.49\textwidth}
         \centering
        \includegraphics[width=\textwidth]{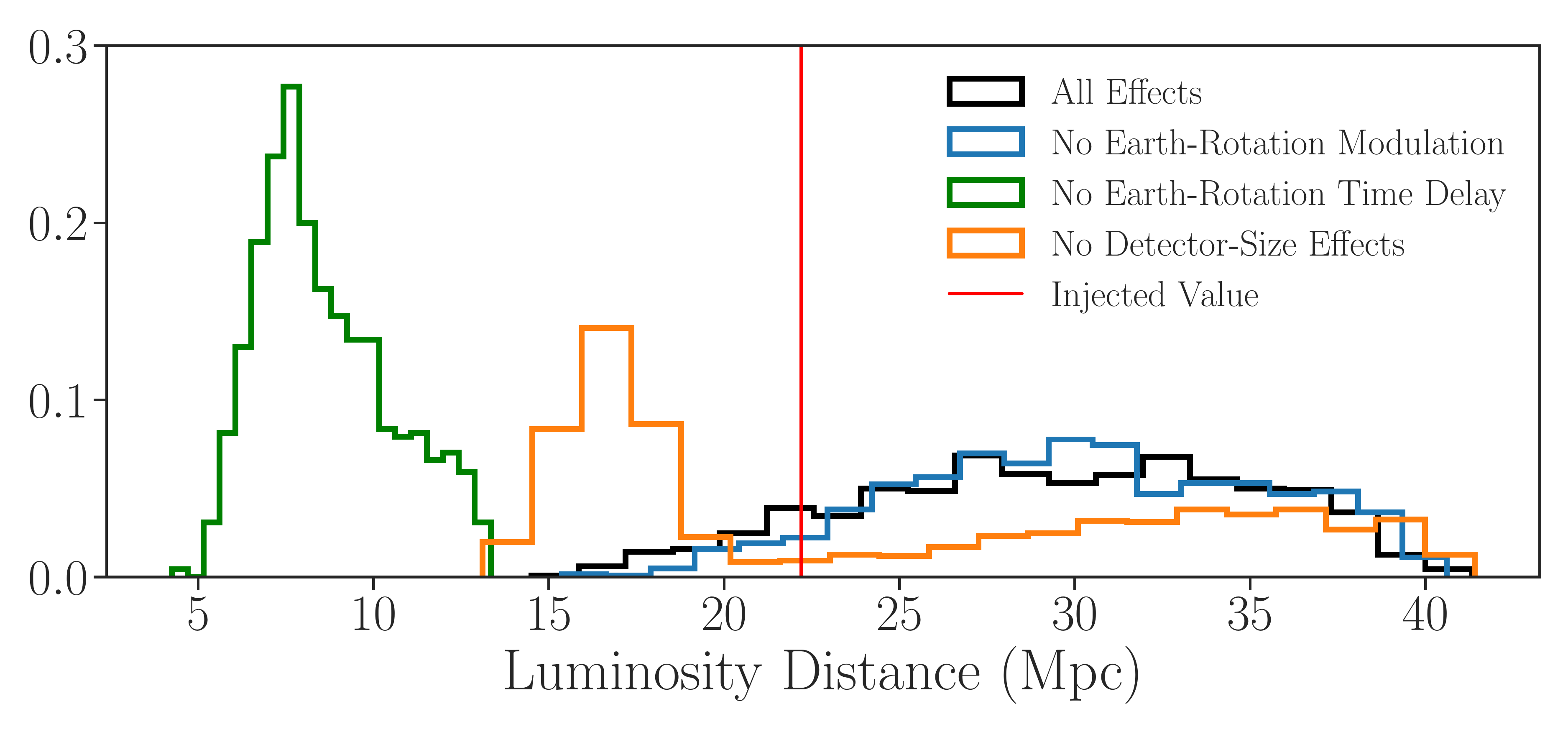}
     \end{subfigure}
     \hfill
     \caption{2D sky localization posterior (left) and inferred luminosity distance posterior (right) for the best-localized event (marked by + in Figure \ref{fig:Sky}) with $\theta_{\rm JN}$ = 1.0. The black contour/ histogram  is for the inference considering all effects. The blue, orange, and green contours/ histograms ignore Earth-rotation amplitude modulation, detector-size amplitude modulation, and Earth-rotation time delay respectively. \phantom{Hope you enjoyed reading it!! Have an enjoyable and productive time ahead. Thank you and goodbye. }}
    \label{fig:b1}
\end{figure*}

\begin{figure*}[h!]
     \centering
     \begin{subfigure}[b]{0.49\textwidth}
         \centering
        \includegraphics[width=\textwidth]{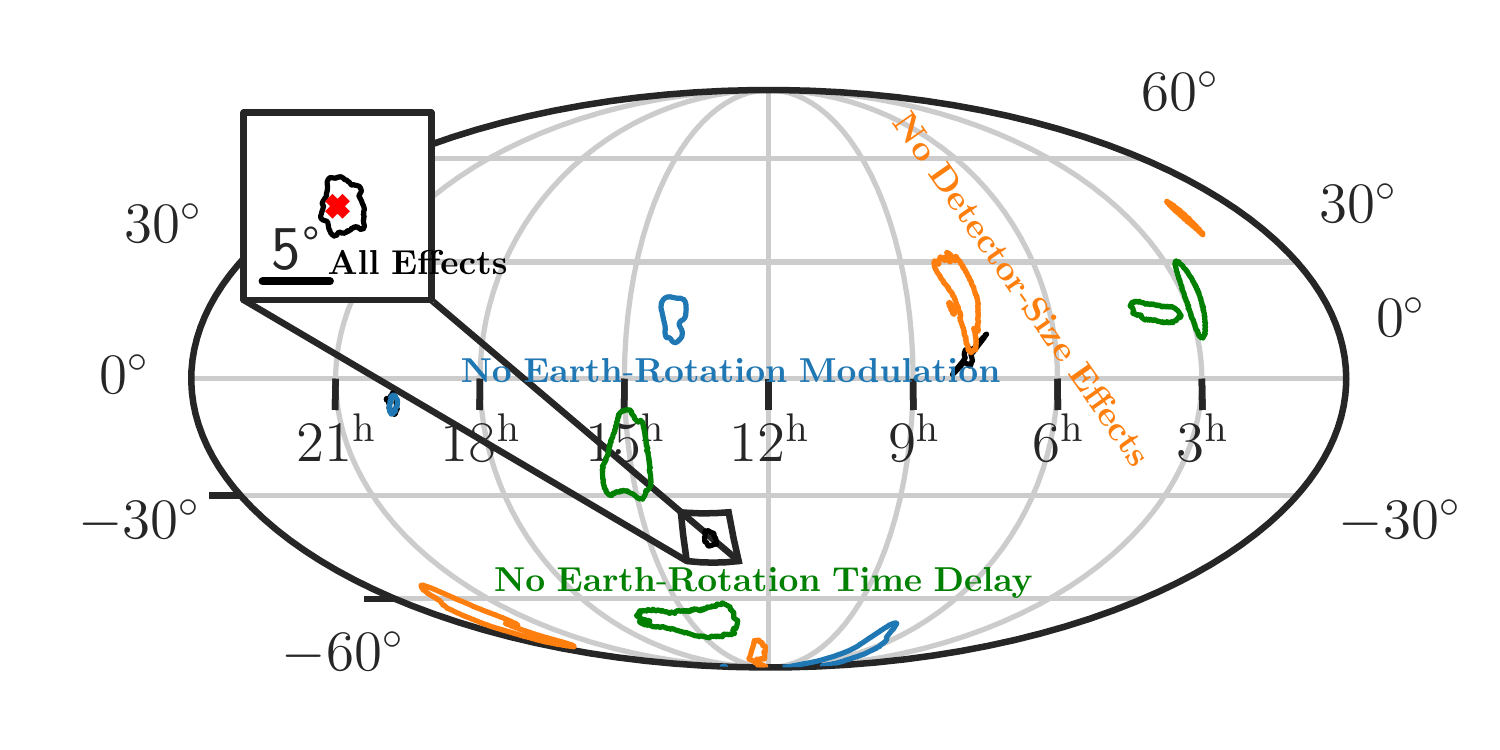}
    \end{subfigure}
     \hfill
     \begin{subfigure}[b]{0.49\textwidth}
         \centering
        \includegraphics[width=\textwidth]{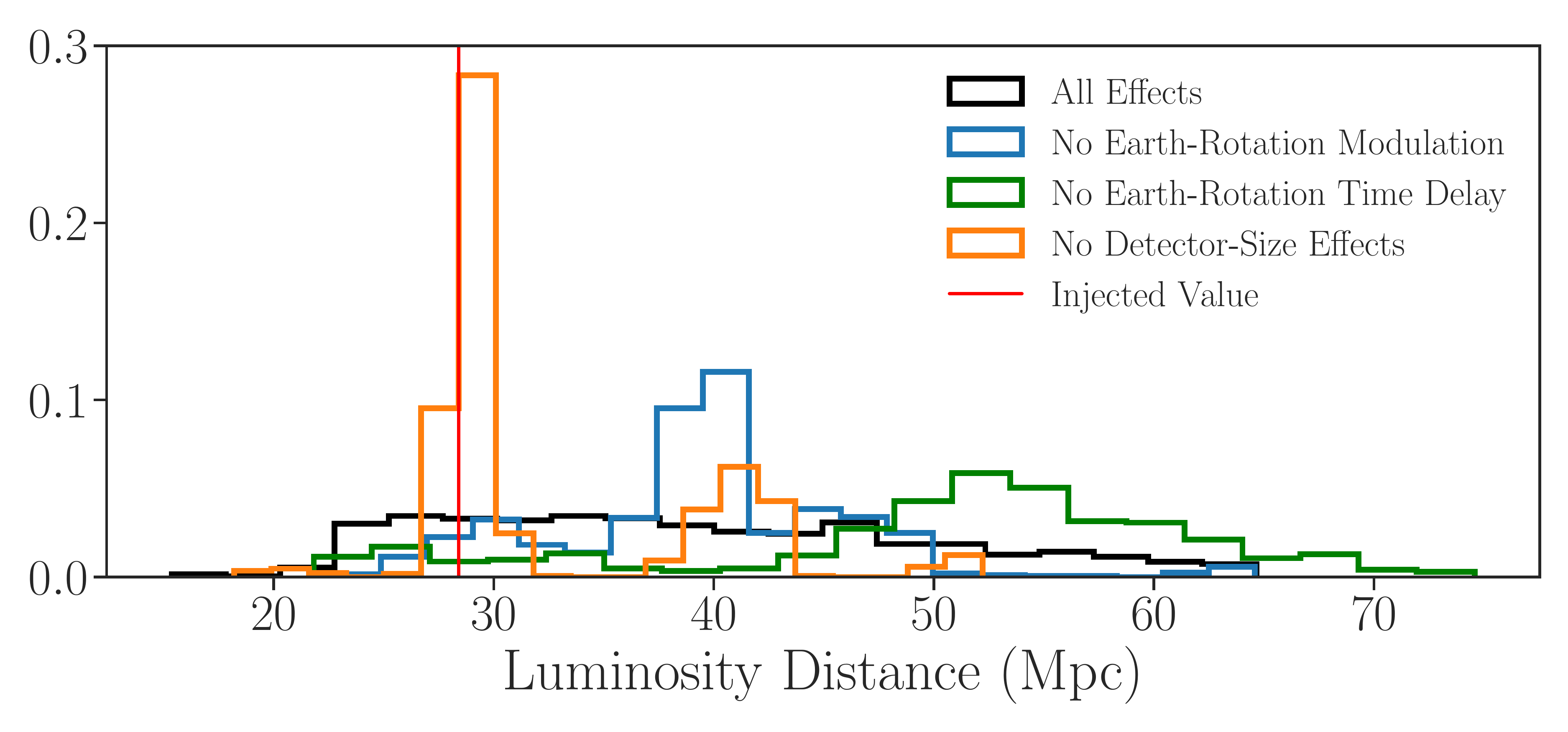}
     \end{subfigure}
     \hfill
     \caption{2D sky localization posterior (left) and inferred luminosity distance posterior (right) for an event on one of the dark spots with $\theta_{\rm JN}$ = 1.0. The black contour/ histogram  is for the inference considering all effects. The blue, orange, and green contours/ histograms ignore Earth-rotation amplitude modulation, detector-size amplitude modulation, and Earth-rotation time delay respectively. \phantom{Hope you enjoyed reading it!! Have an enjoyable and productive time ahead. Thank you and goodbye. Take care. }}
    \label{fig:b3}
\end{figure*}

\begin{figure*}[h!]
     \centering
     \begin{subfigure}[b]{0.49\textwidth}
         \centering
        \includegraphics[width=\textwidth]{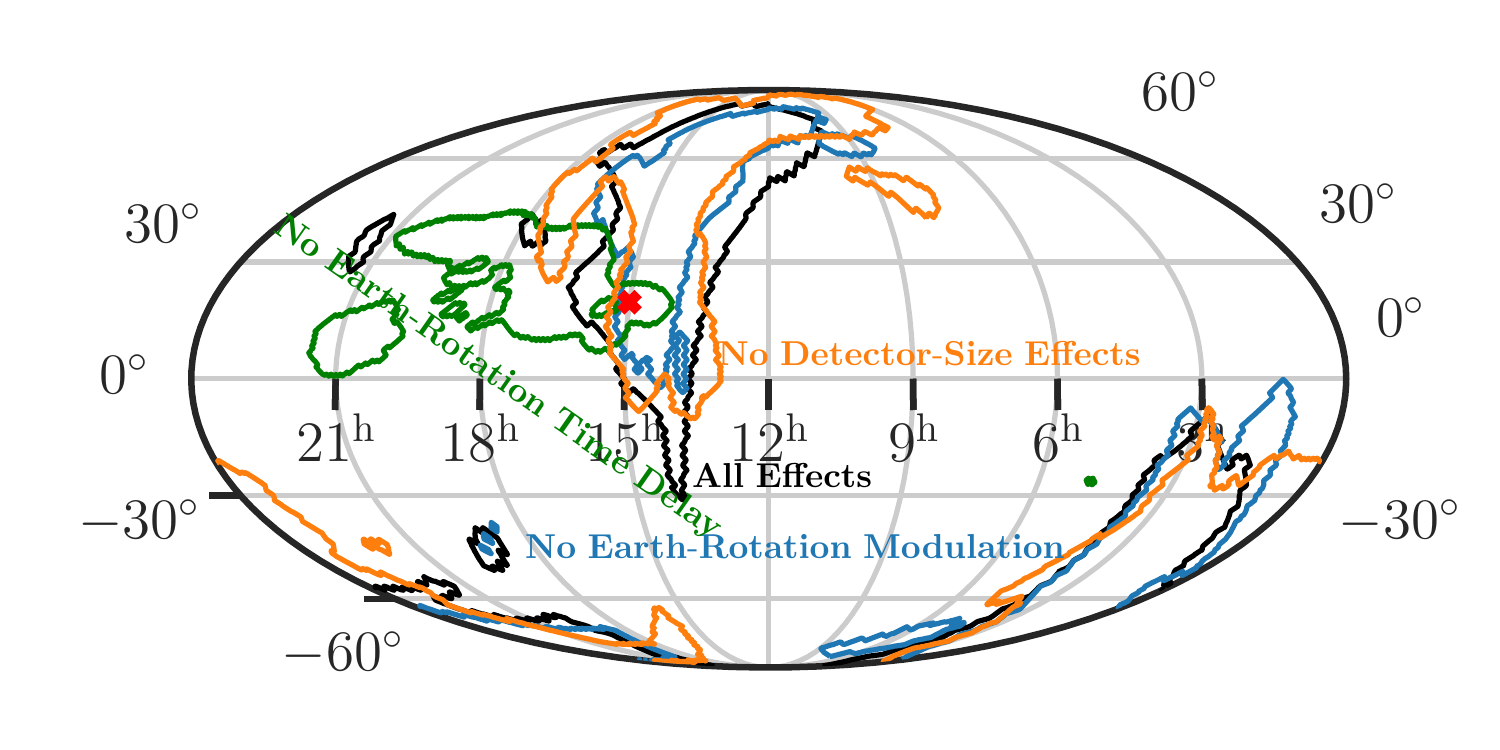}
    \end{subfigure}
     \hfill
     \begin{subfigure}[b]{0.49\textwidth}
         \centering
        \includegraphics[width=\textwidth]{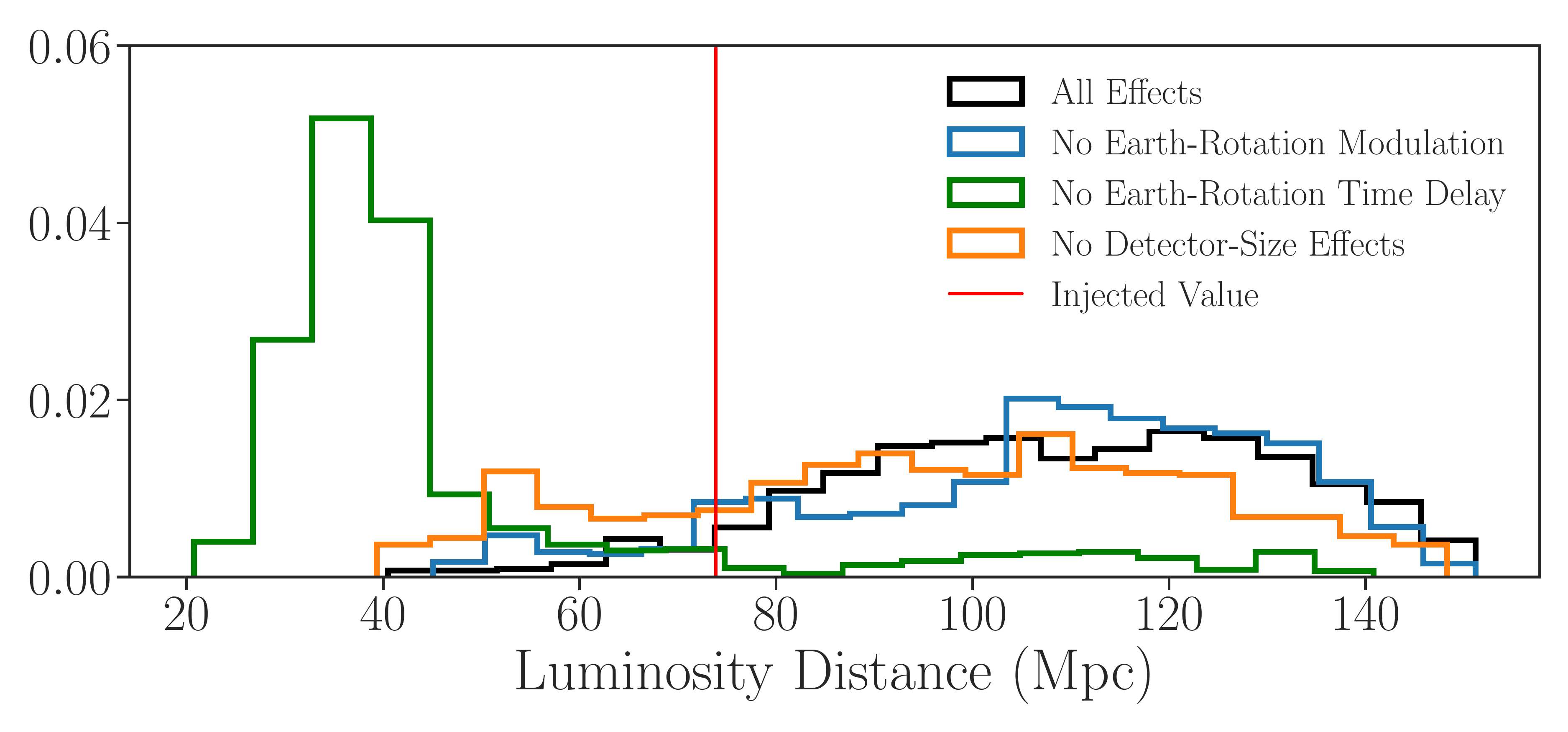}
     \end{subfigure}
     \hfill
     \caption{2D sky localization posterior (left) and inferred luminosity distance posterior (right) for the worst-localized event (marked by $\times$ in Figure \ref{fig:Sky}) with $\theta_{\rm JN}$ = 1.0. The black contour/ histogram  is for the inference considering all effects. The blue, orange, and green contours/ histograms ignore Earth-rotation amplitude modulation, detector-size amplitude modulation, and Earth-rotation time delay respectively. \phantom{Hope you enjoyed reading it!! Have a enjoyable and productive time ahead . Thanks and goodbye. }}
    \label{fig:b4}
\end{figure*}

To understand how the sky-area posteriors are affected due to ignoring a particular effect, we perform PE runs switching off one effect at a time.  The injections take into account all relevant effects as usual. We focus on three events:- two on the dark spot and one on the bright spot. Ignoring the Earth-rotation time delay shifts the posteriors of sky areas and luminosity distance, leading to biases for well-localized sources. Figure \ref{fig:b1} is for the best-localized case which is located on a dark spot along the equatorial plane and Figure \ref{fig:b3} is for a source located on the other dark spot. As discussed earlier the finite size of the detector plays a major role for sources on the dark spot and ignoring it leads to significant biases in the inferred posterior. For Figure \ref{fig:b3} all effects are important and ignoring any one of them leads to biases. For the worst-localized case in Figure \ref{fig:b4} switching off the amplitude modulation effects due to detector size or Earth-rotation creates very little difference as the waveform taking into account all effects and ignoring all effects are very similar.

\newpage
\bibliography{main}{}
\bibliographystyle{yahapj}
\end{document}